\documentclass[a4paper]{article}
\pdfoutput=1
\usepackage{jheppub}

\usepackage[utf8]{inputenc}

\usepackage[table ]{ xcolor}

\usepackage{multirow}

\usepackage{amsmath}

\usepackage{tikz}
\usetikzlibrary{shapes,arrows}
\tikzstyle{startstop} = [rectangle,rounded corners, minimum width=3cm,minimum height=1cm,text centered, draw=black,fill=red!30]
\tikzstyle{io} = [trapezium, trapezium left angle = 70,trapezium right angle=110,minimum width=3cm,minimum height=1cm,text centered,draw=black,fill=blue!30]
\tikzstyle{process} = [rectangle,minimum width=3cm,minimum height=1cm,text centered,text width =3cm,draw=black,fill=orange!30]
\tikzstyle{decision} = [diamond,minimum width=3cm,minimum height=1cm,shape aspect=3,inner sep = 0.4pt,text centered,draw=black,fill=green!30]
\tikzstyle{arrow} = [thick,->,>=stealth]
\tikzstyle{shadow}=[preaction={fill=black,opacity=.5,transform canvas={xshift=0.5mm,yshift=-0.5mm},shading=radial,shading angle=20},fill=red]

\tikzstyle{ellipse}=[draw, rectangle, minimum width=2.8em, rounded corners=6pt,line width=0.5pt]% minimum height=1.5em, fill=red!20 椭圆
\tikzstyle{pxsbx}=[trapezium, trapezium left angle=75, trapezium right angle=105, minimum width=3em, text centered, draw = black, fill=white,line width=0.5pt] %平行四边形
\tikzstyle{lingxing}=[draw,diamond,shape aspect=3,inner sep = 0.4pt,thick,font=\itshape,line width=0.5pt]%,minimum size=8mm 菱形

\usepackage{amssymb}
\usepackage{graphicx,color}

\usepackage{amsmath,amsthm}

\usepackage{mathrsfs}

\usepackage{bm}

\ifx\Ref\undefined
\newcommand{\Ref}[1]{(\ref{#1})}
\fi
\newcommand{\R}{\mathbb{R}}

\newcommand{\ccirc}{\kern0.2ex\vcenter{\hbox{$\scriptstyle\circ$}}\kern0.2ex}

\newcommand{\E}{\mathrm{e}}

\newcommand{\K}{{\text{K}_\gamma}}

\def\be{\begin{eqnarray}}
\def\ee{\end{eqnarray}}

\newcommand{\cc}{\mathcal C}

\newcommand{\ce}{\mathcal E}
\newcommand{\cf}{\mathcal F}

\newcommand{\ch}{\mathcal H}

\newcommand{\ck}{\mathcal K}
\newcommand{\cl}{\mathcal L}

\newcommand{\cp}{\mathcal P}

\newcommand{\cs}{\mathcal S}

\newcommand{\cv}{\mathcal V}

\newcommand{\fh}{\mathfrak{h}}

  \newcommand{\Fs}{\mathfrak{S}}

\renewcommand{\a}{\alpha}
\renewcommand{\b}{\beta}

\newcommand{\eps}{\varepsilon}

\newcommand{\sig}{\sigma}
\newcommand{\Sig}{\Sigma}
\renewcommand{\l}{\lambda}
\renewcommand{\L }{\Lambda}
\renewcommand{\o}{\omega}
\renewcommand{\O}{\Omega}
\renewcommand{\t}{\tau}

\newcommand{\rmd}{\mathrm d}

\newcommand{\lt}{\left}
\newcommand{\rt}{\right}

\newcommand{\lag}{\left\langle}
\newcommand{\rag}{\right\rangle}

\newcommand{\sgn}{\mathrm{sgn}}

\title{Improved Effective Dynamics of Loop-Quantum-Gravity Black Hole and Nariai Limit}

\author[1,2]{Muxin Han}  

\affiliation[1]{Department of Physics, Florida Atlantic University, 777 Glades Road, Boca Raton, FL 33431-0991, USA}

\affiliation[2]{Institut f\"ur Quantengravitation, Universit\"at Erlangen-N\"urnberg, Staudtstr. 7/B2, 91058 Erlangen, Germany}

\author[2]{\ Hongguang Liu}

\emailAdd{hanm(At)fau.edu}
\emailAdd{hongguang.liu(At)gravity.fau.de}
\abstract{We propose a new model of the spherical symmetric quantum black hole in the reduced phase space formulation. We deparametrize gravity by coupling to the Gaussian dust which provides the material coordinates. The foliation by dust coordinates covers both the interior and exterior of the black hole. After the spherical symmetry reduction, our model is a 1+1 dimensional field theory containing infinitely many degrees of freedom. The effective dynamics of the quantum black hole is generated by an improved physical Hamiltonian ${\bf H}_\Delta$. The holonomy correction in ${\bf H}_\Delta$ is implemented by the $\bar{\mu}$-scheme regularization with a Planckian area scale $\Delta$ (which often chosen as the minimal area gap in Loop Quantum Gravity). The effective dynamics recovers the semiclassical Schwarzschild geometry at low curvature regime and resolves the black hole singularity with Planckian curvature, e.g. $R_{\mu\nu\rho\sigma}R^{\mu\nu\rho\sigma}\sim 1/{\Delta}^2$. Our model predicts that the evolution of the black hole at late time reaches the charged Nariai geometry ${\rm dS}_2\times S^2$ with Planckian radii $\sim \sqrt{\Delta}$. This result is similar to the earlier work of the $\bar{\mu}$-scheme black hole \cite{Bohmer:2007wi} but is free of its problems. The Nariai geometry is stable under linear perturbations but may be unstable by nonperturbative quantum effects. Our model suggests the existence of quantum tunneling of the Nariai geometry and a scenario of black-hole-to-white-hole transition. During the transition, the linear perturbations exhibit chaotic dynamics with Lyapunov exponent $\lambda=2\pi T_{\rm dS}\sim \Delta^{-1/2}$ relating to the Hawking temperature $T_{\rm dS}$ of ${\rm dS}_2$. In addition, the Nariai geometry in our model provides an interesting example of Wheeler's bag of gold and contains infinitely many infrared soft modes with zero energy density. These infrared modes span a Hilbert space carrying a representation of 1-dimensional spatial diffeomorphisms (or the Witt/Virasoro algebra). The spatial diffeomorphisms are conserved charges of the effective dynamics by ${\bf H}_\Delta$. 
}


\begin{document}

\maketitle%\vspace{-7mm}

\section{Introduction}

The research on quantum black holes has recently made important progress in the framework of Loop Quantum Gravity (LQG) (see e.g. \cite{Ashtekar:2005qt,Modesto:2005zm,Bohmer:2007wi,Ashtekar:2010qz,Chiou:2012pg,Gambini:2013hna,Bianchi:2018mml,DAmbrosio:2020mut,Olmedo:2017lvt,Ashtekar:2018cay,Bojowald:2018xxu,Bodendorfer:2019cyv,Alesci:2019pbs,Assanioussi:2019twp,Kelly:2020lec,Gambini:2020nsf}, see also \cite{Ashtekar:2020ifw} for a recent review). LQG, as a background-independent and non-perturbative approach of quantum gravity, has the advantage for studying non-perturbative quantum effects in strong gravitational fields such as inside the black hole or the beginning of the universe. In particular, LQG leads to the success of resolving both black hole and big-bang singularities (see e.g. \cite{Bojowald:2001xe,Ashtekar:2006wn} for cosmology). In both cases, the classical curvature singularities are replaced by the non-singular bounces where their curvatures are finite and Planckian.

These developments of LQG black hole are based on symmetry reduced models. Instead of the full quantum theory of gravity, these models of black hole quantize spherical symmetric gravitational degrees of freedom (DOFs). They fall into 2 categories: The models in the first category e.g. \cite{Ashtekar:2005qt,Modesto:2005zm,Bohmer:2007wi,Ashtekar:2010qz,Olmedo:2017lvt,Ashtekar:2018cay,Bodendorfer:2019cyv,Assanioussi:2019twp} quantizes the black hole interior and exterior separately using the homogeneous Kantowski-Sachs slices, and they only quantize a finite number of DOFs due to the spherical symmetry and homogeneity. The models in the second category e.g. \cite{Chiou:2012pg,Gambini:2013hna,Alesci:2019pbs,Kelly:2020lec,Gambini:2020nsf} only reduce the DOFs by spherical symmetry, so these models are 1+1 dimensional field theories which still contain infinitely many DOFs, and they can treat both the black hole interior and exterior in a unified manner. Our approach in this work belongs to the second category.

The effective dynamics is one of important tools for studying black holes in LQG. The effective Hamiltonian of the black hole modifies the classical Hamiltonian in the canonical spherical symmetric gravity by including holonomy corrections to the Ashtekar-Barbero connection $A$, motivated by the LQG quantization. It is shown that in Loop Quantum Cosmology (LQC), the effective dynamics is able to accurately capture key features of the quantum dynamics \cite{Rovelli:2013zaa}. It is expected that the effective dynamics should also be valid for black hole, or at least be a crucial first step toward understanding full quantum effects of black hole. The effective dynamics from all the existing models lead to the black hole singularity resolution and bounce, but their detailed behaviors are different and depend on their schemes of the holonomy corrections. The choice of schemes are involved in these models due to the regularization/discretization of the Hamiltonian in which the curvature of $A$ is discretized by the loop holonomy around a plaquette. Here are a few examples of schemes used in these models: Among models in the first category, the earliest models \cite{Ashtekar:2005qt,Modesto:2005zm} follow the strategy similar to the $\mu_0$-scheme in LQC, and use plaquettes with constant fiducial scale. Their resulting behaviors of the bounce depend on the fiducial scale of plaquette. The model in \cite{Bohmer:2007wi} follows the similar strategy of the improved $\bar{\mu}$-scheme of LQC, where the scale of the plaquette is dynamical. This model removes the dependence on the fiducial scale, and interestingly relates the final state of black hole to the Nariai geometry ${\rm dS}_2\times S^2$ (see e.g. \cite{Bousso:1996pn} for earlier studies of the Nariai solution). But it has two problems (1) it gives large quantum effect near the event horizon where the spacetime curvature is small (this may be the consequence from choosing the Kantowski-Sachs foliation whose spatial slice approaching null near the horizon), and (2) the area of 2-sphere becomes too small in the evolution, making the model self-inconsistent with the $\bar{\mu}$-type regularization (see \cite{Ashtekar:2018cay} for the discussion). The more recent model \cite{Ashtekar:2018cay} applies a new regularization scheme. In this scheme, the fiducial scale is a conserved Dirac observable and remains constant along the trajectory of evolution, but it may change value along different trajectories. The models \cite{Chiou:2012pg,Gambini:2013hna,Kelly:2020lec,Gambini:2020nsf} in the second category applies the $\bar{\mu}$-type regularization, complemented by certain gauge fixing and choice of lapse function.  The foliations of these models are not Kantowski-Sachs but cover both the black hole interior and exterior. 

In this work, we propose a new model of the spherical symmetric LQG black hole and study the effective dynamics. The model is embedded in the framework of the reduced phase formulation. We deparametrize gravity by coupling it to the Gaussian dust \cite{Kuchar:1990vy,Giesel:2012rb}. The dust fields provide a material reference frame of the time and space. The dynamics in the reduced phase space is governed by a physical Hamiltonian, which in our case is identical to the Hamiltonian constraint with unit lapse. When reducing to the sector of spherical symmetric DOFs, we obtain an 1+1 dimensional Hamiltonian field theory describing the spherical symmetric gravity-dust system. Classical solutions of this theory are Lemaitre-Tolman-Bondi spacetimes (classically, our model closely relates to the earlier spherical symmetric reduced phase space model \cite{Giesel:2009jp}, see also \cite{Kelly:2020lec,Munch:2020czs} for recent works on coupling dust to black hole). We prefer the reduced phase space formulation because the dynamics is generated by the Hamiltonian which can make sense the unitarity when we pass to the quantum theory.

Our model applies the $\bar{\mu}$-scheme regularization (following \cite{Chiou:2012pg}) to include the holonomy correction to the physical Hamiltonian. The improved effective Hamiltonian ${\bf H}_\Delta$ depends on the Planckian area scale $\Delta\sim\ell_P$ which usually is set to be the minimal LQG area gap. The improved effective dynamics is given by solving the Hamiltonian equations of motion (EOMs) of ${\bf H}_\Delta$. The EOMs are a set of partial differential equations (PDEs) which we call the effective EOMs. The corrections of $O(\Delta)$ are understood as quantum corrections to the classical EOMs. As a reason of choosing the $\bar{\mu}$-scheme regularization, it has the nice properties that the improved effective dynamics from ${\bf H}_\Delta$ has infinitely many conserved charges corresponding to spatial diffeomorphisms, which play an interesting role in our discussion.

An important advantage of our model is that it includes infinitely many DOFs and describes the effective dynamics of both interior and exterior of the black hole in one set of effective EOMs. Indeed, in the case of zero dust density and classical limit $\Delta\to0$, the solution of effective EOMs is the Schwarzschild geometry in Lemaitre coordinate, which covers both the interior and exterior of the black hole. Classically the spatial slice of Lemaitre coordinate starts from the spatial infinity, crosses the event horizon, and ends at the singularity (the slice is further extended to another infinity by the singularity resolution in our model). Our model based on the reduced phase formulation is different from other ones in the second category. ${\bf H}_\Delta$ corresponds to the unit lapse function, and does not use the areal gauge fixing, as other 2 differences from \cite{Gambini:2013hna,Kelly:2020lec,Gambini:2020nsf}.

When solving the effective EOMs of ${\bf H}_\Delta$, we mainly focus on the situation with small dust density $\rho$. We firstly derive non-perturbatively the vacuum solution in the case of negligible $\rho$, then turn on nontrivial $\rho$ by including linear perturbations. The vacuum solution is obtained by the numerical method with high precision. Some features of this solution are highlighted below, while the details is given in Section \ref{From Schwarzschild black hole to dS2S2}.

\begin{enumerate}

\item The solution satisfies the semiclassical boundary condition, i.e. reduces to classical Schwarzschild geometry in Lemaitre coordinate near spatial infinity. The quantum correction are negligible in the low curvature regime.  

\item The black hole singularity is resolved and replaced by the non-singular bounce of the spatial volume element. The Kretschmann scalar $\ck\sim\Delta^{-2}$ at the bounce is Planckian as $\Delta\sim \ell_P^2$. The Lemaitre spatial slice, classically ending at the singularity, is now further extended to another infinity. 

\item After the bounce, the evolution stabilizes asymptotically to the charged Nariai limit ${\rm dS}_2\times S^2$ with different dS and $S^2$ radii. It is interesting that we obtain the Nariai limit similar to the result in \cite{Bohmer:2007wi}, although our model is very different from theirs (the model in \cite{Bohmer:2007wi} belongs to the first category while ours belongs to the second category). 

\item Although we have the similar result as in \cite{Bohmer:2007wi}, our model is free of its problems: Thanks to the foliation used in our approach, the geometry near the event horizon is semiclassical with negligible quantum correction. The area of 2-sphere never becomes smaller than $\Delta$ in the evolution, consistent with the $\bar{\mu}$-scheme regularization. 

\end{enumerate}

The charged Nariai limit ${\rm dS}_2\times S^2$ in our model is due to quantum effect. Both radii of dS and $S^2$ are of $O(\sqrt{\Delta})$. The appearance of the Nariai geometry is an interesting feature of the model, given that extensive studies in 90s suggesting the relation between the Nariai geometry and the black hole remanent e.g. \cite{Hawking:1995ap,Bousso:1999ms,Bousso:1996pn,Bousso:1996wz}. As a difference from early results on the Nariai limit, in our model the Nariai geometry ${\rm dS}_2\times S^2$ is stable when we turn on linear perturbations. This result is similar to \cite{Bohmer:2007wi,Boehmer:2008fz}. But we find evidence that the Nariai geometry is unstable by taking into account the non-perturbative quantum effect. The quantum tunneling effect may send ${\rm dS}_2\times S^2$ to $\widetilde{{\rm dS}_2\times S^2}$ with opposite time orientation. Then following the effective EOMs, $\widetilde{{\rm dS}_2\times S^2}$ decays to the Schwarzschild white hole spacetime with complete future timelike and null infinities. The solution evolving from $\widetilde{{\rm dS}_2\times S^2}$ to the white hole is the time reversal of the vacuum solution discussed above. The entire picture taking into account the black hole evaporation and the quantum tunneling proposes a scenario of black-hole-to-white-hole transition similar to \cite{DAmbrosio:2020mut,Bianchi:2018mml,Rovelli:2014cta}. The detailed discussions of the black-hole-to-white-hole transition and evidences of quantum tunneling is given in Sections \ref{Picture of black hole evaporation}, \ref{Black hole to white hole transition}, and \ref{Evidence of quantum tunneling}.

When we turn on linear perturbations on $\widetilde{{\rm dS}_2\times S^2}$, the perturbations exhibit chaotic dynamics where we can extract the Lyapunov exponent $\l=2\pi T_{\rm dS}$ relating to the Hawking temperature $T_{\rm dS}$ at the cosmological horizon of ${\rm dS}_2$ (see Section \ref{Chaos}). Moreover $\l\sim \Delta^{-1/2}$ indicates that this chaos is due to the quantum gravity effect. The relation between $\l$ and $ T_{\rm dS}$ resembles the known relation in the black hole butterfly effect \cite{Shenker:2013pqa,Maldacena:2015waa}.%, although here it gives a much stronger quantum scrambling since $\Delta\sim\ell_P^2$. 

As another interesting aspect of the Nariai limit ${\rm dS}_2\times S^2$, it is an example of Wheeler's bag of gold (see Section \ref{Infinitely many infrared states}). The foliation corresponds to the inflationary coordinate of ${\rm dS}_2$ and gives large space volume behind the event horizon of small area at the late time of the evaporation. When we turn on perturbations, we find that all perturbations become infinitely many infrared modes with zero energy density in ${\rm dS}_2\times S^2$. It is also the reason why ${\rm dS}_2\times S^2$ geometry is perturbatively stable. This result seems to suggests that ${\rm dS}_2\times S^2$ should have quantum degeneracy, and all the infrared modes should span a Hilbert space $\ch_{{\rm dS}_2\times S^2}$. In the reduced phase space formulation, the spatial diffeomorphisms are the symmetries of the physical Hamiltonian ${\bf H}_\Delta$ and give infinitely many conserved charges. Then $\ch_{{\rm dS}_2\times S^2}$ is the representation space of the spatial diffeomorphism group ${\rm Diff}(S^1)$ on the space of ${\rm dS}_2$. The Lie algebra of ${\rm Diff}(S^1)$ is the Witt algebra, or the Virasoro algebra if the central extension is considered.

Here is the organization of this paper: Section \ref{Reduced phase space formulation} discusses the preliminaries including the reduced phase space formulation with Gaussian dust and the spherical symmetric reduction. Section \ref{Improved Hamiltonian and equations of motion} constructs the improved effective Hamiltonian by the $\bar{\mu}$-scheme regularization. Section \ref{Effective equations of motion} discusses the effective EOMs. Section \ref{From Schwarzschild black hole to dS2S2} discusses the numerical black hole solution of the effective EOMs and the Nariai limit. Section \ref{Picture of black hole evaporation} takes into account the black hole evaporation and motivates the extension of the effective spacetime. \ref{Black hole to white hole transition} discusses the scenario of the black-hole-to-white-hole transition. \ref{Evidence of quantum tunneling} discusses the evidence of quantum tunneling from ${\rm dS}_2\times S^2$ to $\widetilde{{\rm dS}_2\times S^2}$. Section \ref{Chaos} demonstrates the chaos of perturbations on $\widetilde{{\rm dS}_2\times S^2}$. Section \ref{Infinitely many infrared states} discusses the relation between the Nariai limit and Wheeler's bag of gold, and shows that ${\rm dS}_2\times S^2$ contains infinitely many infrared states. Section \ref{outlook} discusses some future perspectives and open questions.

\section{Reduced phase space formulation}\label{Reduced phase space formulation}

\subsection{Deparametrized gravity with Gaussian dust} 

The reduced phase space formulation couples gravity to clock fields at classical level. In this paper, we mainly focus on the scenario of gravity coupled to Gaussian dust \cite{Kuchar:1990vy,Giesel:2012rb}. The action is given by 
\be
    S  = S_{\rm GR} + S_{\rm GD},
\ee
where $S_{\rm GR}$ is the Host action of gravity \cite{holst}
\be 
S_{\rm GR}\lt[e^\mu_{I},\Omega_{\mu \nu}^{IJ}\rt] =\frac{1}{16\pi G} \int_{M} d^{4}y\, \sqrt{|\operatorname{det}(g)|}\, e_{I}^{\mu} e_{J}^{\nu}\left(\Omega_{\mu \nu}^{IJ}+\frac{1}{2 \beta} \epsilon_{\ \ \ KL}^{IJ} \Omega_{\mu \nu}^{KL}\right)
\ee 
where the tetrad $e^\mu_{I}$ determines the 4-metric by $g_{\mu\nu}=\eta_{IJ}e^\mu_{I}e^\nu_{J}$, and $\Omega_{\mu \nu}^{IJ}$ is the curvature of the so(1,3) connection $\o_\mu^{IJ}$. $\b$ is the Barbero-Immirzi parameter. 
$S_{\rm GD}$ is the action of the Gaussian dust:
\be
S_{\rm GD}\left[\rho, g_{\mu \nu}, T, S^{j}, W_{j}\right]=-\int_M \mathrm{d}^{4} y \sqrt{|\operatorname{det}(g)|}\left[\frac{\rho}{2}\left(g^{\mu \nu} \partial_{\mu} T \partial_{\nu} T+1\right)+g^{\mu \nu} \partial_{\mu} T\left(W_{j} \partial_{\nu} S^{j}\right)\right]
\ee
where $T,S^{j=1,2,3}$ are clock fields and defines time and space coordinates in the dust reference frame. $\rho,W_{j}$ are Lagrange multipliers. The energy-momentum tensor of the Gaussian dust is 
\be
T_{\mu\nu}=\rho U_\mu U_\nu-U_{(\mu}  W_{\nu)},\quad U_\mu=-\partial_{\mu} T,\quad W_{\nu}=W_{j}\partial_{\nu} S^{j}.
\ee
which indicates that $\rho$ is the energy density and $W_\mu$ relates to the heat-flow \cite{Kuchar:1990vy}.

We assume $M\simeq \R\times \Sig$ and and make Legendre transform of dust variables:
\be
\begin{aligned}
P &:=\frac{\delta{S}_{\mathrm{GD}}}{\delta \dot{T}}=\sqrt{\operatorname{det}(q)}\left\{\rho\left[\cl_{n} T\right]+W_{j}\left[\cl_{n} S^{j}\right]\right\} \\
P_{j} &:=\frac{\delta{S}_{\mathrm{GD}}}{\delta \dot{S}^{j}}=\sqrt{\operatorname{det}(q)} W_{j}\left[\cl_{n} T\right] \\
\pi &:=\frac{\delta S_{\mathrm{GD}}}{\delta \dot{\rho}}=0 \\
\pi^{j} &:=\frac{\delta S_{\mathrm{GD}}}{\delta \dot{W}_{j}}=0
\end{aligned}
\ee
where $q_{\a\b}$ ($\a,\b=1,2,3$) is the 3-metric and $\cl_n$ denotes the Lie derivative along the normal to the hypersurface $\Sig$. The constraint analysis \cite{Kuchar:1990vy,Giesel:2012rb} results in Hamiltonian and diffeomorphism constraints $\cc^{\rm tot},\cc^{\rm tot}_\a$ which are first-class constraints, and 8 second-class constraints $z,z^j,\zeta_1,\zeta_2,s,K$:
\be
z&=&\pi,\quad z^{j}=\pi^{j},\quad\zeta_{1}=W_{1}P_{2}-{W_{2}} P_{1},\quad \zeta_{2}:=W_{1}P_{3}-W_{3} P_{1},\\
s&=&-\frac{1}{\sqrt{\det(q)}}P_{1}{}^{2}+\sqrt{\det(q)}\left(q^{\a\b} T_{, \a} T_{, \b}+1\right)W_{1}{}^2,\\
K&=&-\frac{P P_{1}{}^2W_{1}}{ \sqrt{\det(q)}}+\frac{\rho}{\sqrt{\det(q)}} P_{1}{}^{3}+\sqrt{\det(q)} W_1{}^3 q^{\a \b} T_{, \a}\left(P_{j} S_{, \b}^{j}\right)
\ee
where $T_{,\a}\equiv\partial_\a T$. Solving second-class constraints gives 
\be
W_j&=&\frac{P_j}{\sqrt{\det(q)}\left(q^{\alpha\beta}T_{,\alpha}T_{,\beta}+1\right)^{1/2}}\\
\rho&=&\frac{P}{\sqrt{\det(q)}\left(q^{\alpha\beta}T_{,\alpha}T_{,\beta}+1\right)^{1/2}}-\frac{q^{\alpha\beta}T_{,\alpha}\left(P_{j}S_{,\beta}^{j}\right)}{\sqrt{\det(q)}\left(q^{\alpha\beta}T_{,\alpha}T_{,\beta}+1\right)^{3/2}}\label{rho111}
\ee
by a choice of sign in the ratio between $W_j,P_j$. These relations simplifies $\cc^{\rm tot},\cc^{\rm tot}_\a$ to equivalent forms:
\be
\cc^{\mathrm{tot}}&=&P+ h,\quad h= \cc \sqrt{1+q^{\a \b} T_{, \a} T_{, \b}}-q^{\a \b} T_{, \a} \cc_{\b},\label{cc111}\\
\cc_{\a}^{\mathrm{tot}}&=&\cc_{\a}+P T_{, \a}+P_{j} S_{, \a}^{j}
\ee
where $\cc,\cc_a$ are pure gravity Hamiltonian and diffeomorphism constraints from $S_{\rm GR}$.

$S_{\rm GR}$ leads to gravity canonical variables $A^a_\a(y),E^\a_a(y)$, where $A^a_\a(y)$ is the Ashtekar-Barbero connection and $E^\a_a(y)=\sqrt{\det q}\, e^\a_a(y)$ is the densitized triad. $a=1,2,3$ is the Lie algebra index of su(2). Dirac observables are constructed relationally by parametrizing $(A,E)$ with values of dust fields $T(y)\equiv t,S^j(y)\equiv\sig^j$, i.e. $A_j^a(\sig,t)=A_j^a(y)|_{T(y)\equiv t,\,S^j(y)\equiv\sig^j}$ and $E^j_a(\sig,t)=E^j_a(y)|_{T(y)\equiv t,\,S^j(y)\equiv\sig^j}$, where $\sig, t$ are physical space and time coordinates in the dust reference frame. Here $j=1,2,3$ is the dust coordinate index (e.g. $A_j(y)=A_\a(y) \partial y^\a/\partial\sig^j$). $A_j^a(\sig,t),E^j_a(\sig,t)$ depending only on values of dust fields are independent of gauge choices of coordinates $y$. They are proven to be invariant (on the constraint surface) under gauge transformations generated by diffeomorphism and Hamiltonian constraints \cite{Dittrich:2004cb,Thiemann:2004wk,Giesel:2007wn}. Moreover $A_j^a(\sig,t),E^j_a(\sig,t)$ satisfy the standard Poisson bracket in the dust frame: 
\be
\{E^i_a(\sig,t),A_j^b(\sig',t)\}=-\frac{1}{2}\kappa \b\ \delta^{i}_j\delta^b_a\delta^{3}(\sig,\sig')
\ee 
where $\b$ is the Barbero-Immirzi parameter and $\kappa=16\pi G$. $A_j^a(\sig,t),E^j_a(\sig,t)$ are the conjugate pair in the reduced phase space $\cp_{red}$.

The evolution in physical time $t$ is generated by the physical Hamiltonian ${\bf H}_0$ given by integrating $h$ on the constant $T(y)=t$ slice $\cs$. The constant $T$ slice $\cs$ is coordinated by the value of dust scalars $S^j=\sig^j$ thus is referred to as the dust space \cite{Giesel:2007wn,Giesel:2012rb}. $T_{, \a}=0$ on $\cs$ leads to 
\be
{\bf H}_0=\int_\cs\rmd^3\sig\, \cc(\sig)
\ee 
${\bf H}_0$ formally coincides with smearing the gravity Hamiltonian $\cc$ with the unit lapse, while here $\cc(\sig)$ is in terms of Dirac observables $A_j^a(\sig),E^j_a(\sig)$ and $\sig^{j=1,2,3}$ are dust coordinates on $\cs$. The dynamics is governed by
\be
\frac{\rmd f}{\rmd t}=\{f,\,{\bf H}_0\}.
\ee 
for all functions $f$ on the reduced phase space $\cp_{red}$.

The recent result in \cite{Han:2020chr} leads to an understanding of dusts or other clock fields from the LQG point of view, particularly about if dusts are valid in the quantum regime. The altitude is that when we quantize $\cp_{red}$ to obtain the physical Hilbert space and define $\hat{\bf H}$ as the quantization of ${\bf H}_0$, the quantum theory should be taken as the fundamental theory and starting point of discussions. Although the quantum theory is formally obtained by quantizing the classical theory, the classical theory is not fundamental but emergent from the fundamental quantum theory. From the quantum point of view, both classical gravity and dust are low-energy effective degrees of freedom produced from the quantum theory via the semiclassical approximation, as demonstrated in \cite{Han:2020chr}. Both classical gravity and dusts are not fundamental and not valid in the quantum regime but emergent at low energy, while what valid in the quantum regime are $\hat{\bf H}$ and the physical Hilbert space. ${\bf H}_\Delta$ to be constructed in Section \ref{Improved Hamiltonian and equations of motion} is expected to describe the quantum effective theory of $\hat{\bf H}$.

\subsection{Spherical symmetric reduction}

We assume the dust space $\cs\simeq \R\times S^2$ and define the spherical coordinate ${\sig}=(x,\theta,\phi)$. When the reduced phase space $\cp_{red}$ is further reduced to the phase space $\cp_{sph}$ of spherical symmetric field configurations, the symplectic structure reduces to 
\be
\O&=&-\frac{2}{\kappa\b}\int \rmd^3\sig\lt[\delta A_j^a(\sig)\wedge\delta E^j_a(\sig)\rt]\nonumber\\
&=&-\frac{2}{\kappa\b}\int \rmd^3\sig\lt[\delta A_1^1(\sig)\wedge\delta E^1_1(\sig)+\delta A_2^2(\sig)\wedge\delta E^2_2(\sig)+\delta A_3^3(\sig)\wedge\delta E^3_3(\sig)\rt]\nonumber\\
&=&-\frac{2}{\kappa\b}\int \rmd x\int_0^\pi\rmd \theta\int_0^{2\pi}\rmd\varphi \sin(\theta)\lt[2\b \delta  K_x(x)\wedge\delta E^{x}(x) + 2\b \delta K_\varphi(x)\wedge\delta E^{\varphi}(x) \rt]\nonumber\\
&=&-\frac{16\pi}{\kappa}\int \rmd x \lt[\delta  K_x(x)\wedge\delta E^{x}(x) + \delta K_\varphi(x)\wedge\delta E^{\varphi}(x) \rt],\label{symplectic}
\ee
where in the second step we fix the gauge such that $E^j_a(\sig)$ is diagonal (details of this gauge fixing and solving Gauss constraint can be found in \cite{Chiou:2012pg,Gambini:2013hna}). $E^{x}(x),E^{\varphi}(x),K_{x}(x),K_{\varphi}(x)$ relates to nonvanishing components of $E^j_a(\sig), A_j^a(\sig)$ by 
\be
&E_{1}^{1}(\sig)=E^{x}(x) \sin (\theta), \quad E_{2}^{2}(\sig)=E^{\varphi}(x) \sin (\theta), \quad E_{3}^{3}(\sig)=E^{\varphi}(x),&\\
&A^1_1(\sig)=2\b K_x(x),\quad A^2_2(\sig)=\b K_\varphi(x),\quad A^3_3(\sig)=\b K_\varphi(x)\sin(\theta),&\\
&A^1_3(\sig)=\cos(\theta),\quad A^2_3(\sig)=-\sin(\theta)\frac{E^{x}{}'(x)}{2E^{\varphi}(x)},\quad A^3_2(\sig)=\frac{E^{x}{}'(x)}{2E^{\varphi}(x)}.&
\ee
Here $K_x$ is $1/2$ of the extrinsic curvature along $x$. The canonical conjugate pairs in $\cp_{sph}$ are \cite{Gambini:2013hna}
\be
\left\{K_{x}(x), E^{x}\left(x^{\prime}\right)\right\}&=&G \delta\left(x-x^{\prime}\right) \\
\left\{K_{\varphi}(x), E^{\varphi}\left(x^{\prime}\right)\right\}&=&G \delta\left(x-x^{\prime}\right).
\ee

The physical Hamiltonian reduced to $\cp_{sph}$ is expressed as
\be
\mathbf{H}_0&=&\int dx\, \cc(x),%+{\bf H}_{bdy}
\label{hplusboundary}\\
\cc(x)&=&\frac{4\pi}{\kappa}\frac{\text{sgn}(E^{\varphi}{})}{\sqrt{\left|E^{x}{}\right|}}\Bigg(-\frac{2E^{x}{}E^{x}{}^\prime{}E^{\varphi}{}^\prime{}}{E^{\varphi}{}^{2}}+\frac{4E^{x}{}E^{x}{}^{\prime\prime}{}+E^{x}{}^\prime{}^{2}}{2E^{\varphi}{}} -8E^{x}{}K_{x}{}K_{\varphi}{}-2E^{\varphi}{}\left[K_{\varphi}{}^{2}+1\right]\Bigg).\label{hplusboundary1} 
\ee
where e.g. $E^{x}{}^\prime\equiv \partial_x E^{x}$. 

The time evolution of any observable $f$ on $\cp_{sph}$ is given by $\rmd f/\rmd t=\{f,{\bf H}_0\}$ where the Poisson bracket is from (\ref{symplectic}). Comparing $\delta {\bf H}_0=\int_\cs\rmd^3\sig \delta\cc(\sig)$ to the standard pure gravity Hamiltonian $\delta H_{\rm GR}=\int_\cs\rmd^3\sig( N\delta\cc+N^j\delta\cc_j )$. The dust coordinates correspond to the lapse function $N=1$ and zero shift. Any solution $E^x(x),E^\varphi(x)$ of equations of motion (EOMs) can construct a 4d metric by expressing in the dust frame ($t,x,\theta,\varphi$)
\be
\rmd s^2=-\rmd t^2+\L(t,x)^2\rmd x^2+R(t,x)^2\lt[\rmd\theta^2+\sin^2(\theta)\rmd\varphi^2\rt],\quad \L=\frac{E^\varphi}{\sqrt{|E^x|}},\quad R=\sqrt{|E^x|}.\label{dustmetric}
\ee
General solutions of the EOMs by ${\bf H}_0$ give the Lema\^{\i}tre-Tolman-Bondi (LTB) metric \cite{Giesel:2009jp}
\be
\L(x)^2=\frac{\lt[\partial_xR(t,x)\rt]^2}{1+\ce(x)},\quad \partial_t R(t,x)=\pm\sqrt{\ce(x)+\frac{\cf(x)}{R(t,x)}}
\ee
with arbitrary real functions $\ce(x),\cf(x)$. When $\ce(x)>0$, it relates to the energy per unit mass of the dust particles. $\cf(x)>0$ relates to the gravitational mass within the sphere at radius $x$. Solutions can be classified into 3 cases $\ce(x)>0$, $\ce(x)=0$, or $\ce(x)<0$:
\be
\ce(x)>0:&&\quad R\left(t, x\right) =\frac{\cf\left(x\right)}{2 \ce\left(x\right)}\lt[\cosh (\eta)-1\rt], \quad\sinh (\eta)-\eta =\frac{2\left[\ce\left(x\right)\right]^{\frac{3}{2}}\left(\beta\left(x\right)-t\right)}
{\cf\left(x\right)},\label{positiveE}\\
\ce(x)=0:&&\quad R\left(t, x\right)=\left[\frac{3}{2} \sqrt{\cf(x)}\left(\beta\left(x\right)-t\right)\right]^{2 / 3},\label{zeroE}\\
\ce(x)<0:&&\quad R\left(\tau, x\right) =\frac{\cf\left(x\right)}{2|\ce\left(x\right)|}\lt[1-\cos (\eta)\rt], \quad \eta-\sin (\eta) =\frac{2|\ce\left(x\right)|^{\frac{3}{2}}\left(\beta\left(x\right)-t\right)}{\cf\left(x\right)},\label{negativeE}
\ee
where $\b(x)$ is an arbitrary function. $t=\b(x)$ is the singularity of the metric. The solution at $\ce=0$ and constant $\cf=R_s=2GM$ ($\b(x)=x$) is the Schwarzschild metric in Lema\^{\i}tre coordinates.

It is necessary to discuss the boundary condition and boundary term in the physical Hamiltonian, since the dust space is noncompact. But we postpone this discussion to Section \ref{Improved Hamiltonian and equations of motion}.

The time evolution by ${\bf H}_0$ has infinitely many conserved charges $\cv(N)$ from spatial diffeomorphisms,
\be
\cv(N)&=&\int\rmd x N(x)\cc_x(x)%+\cv_{bdy},
\quad \lt\{{\bf H}_0,\,\cv(N)\rt\}=0\\
\cc_x(x)&=&E^\varphi(x) K_\varphi'(x)-K_x(x) {E^x}'(x),%\quad \cv_{bdy}=-NE^\varphi K_\varphi|_{bdy}
\ee
for all $N(x)$ (vanishing at boundary). Moreover since the poisson bracket $\{{\bf H}_0,\cc(x)\}$ is proportional to $\cc_x(x)$, $\cc(x)$ becomes infinitely many additional conserved charges when the initial value of the dynamics satsifies $\cc_x(x)=0$.

%By the boundary condition $\delta E^x|_{bdy}=\delta E^\varphi|_{bdy}=0$, the boundary term $\cv_{bdy}$ cancels the total derivative in $\delta\cv(N)$. 

%\section{Holonomy corrections and improved effective dynamics}

\section{Improved Hamiltonian}\label{Improved Hamiltonian and equations of motion}

Recall relations between $K_x,K^\varphi$ and Ashtekar-Barbero connection: $A^1_1(\sig)=2\b K_x(x),\  A^2_2(\sig)=\b K_\varphi(x),\ A^3_3(\sig)=\b K_\varphi(x)\sin(\theta)$, we following \cite{Gambini:2020nsf,Kelly:2020lec,Chiou:2012pg} to modify $K_\varphi,K_x(x)$ by including a deformation parameter $\Delta\sim l_P^2=G\hbar$ which may be chosen as the minimal area gap in LQG: 
\be
K_\varphi(x)\to \frac{\sqrt{|E^x|}}{\b\sqrt{\Delta}}\sin\lt[\frac{\sqrt{\Delta}}{\sqrt{|E^x|}}\b K_\varphi(x)\rt], \quad
K_x(x)\to \frac{E^\varphi}{2\b\sqrt{\Delta}\sqrt{|E^x|}}\sin\lt[\frac{\sqrt{\Delta}\sqrt{|E^x|}}{{E^\varphi}}2\b K_x(x)\rt].\label{mubarKx}
\ee
Clearly as $\Delta\to0$, \Ref{mubarKx} reduces to $K_x$. To motivate these modifications, let's construct SU(2) holonomies of $A_1^1$, $A_2^2$ and $A^3_3$ along edges in $x$, $\theta$, and $\varphi$ directions:

Firstly let $e$ be an edge along the $x$-axis toward the positive direction. We define the holonomy $h_\Delta(A_1^1)$ along $e$ as below, and change parametrization of $e$ from $x$ to $u$ such that $u\in [0,1]\mapsto e(u)\in\cs$.
\be
h_\Delta(A_1^1)=\cp\exp\lt[-i\int_e\rmd x\,A_1^1(x)\frac{\sig^1}{2}\rt]=\cp\exp\lt[-i\int_0^1\rmd u\frac{\rmd x}{\rmd u}
\,A_1^1(x(u))\frac{\sig^1}{2}\rt], 
\ee
where $\sigma^{a=1,2,3}$ are Pauli matrices. If we set $\frac{\rmd x}{\rmd u}=|\sqrt{\Delta}\L^{-1}|=\frac{\sqrt{\Delta}\sqrt{|E^x|}}{{|E^\varphi|}}\equiv\bar{\mu}_x$ (recall Eq.\Ref{dustmetric}), the length of $e$
\be
\int_0^1\rmd u\lt|\frac{\rmd x}{\rmd u}\rt|\sqrt{q_{xx}}=\sqrt{\Delta}\sim l_P, \quad q_{xx}=\L^2.
\ee
is fixed to $\sqrt{\Delta}$. Therefore $h_\Delta(A_1^1)$ along the fixed-length edge $e$ in the $x$-direction can be written as
\be
h_\Delta(A_1^1)=\cp\exp\lt[-i\int_0^1\rmd u\frac{\sqrt{\Delta}\sqrt{|E^x|}}{{|E^\varphi|}}\,2\b K_x(x(u))\frac{\sig^1}{2}\rt]\simeq \exp\lt[-i\frac{\sqrt{\Delta}\sqrt{|E^x|}}{{|E^\varphi|}}\,2\b K_x\frac{\sig^1}{2}\rt],
\ee
where we assume the fields are approximately constant along $e$ whose length is $\sqrt{\Delta}\sim l_P$. In the result of $h_\Delta(A_1^1)$, $E^x,E^\varphi,K_x$ are evaluated at the starting point of $e$. %we have
%\be
%\sin\lt[\int_e\rmd x\,A_1^1(x)\rt]\simeq \sin\lt[\frac{\sqrt{\Delta}\sqrt{|E^x|}}{{E^\varphi}} 2\b K_x\rt]
%\ee
%which relates to Eq.\Ref{mubarKx}. 

We define the holonomy $h_\Delta(A^2_2)$ to be along an edge $e$ toward the $\theta$-direction. Again by changing parametrization of $e$ from $\theta$ to $s$ such that $s\in [0,1]\mapsto e(v)\in\cs$, and noticing that $A_2^2=\b K_\varphi$ is independent of $\theta$
\be
h_\Delta(A^2_2)=\exp\lt[-i\int_e\rmd \theta\,A_2^2\frac{\sig^2}{2}\rt]=\exp\lt[-i\int_0^1\rmd s\frac{\rmd \theta}{\rmd s}\,A_2^2\frac{\sig^2}{2}\rt], 
\ee
If we let $\frac{\rmd \theta}{\rmd s}=\sqrt{\Delta} R^{-1}={\sqrt{\Delta}}/\sqrt{|E^x|}\equiv\bar{\mu}_\theta$, the length of $e$ is fixed to be $\sqrt{\Delta}$:
\be
\int_0^1\rmd s\lt|\frac{\rmd \theta}{\rmd s}\rt|\sqrt{q_{\theta\theta}}=\sqrt{\Delta},\quad q_{\theta\theta}=R^2.
\ee
Therefore the U(1) holonomy along a fixed-length edge $e$ in the $\theta$-direction can be written as
\be
h_\Delta(A^2_2)=\exp\lt[-i\frac{\sqrt{\Delta}}{\sqrt{|E^x|}}\b K_\varphi\frac{\sig^2}{2}\rt]
\ee
where $E^x,E^\varphi,K_x$ are evaluated at the starting point of $e$. %Assuming fields are approximately constant along $e$ whose length is $O(l_P)$, 
%\be
%\sin\lt[\int_e\rmd \theta\,A_2^2(\theta)\rt]\simeq \sin\lt[\frac{\sqrt{\Delta}}{\sqrt{|E^x|}} \b K_\varphi\rt].\label{kvarphimubar}
%\ee
The holonomy $h_\Delta(A^3_3)$ along $\varphi$ direction can be derived similarly. We let $\frac{\rmd \varphi}{\rmd s}=\sqrt{\Delta}/\sqrt{q_{\varphi\varphi}} ={\sqrt{\Delta}}/(\sqrt{|E^x|}\sin(\theta))\equiv\bar{\mu}_\varphi/\sin(\theta)$ where $\bar{\mu}_\varphi=\bar{\mu}_\theta$ is obtained.
\be
&&h_\Delta(A^3_3)=\exp\lt[-i\int_e\rmd \varphi\,A_3^3\frac{\sig^3}{2}\rt]%=\exp\lt[i\int_0^1\rmd s\frac{\sqrt{\Delta}}{\sqrt{|E^x|}\sin(\theta)}\,A_3^3(\varphi(s))\rt]
=\exp\lt[-i\frac{\sqrt{\Delta}}{\sqrt{|E^x|}}\,\b K_\varphi\frac{\sig^2}{2}\rt].%\\
%&&\sin\lt[\int_e\rmd \varphi\,A_3^3(\varphi)\rt]\simeq \sin\lt[\frac{\sqrt{\Delta}}{\sqrt{|E^x|}} \b K_\varphi\rt],
\ee
%give the same result as in Eq.\Ref{kvarphimubar}.
These fixed-edge-length holonomies are analogs of dressed holonomies used in $\bar{\mu}$-scheme in LQC \cite{Ashtekar:2006wn,Han:2019feb}. Therefore the modification \Ref{mubarKx} is called the ``$\bar{\mu}$-scheme'' of the spherical symmetric LQG. Note that the holonomies $h_\Delta(A^1_1)$, $h_\Delta(A^2_2)$, $h_\Delta(A^3_3)$ only capture components relating to $K_x,K_\varphi$ in the Ashtekar-Barbero connection, but do not capture the spin-connection components $A^1_3,A^2_3,A^3_2$. This choice of treatment is often referred to as the K-holonomy regularization, and has often been used in LQC and black holes \cite{Singh:2013ava,Bojowald:2005cb,Gambini:2013hna,Kelly:2020lec}.

The modification \Ref{mubarKx} can be obtained from regularizing the curvature $F(A)=\rmd A+A\wedge A$ in the LQG Hamiltonian with $ F\simeq\frac{1}{\Delta}[h_\Delta(\Box)-1]$ where $\Box$ is the elementary plaquette with fixed area $\Delta$ \cite{Chiou:2012pg,Bojowald:2005cb}. The plaquette is in a cubic lattice adapted to the dust coordinate $(x,\theta,\varphi)$. Let coordinate lengths of edges of $\Box$ are $\bar{\mu}_x,\bar{\mu}_\theta,\bar{\mu}_\varphi/\sin(\theta)$ (the edge along $\varphi$ has constant $\theta$) as defined above, they indeed give the fixed area to every plaquette $\Box$:
\be
\left(\bar{\mu}_{x} \sqrt{q_{x x}}\right)\left(\bar{\mu}_{\theta} \sqrt{q_{\theta\theta}}\right) =\Delta, \quad
\left(\bar{\mu}_{\theta} \sqrt{q_{\theta\theta}}\right)\lt(\bar{\mu}_{\varphi} \sqrt{q_{\varphi\varphi}}/\sin(\theta)\rt) =\Delta,\quad \left(\bar{\mu}_{x} \sqrt{q_{xx}}\right)\lt(\bar{\mu}_{\varphi} \sqrt{q_{\varphi\varphi}}/\sin(\theta)\rt) =\Delta.\nonumber
\ee
$h_\Delta(\Box) $ is made by 4 holonomies along edges with fixed length $\sqrt{\Delta}$:
\be
h_\Delta(\Box_{jk})= h_\Delta(A^j_j)h_\Delta(A^k_k)h_\Delta(A^j_j)^{-1}h_\Delta(A^k_k)^{-1},\quad j,k=1,2,3.
\ee
Then the physical Hamiltonian ${\bf H}_0$ can be regularized by inserting $h(\Box)$ in the Euclidean Hamiltonian $\mathrm{Tr}(F_{ij}[E^j,E^k])/\sqrt{\det(q)}$ (where $F_{ij}=F_{ij}^a\frac{-i\sigma^a}{2},\ E^j=E^j_a\frac{-i\sigma^a}{2}$):
\be
{\bf H}_0&\to&\frac{2}{\b^2\kappa\Delta}\int\rmd^3x\sum_{i,j}e(\Box_{jk})\mathrm{Tr}\lt(h_\Delta(\Box_{jk})\frac{[E^j,E^k]}{\sqrt{\det(q)}}\rt)+\text{terms indep. of}\ K\nonumber\\
&=&\int\rmd x\frac{4\pi}{\kappa}\frac{1}{\sqrt{\left|E^{x}{}\right|}}\Bigg(-\frac{4E^{x}E^\varphi}{\b^2\Delta}\sin\lt[\frac{\sqrt{\Delta}\sqrt{|E^x|}}{{|E^\varphi|}}2\b K_x(x)\rt]\sin\lt[\frac{\sqrt{\Delta}}{\sqrt{|E^x|}}\b K_\varphi(x)\rt]\nonumber\\
&&-\frac{2|E^{\varphi}|{|E^x|}}{\b^2{\Delta}}\sin^2\lt[\frac{\sqrt{\Delta}}{\sqrt{|E^x|}}\b K_\varphi(x)\rt]\Bigg)+\text{terms indep. of}\ K,\label{Kregularization}
\ee
where $e(\Box_{jk})$ denotes the area element on $\Box_{ij}$ (see \cite{Han:2019feb}). The terms independent of $K$ in Eq.\Ref{hplusboundary1} are contributions from the spin-connection compatible to $E_a^j$, and are kept unchange. Comparing the above result to the terms depending on $K$ in Eq.\Ref{hplusboundary1} justifies the replacement \Ref{mubarKx}. 

We define the improved Hamiltonian ${\bf H}_\Delta$ to be ${\bf H}_0$ modified by the replacement \Ref{mubarKx}:
\be
\mathbf{H}_\Delta&=&\int_{-\infty}^\infty \rmd x\, \cc_\Delta(x),%+{\bf H}_{bdy}
\label{HDelta1}\\
\cc_\Delta(x)&=&\frac{4\pi}{\kappa}\frac{\text{sgn}(E^{\varphi}{})}{\sqrt{\left|E^{x}{}\right|}}\Bigg(-\frac{2E^{x}{}E^{x}{}^\prime{}E^{\varphi}{}^\prime{}}{E^{\varphi}{}^{2}}+\frac{4E^{x}{}E^{x}{}^{\prime\prime}{}+E^{x}{}^\prime{}^{2}}{2E^{\varphi}{}} \nonumber\\
&&-\frac{4E^{x}E^\varphi}{\b^2\Delta}\sin\lt[\frac{\sqrt{\Delta}\sqrt{|E^x|}}{{E^\varphi}}2\b K_x(x)\rt]\sin\lt[\frac{\sqrt{\Delta}}{\sqrt{|E^x|}}\b K_\varphi(x)\rt]\nonumber\\
&&-\frac{2E^{\varphi}{|E^x|}}{\b^2{\Delta}}\sin^2\lt[\frac{\sqrt{\Delta}}{\sqrt{|E^x|}}\b K_\varphi(x)\rt]-2E^{\varphi}\Bigg). \label{HDelta2}
\ee
Entries of the above sine functions are restricted into a single period $(-\pi,\pi]$. $\mathbf{H}_\Delta$ is the same as \Ref{Kregularization}. %as $E^\varphi>0$ which we always assume in this paper\footnote{Rigorously speaking, when we consider $E^\varphi<0$, \Ref{Kregularization} and \Ref{HDelta2} are not equivalent, then adopting \Ref{Kregularization} or \Ref{HDelta2} is a choice. But this choice doesn't affect our discussion. }. 
The dynamics generated by ${\bf H}_\Delta$ is referred to as the improved effective dynamics.

In deriving EOMs from ${\bf H}_\Delta$, variations $\delta_{E^x(x)}\int_{-\infty}^\infty\rmd x\, \cc_\Delta(x)$ and $\delta_{E^\varphi(x)}\int_{-\infty}^\infty\rmd x\, \cc_\Delta(x)$ and integrations by part result in boundary terms respectively
\be
\frac{8 \pi  E^x{} \delta E^x{}'{}}{\kappa  \sqrt{\left| E^x{}\right| } \left| E^\varphi{}\right| },\quad-\frac{8 \pi  E^x{} \delta E^\varphi {} \left| E^\varphi{}\right|  E^x{}'{}}{\kappa  E^\varphi{}^3 \sqrt{\left| E^x{}\right| }}.\label{bdyterm}
\ee
We assume that $E^x,E^\varphi$ behaves asymptotically the same as in the Lema\^{\i}tre coordinates of the Schwarzschild spacetime as $x\to\infty$:
\be
&&E^x\sim \left(\frac{3}{2} \sqrt{R_s}\, x\right)^{4 / 3},\quad E^\varphi\sim  \sqrt{{R_s}} \lt(\frac{3}{2}{\sqrt{{R_s}}\, x}\rt)^{1/3},\label{bdyschw1}\\
&&K_x\sim  \frac{R_s}{3\times 2^{2/3} {3}^{1/3}
   \left(\sqrt{R_s} x\right)^{4/3}},\quad K_\varphi\sim-\frac{\lt(\frac{2}{3}\rt)^{1/3}
   \sqrt{R_s}}{\lt({\sqrt{R_s} x}\rt)^{1/3}},\label{bdyschw2}
\ee
and asymptotically $\delta E^\varphi\sim \delta R_s\partial_{R_s}E^\varphi$, $\delta E^x{}'\sim \delta R_s\partial_{R_s}E^x{}'$ (we allow $R_s=2G M$ to vary). This boundary condition corresponds to the asymptotically flatness formulated in the dust coordinate, given that $E^x,E^\varphi$ reduce to the Schwarzschild spacetime at infinity. 

We add the following boundary Hamiltonian to ${\bf H}_\Delta$ at $x\to\infty$
\be
{\bf H}_{\infty}=-\frac{8\pi}{\kappa}\lt(\frac{\sqrt{E^{x}}E^{x}{}^\prime{}}{E^{\varphi}}-2\sqrt{E^x}\rt)\Bigg|_{x\to\infty}
\ee 
$\delta_{E^x(x)}{\bf H}_{\infty}$ and $\delta_{E^\varphi(x)}{\bf H}_{\infty}$ cancel the boundary terms in \Ref{bdyterm} respectively, with the boundary condition \Ref{bdyschw1}. However, ${\bf H}_{\infty}$ vanishes at $x\to\infty$ by \Ref{bdyschw1}, so we end up with zero boundary Hamiltonian at $x\to\infty$.

Alternatively, instead of the Schwarzschild boundary condition \Ref{bdyschw1} and \Ref{bdyschw2}, we may make an infrared cut-off of the dust space at ${bdy}=\{x=L\gg1\}$ and impose the Dirichlet boundary condition $\delta E^x|_{bdy}=0$. In this case, we have to add the boundary term to the physical Hamiltonian
\be
{\bf H}_\Delta=\int_{-\infty}^\infty\rmd x\, \cc_\Delta(x)+{\bf H}_{bdy},\quad{\bf H}_{bdy}=-\frac{8\pi}{\kappa}\lt(\frac{\sqrt{E^{x}}E^{x}{}^\prime{}}{E^{\varphi}}-2\sqrt{E^x}\rt)\Bigg|_{bdy}.
\ee 
$\delta_{E^\varphi(x)}{\bf H}_{bdy}$ cancels the boundary terms from $\delta_{E^\varphi(x)}\int_{-\infty}^\infty\rmd x\, \cc_\Delta(x)$, while $\delta_{E^x(x)}{\bf H}_{bdy}$ cancels the boundary term from $\delta_{E^x(x)}\int_{-\infty}^\infty\rmd x\, \cc_\Delta(x)$ up to a term proportional to $\delta E^x$ which vanishes by the Dirichlet boundary condition.

On the other side $x\to -\infty$, we impose the Neumann boundary condition
\be
E^x{}'\sim 0 \quad \text{as}\quad x\to-\infty,\label{otherboundary}
\ee
so that both terms in \Ref{bdyterm} vanish at the other asymptotic boundary $x\to-\infty$. The reason to choose this boundary condition at $x\to-\infty$ will become clear when we analyze solutions in Section \ref{Properties of solutions}.

%We still impose boundary condition \Ref{bdyschw1} and \Ref{bdyschw1} at the boundary $x\to\infty$. The result of boundary terms of ${\bf H}_0$ is carried over to ${\bf H}_\Delta$, since $\cc_\Delta$ doesn't contain any derivative of $K_x,K_\varphi$.

As an advantage of the improved Hamiltonian ${\bf H}_\Delta$, $\cv(N^x)$ are still conserved by ${\bf H}_\Delta$, i.e. for all $N^x(x)$ vanishing at boundary,
\be
\cv(N^x)&=&\int\rmd x N^x(x)\cc_x(x)%+\cv_{bdy},
\quad \lt\{{\bf H}_\Delta,\,\cv(N)\rt\}=0\label{charge1111}\\
\cc_x(x)&=&E^\varphi(x) K_\varphi'(x)-K_x(x) {E^x}'(x),%\quad \cv_{bdy}=-NE^\varphi K_\varphi|_{bdy}
\ee
where $K_x,K_\varphi$ are not modified in $\cc_x$.

If we consider the standard formulation of pure gravity in terms of constraints, and understand $\cc_\Delta(x)$ as the improved Hamiltonian constraint, $\{\int\rmd x N(x) \cc_\Delta(x),\ \cv(N^x)\}=-\int\rmd x N^x\partial_xN(x) \cc_\Delta(x)$ nicely resembles the poisson bracket between classical Hamiltonian and diffeomorphism constraints \cite{Chiou:2012pg}. However, the poisson bracket between 2 Hamiltonian constraints $\{\int\rmd x N(x) \cc_\Delta(x),\ \int\rmd x' M(x') \cc_\Delta(x')\}$ does not vanish when the diffeomorphism constraint is satisfied. The physical implication is that the resulting dynamics may depend on foliation of the spacetime. This issue is relieved in the reduced phase space formulation where the dust clock fields provide a physical foliation of the spacetime. $\cc_\Delta(x)$ and $\cc_x(x)$ are not understood as constraints. Both Hamiltonian and diffeomorphism constraints have been resolved before we arrive ${\bf H}_\Delta$ and $\cv(N^x)$. The difference between $\{\int\rmd x N(x) \cc_\Delta(x),\ \int\rmd x' M(x') \cc_\Delta(x')\}$ and the classical counterpart only indicates that $\cc_\Delta(x)$ are not anymore conserved in the dynamics when the initial value satisfies $\cv(N^x)=0$, although its classical version $\cc(x)$ is conserved. Since $\cc(x)$ is anyway not a conserved charge on the entire reduced phase space, breaking its conservation in some special cases (dynamics with initial value satisfying $\cv(N^x)=0$) by quantum effect should not make the full dynamics problematic. 

Note that $\cc_\Delta(x)$ only takes into account the holonomy correction while neglecting the inverse triad correction. It was argued in \cite{Ashtekar:2007em} that at least for LQC, the effect from the inverse triad correction may be negligible in the effective dynamics on physical grounds. 

\section{Effective equations of motion}\label{Effective equations of motion}

The signs of $E^x,E^\varphi$ has to be fixed in order to derive the Hamiltonian EOMs from ${\bf H}_\Delta$. Here we focus on 2 cases: both $E^x,E^\varphi>0$ or $E^x<0,E^\varphi>0$. 

When both $E^x,E^\varphi>0$, the Hamiltonian EOMs from ${\bf H}_\Delta$ are given below:
\be
\partial_t K_x&=&-\frac{\partial_x E^x \partial_x E^\varphi}{4 \sqrt{E^x{}} E^\varphi{}^2}-\frac{\left(\partial_x E^x\right)^2}{16 E^x{}^{3/2} E^\varphi{}}+\frac{\partial^2_x E^x}{4
   \sqrt{E^x{}} E^\varphi{}}+\frac{E^\varphi{}}{4 E^x{}^{3/2}}\nonumber\\
&&-\frac{E^\varphi{} \sin
   \left(\frac{\beta  \sqrt{\Delta } K_\varphi{}}{\sqrt{E^x{}}}\right) \sin \left(\frac{2 \beta 
   \sqrt{\Delta } \sqrt{E^x{}} K_x{}}{E^\varphi{}}\right)}{2 \beta ^2 \Delta 
   \sqrt{E^x{}}}-\frac{K_x{} \sin \left(\frac{\beta  \sqrt{\Delta } K_\varphi{}}{\sqrt{E^x{}}}\right) \cos \left(\frac{2 \beta  \sqrt{\Delta } \sqrt{E^x{}} K_x{}}{E^\varphi{}}\right)}{\beta  \sqrt{\Delta }}\nonumber\\
&&+\frac{E^\varphi{} K_\varphi{} \cos \left(\frac{\beta  \sqrt{\Delta }
   K_\varphi{}}{\sqrt{E^x{}}}\right) \sin \left(\frac{2 \beta  \sqrt{\Delta } \sqrt{E^x{}}
   K_x{}}{E^\varphi{}}\right)}{2 \beta  \sqrt{\Delta } E^x{}}-\frac{E^\varphi{} \sin
   ^2\left(\frac{\beta  \sqrt{\Delta } K_\varphi{}}{\sqrt{E^x{}}}\right)}{4 \beta ^2 \Delta 
   \sqrt{E^x{}}}\nonumber\\
&&+\frac{E^\varphi{} K_\varphi{} \sin \left(\frac{\beta  \sqrt{\Delta } K_\varphi
   }{\sqrt{E^x{}}}\right) \cos \left(\frac{\beta  \sqrt{\Delta } K_\varphi{}}{\sqrt{E^x{}}}\right)}{2
   \beta  \sqrt{\Delta } E^x{}},\label{effeom1}\\
\partial_t K_\varphi&=&\frac{\left(\partial_x E^x\right)^2}{8
   \sqrt{E^x{}} E^\varphi{}^2}-\frac{\sqrt{E^x{}} \sin \left(\frac{\beta  \sqrt{\Delta } K_\varphi{}}{\sqrt{E^x{}}}\right) \sin \left(\frac{2 \beta  \sqrt{\Delta } \sqrt{E^x{}} K_x{}}{E^\varphi{}}\right)}{\beta ^2 \Delta }\nonumber\\
   &&+\frac{2 E^x{} K_x{} \sin \left(\frac{\beta  \sqrt{\Delta } K_\varphi{}}{\sqrt{E^x{}}}\right) \cos \left(\frac{2 \beta  \sqrt{\Delta } \sqrt{E^x{}} K_x{}}{E^\varphi{}}\right)}{\beta  \sqrt{\Delta } E^\varphi{}}-\frac{\sqrt{E^x{}} \sin ^2\left(\frac{\beta  \sqrt{\Delta
   } K_\varphi{}}{\sqrt{E^x{}}}\right)}{2 \beta ^2 \Delta }-\frac{1}{2
   \sqrt{E^x{}}},\label{effeom2}\\
\partial_t E^x&=&\frac{2 E^x{} \sin \left(\frac{\beta  \sqrt{\Delta } K_\varphi{}}{\sqrt{E^x{}}}\right) \cos \left(\frac{2 \beta  \sqrt{\Delta } \sqrt{E^x{}} K_x{}}{E^\varphi{}}\right)}{\beta  \sqrt{\Delta }},\label{effeom3}\\
\partial_t E^\varphi&=&\frac{E^\varphi{} \cos \left(\frac{\beta 
   \sqrt{\Delta } K_\varphi{}}{\sqrt{E^x{}}}\right) \sin \left(\frac{2 \beta  \sqrt{\Delta }
   \sqrt{E^x{}} K_x{}}{E^\varphi{}}\right)}{\beta  \sqrt{\Delta }}+\frac{E^\varphi{} \sin
   \left(\frac{\beta  \sqrt{\Delta } K_\varphi{}}{\sqrt{E^x{}}}\right) \cos \left(\frac{\beta  \sqrt{\Delta
   } K_\varphi{}}{\sqrt{E^x{}}}\right)}{\beta  \sqrt{\Delta }}\label{effeom4}.
\ee
${\bf H}_\Delta$ and above EOMs reduces to ${\bf H}_0$ and corresponding EOMs in the limit $\Delta\to0$. We refer to the dynamics determined by above EOMs with nonzero $\Delta$ as the improved effective dynamics in analogy with LQC and various models of black holes \cite{Ashtekar:2006wn,Han:2019feb,Gambini:2020nsf,Kelly:2020lec,Chiou:2012pg}.

When $E^x<0,E^\varphi>0$, ${\bf H}_\Delta$ gives a different set of EOMs. We express the coordinates to be $(\tilde{t},\tilde{x},\tilde{\theta},\tilde{\varphi})$ and write $E^{\tilde{x}}<0,E^{\tilde \varphi}>0$ to distinguish from the above case of both $E^x,E^\varphi>0$. The EOMs in this case is given by
\be
\partial_{\tilde{t}} K_{\tilde{x}}&=&-\frac{ \partial_{\tilde{x}} E^{\tilde{x}} \partial_{\tilde{x}} E^{\tilde{\varphi}}}{4 \sqrt{-E^{\tilde{x}}}  E^{\tilde{\varphi}}{}^2}
+\frac{( \partial_{\tilde{x}} E^{\tilde{x}}{})^2}{16 (-E^{\tilde{x}}{})^{3/2}  E^{\tilde{\varphi}}{}}
+\frac{ \partial_{\tilde{x}}^2E^{\tilde{x}}}{4 \sqrt{-E^{\tilde{x}}}   E^{\tilde{\varphi}}{}}
-\frac{  E^{\tilde{\varphi}}{}}{4 (-E^{\tilde{x}}{})^{3/2}}\nonumber\\
&&-\frac{  E^{\tilde{\varphi}}{} \sin \left(\frac{\beta  \sqrt{\Delta } K_{\tilde{\varphi}}{}}{\sqrt{-E^{\tilde{x}}}}\right) \sin \left(\frac{2 \beta  \sqrt{\Delta } \sqrt{-E^{\tilde{x}}} K_{\tilde{x}}{}}{ E^{\tilde{\varphi}}{}}\right)}{2 \beta ^2 \Delta  \sqrt{-E^{\tilde{x}}}}
-\frac{K_{\tilde{x}}{} \sin \left(\frac{\beta  \sqrt{\Delta } K_{\tilde{\varphi}}{}}{\sqrt{-E^{\tilde{x}}}}\right) \cos \left(\frac{2 \beta  \sqrt{\Delta } \sqrt{-E^{\tilde{x}}} K_{\tilde{x}}{}}{ E^{\tilde{\varphi}}{}}\right)}{\beta  \sqrt{\Delta }}\nonumber\\
&&-\frac{ E^{\tilde{\varphi}}{} K_{\tilde{\varphi}}{} \cos \left(\frac{\beta  \sqrt{\Delta } K_{\tilde{\varphi}}{}}{\sqrt{-E^{\tilde{x}}}}\right) \sin \left(\frac{2 \beta  \sqrt{\Delta } \sqrt{-E^{\tilde{x}}} K_{\tilde{x}}{}}{ E^{\tilde{\varphi}}{}}\right)}{2 \beta  \sqrt{\Delta } E^{\tilde{x}}}+\frac{  E^{\tilde{\varphi}}{} \sin ^2\left(\frac{\beta  \sqrt{\Delta } K_{\tilde{\varphi}}{}}{\sqrt{-E^{\tilde{x}}}}\right)}{4 \beta ^2 \Delta  \sqrt{-E^{\tilde{x}}}}\nonumber\\
&&+\frac{ E^{\tilde{\varphi}}{} K_{\tilde{\varphi}}{} \sin \left(\frac{\beta  \sqrt{\Delta } K_{\tilde{\varphi}}{}}{\sqrt{-E^{\tilde{x}}}}\right) \cos \left(\frac{\beta  \sqrt{\Delta } K_{\tilde{\varphi}}{}}{\sqrt{-E^{\tilde{x}}}}\right)}{2 \beta  \sqrt{\Delta } E^{\tilde{x}}}\label{effeomop1}\\
\partial_{\tilde{t}}K_{\tilde{\varphi}}&=&\frac{ (\partial_{\tilde{x}} E^{\tilde{x}}{})^2}{8 \sqrt{-E^{\tilde{x}}}  E^{\tilde{\varphi}}{}^2}+\frac{\sqrt{-E^{\tilde{x}}} \sin \left(\frac{\beta  \sqrt{\Delta } K_{\tilde{\varphi}}{}}{\sqrt{-E^{\tilde{x}}}}\right) \sin \left(\frac{2 \beta  \sqrt{\Delta } \sqrt{-E^{\tilde{x}}} K_{\tilde{x}}{}}{ E^{\tilde{\varphi}}{}}\right)}{\beta ^2 \Delta }\nonumber\\
&&+\frac{2 E^{\tilde{x}} K_{\tilde{x}}{} \sin \left(\frac{\beta  \sqrt{\Delta } K_{\tilde{\varphi}}{}}{\sqrt{-E^{\tilde{x}}}}\right) \cos \left(\frac{2 \beta  \sqrt{\Delta } \sqrt{-E^{\tilde{x}}} K_{\tilde{x}}{}}{ E^{\tilde{\varphi}}{}}\right)}{\beta  \sqrt{\Delta }  E^{\tilde{\varphi}}{}}-\frac{\sqrt{-E^{\tilde{x}}} \sin ^2\left(\frac{\beta  \sqrt{\Delta } K_{\tilde{\varphi}}{}}{\sqrt{-E^{\tilde{x}}}}\right)}{2 \beta ^2 \Delta }-\frac{1}{2 \sqrt{-E^{\tilde{x}}}}\label{effeomop2}\\
\partial_{\tilde{t}} E^{\tilde{x}}&=&\frac{2 E^{\tilde{x}} \sin \left(\frac{\beta  \sqrt{\Delta } K_{\tilde{\varphi}}{}}{\sqrt{-E^{\tilde{x}}}}\right) \cos \left(\frac{2 \beta  \sqrt{\Delta } \sqrt{-E^{\tilde{x}}} K_{\tilde{x}}{}}{ E^{\tilde{\varphi}}{}}\right)}{\beta  \sqrt{\Delta }},\label{effeomop3}\\
\partial_{\tilde{t}}E^{\tilde{\varphi}}&=&\frac{ E^{\tilde{\varphi}}{} \sin \left(\frac{\beta  \sqrt{\Delta } K_{\tilde{\varphi}}{}}{\sqrt{-E^{\tilde{x}}}}\right) \cos \left(\frac{\beta  \sqrt{\Delta } K_{\tilde{\varphi}}{}}{\sqrt{-E^{\tilde{x}}}}\right)}{\beta  \sqrt{\Delta }}-\frac{ E^{\tilde{\varphi}}{} \cos \left(\frac{\beta  \sqrt{\Delta } K_{\tilde{\varphi}}{}}{\sqrt{-E^{\tilde{x}}}}\right) \sin \left(\frac{2 \beta  \sqrt{\Delta } \sqrt{-E^{\tilde{x}}} K_{\tilde{x}}{}}{ E^{\tilde{\varphi}}{}}\right)}{\beta  \sqrt{\Delta }}\label{effeomop4}
\ee
Two set of EOMs \Ref{effeom1} - \Ref{effeom4} and \Ref{effeomop1} - \Ref{effeomop4} can be related by the spacetime inversion
\be
\tilde{x}=-x,\quad \tilde{t}=-t,\label{tran00}
\ee
with the fields transforming as follows
\be
{K}_{\tilde{x}}(\tilde{t},\tilde{x})&=&{K}_{\tilde{x}}(-{t},-{x})=K_x(t,x),\label{tran01}\\
{K}_{\tilde{\varphi}}(\tilde{t},\tilde{x})&=&{K}_{\tilde{\varphi}}(-{t},-{x})=-K_\varphi(t,x),\label{tran02}\\
{E}^{\tilde{x}}(\tilde{t},\tilde{x})&=&{E}^{\tilde{x}}(-{t},-{x})= -E^x(t,x),\label{tran03}\\
{E}^{\tilde{\varphi}}(\tilde{t},\tilde{x})&=&{E}^{\tilde{\varphi}}(-{t},-{x})= E^\varphi(t,x)\label{tran04}.
\ee
This transformation can be used as a solution generating map. Given a solution with e.g. $E^x(t,x)$, $-E^x(-\tilde{t},-\tilde{x})\equiv{E}^{\tilde{x}}(\tilde{t},\tilde{x}) $ is a solution to EOMs in terms of $\tilde{x},\tilde{t}$. $E^x$ flips sign in \Ref{tran03} as $x\to -x$ can be understood from the definition $E^j_a=\sqrt{\det (q)} e^j_a$.

In order to obtain special solutions, we impose the following ansatz to simplify the nonlinear partial differential equations (PDEs) \Ref{effeom1} - \Ref{effeom4} (or \Ref{effeomop1} - \Ref{effeomop4}),
\be
&&E^x(t,x)=E^x(z),\quad E^\varphi(t,x)=E^\varphi(z),\nonumber\\
&& K_x(t,x)=K_x(z),\quad K_\varphi(t,x)=K_\varphi(z),\quad z=x-t.\label{ansztzz}
\ee
$z$ in the Lema\^{\i}tre-type coordinates parametrizes the spatial slice when fixing $t$, while parametrizing the time evolution if $x$ is fixed. In the case $\Delta\to0$, solutions from the ansatz corresponds to Eqs.\Ref{positiveE} - \Ref{negativeE} with constant $\cf,\ce$ and $\b=x$%\footnote{$\b=x$ does not lose generality since it can be viewed as a change of variable.} (or constant $\b$ by re-defining $z=\b -t$)
. The ansatz reduces \Ref{effeom1} - \Ref{effeom4} to nonlinear ordinary differential equations (ODEs) of $E^x(z),E^\varphi(z),K_x(z),K_\varphi(z)$. The resulting ODEs are 1st order in $E^\varphi(z),K_x(z),K_\varphi(z)$ and 2nd order in $E^x(z)$ (resulting from $\partial_x^2 E^x$ in Eq.\Ref{effeom1}). But the 2nd order derivative in $E^x(z)$ can be eliminated by using $-\frac{\rmd}{\rmd z}E^x=\frac{2 E^x{}}{\beta  \sqrt{\Delta }} \sin \left(\frac{\beta  \sqrt{\Delta } K_\varphi{}}{\sqrt{E^x{}}}\right) \cos \left(\frac{2 \beta  \sqrt{\Delta } \sqrt{E^x{}} K_x{}}{E^\varphi{}}\right)$ from Eq.\Ref{effeom3}. Moreover for the convenience of solving equations numerically, we make a change of variable:
\be
K_1=\frac{\sqrt{E^x{}} K_x{}}{E^\varphi{}},\quad K_2=\frac{K_\varphi{}}{\sqrt{E^x{}}}\label{varK12},
\ee 
As a result, the EOMs \Ref{effeom1} - \Ref{effeom4} reduce to a standard form of 1st order ODE:
\be
\frac{\rmd}{\rmd z}\left[\begin{array}{c}E^x \\E^\varphi  \\K_1  \\K_2 \end{array}\right]=\left[\begin{array}{c}f^x\lt(E^x,E^\varphi,K_1,K_2\rt)\\ f^\varphi\lt(E^x,E^\varphi,K_1,K_2\rt)\\ f_1\lt(E^x,E^\varphi,K_1,K_2\rt)\\ f_2\lt(E^x,E^\varphi,K_1,K_2\rt)\end{array}\right].\label{ODE1}
\ee
Similar change of variables and reduction can be applied to \Ref{effeomop1} - \Ref{effeomop4} and result in
\be
\frac{\rmd}{\rmd \tilde{z}}\left[\begin{array}{c}E^{\tilde x} \\E^{\tilde \varphi}  \\\tilde{K}_1  \\\tilde{K}_2 \end{array}\right]=\left[\begin{array}{c}f^{\tilde x}\lt(E^{\tilde x},E^{\tilde \varphi},\tilde{K}_1,\tilde{K}_2\rt)\\ f^{\tilde \varphi}\lt(E^{\tilde x},E^{\tilde \varphi},\tilde{K}_1,\tilde{K}_2\rt)\\ \tilde{f}_1\lt(E^{\tilde x},E^{\tilde \varphi},\tilde{K}_1,\tilde{K}_2\rt)\\ \tilde{f}_2\lt(E^{\tilde x},E^{\tilde \varphi},\tilde{K}_1,\tilde{K}_2\rt)\end{array}\right].\label{ODEop1}
\ee
where 
\be
\tilde{K}_1=\frac{\sqrt{-E^{\tilde x}} K_{\tilde x}}{E^{\tilde \varphi}},\quad \tilde{K}_2=\frac{K_{\tilde \varphi}}{\sqrt{-E^{\tilde x}}}.
\ee
Explicit expressions of $f^x,f^\varphi,f_1,f_2$ and $f^{\tilde x},f^{\tilde \varphi},{\tilde f}_1,{\tilde f}_2$ contain long formulae and can be found in \cite{github}.

Eqs.\Ref{ODEop1} can transform to Eqs.\Ref{ODE1} by $\tilde{z}=-z$ and
\be
{K}_{\tilde{x}}(\tilde{z})&=&{K}_{\tilde{x}}(-{z})=K_x(z),\label{transf01}\\
{K}_{\tilde{\varphi}}(\tilde{z})&=&{K}_{\tilde{\varphi}}(-z)=-K_\varphi(z),\label{transf02}\\
{E}^{\tilde{x}}(\tilde{z})&=&{E}^{\tilde{x}}(-z)= -E^x(z),\label{transf03}\\
{E}^{\tilde{\varphi}}(\tilde{z} )&=&{E}^{\tilde{\varphi}}(-z)= E^\varphi(z),\label{transf04}
\ee
deduced from \Ref{tran00} - \Ref{tran04}.

\section{From Schwarzschild black hole to charged Nariai limit}\label{From Schwarzschild black hole to dS2S2}

\subsection{Strategies}

When $z\gg1$, we are far from the classical singularity, $K_x,K_\varphi$ are small thus \Ref{effeom1} - \Ref{effeom4} (or Eq.\Ref{ODE1}) reduce to the classical EOMs from ${\bf H}_0$. We focus on a solution which reduces to the Schwarzschild spacetime when far away from the singularity, by imposing \Ref{zeroE} as the initial condition at $z=z_0\gg1$, i.e. 
\be
&&E^x(z_0)=  \left(\frac{3}{2}\sqrt{R_s}
   z_0\right)^{4/3},\quad E^\varphi(z_0)=  \sqrt{R_s}
   \lt(\frac{3}{2}\sqrt{R_s} z_0\rt)^{1/3},\label{bc0}\\
 &&  K_x(z_0)= \frac{R_s}{3\times 2^{2/3} {3}^{1/3}
   \left(\sqrt{R_s} z_0\right)^{4/3}},\quad K_\varphi(z_0)=-\frac{\lt(\frac{2}{3}\rt)^{1/3}
   \sqrt{R_s}}{\lt({\sqrt{R_s} z_0}\rt)^{1/3}}\label{bc1}
% &&  E^x{}'(z_0)=\frac{3 \lt(\frac{3}{2}\rt)^{1/3} R_s^{2/3} z_0^{4/3} \sin \left(\frac{2 \beta  \sqrt{\Delta }}{3 z_0}\right) \cos \left(\frac{\beta  \sqrt{\Delta }}{3 z_0}\right)}{\beta \sqrt{\Delta }}\label{bc2}
\ee  
where $E^x(z_0),E^\varphi(z_0),K_x(z_0),K_\varphi(z_0)$ are obtained from the solution of classical EOMs from ${\bf H}_0$. They give
\be
K_1(z_0)=\frac{1}{6 z_0},\quad K_2(z_0)=-\frac{2}{3 z_0}.
\ee
An example of $z_0$ in numerically solving EOMs is $z_0=3\times 10^8$ which leads to $K_1\sim 10^{-10},\ K_2\sim 10^{-9}$ which guarantee that \Ref{effeom1} - \Ref{effeom4} approximates the classical EOMs at $z_0$. 

\begin{figure}[t]
\begin{center}
\includegraphics[width = 0.9\textwidth]{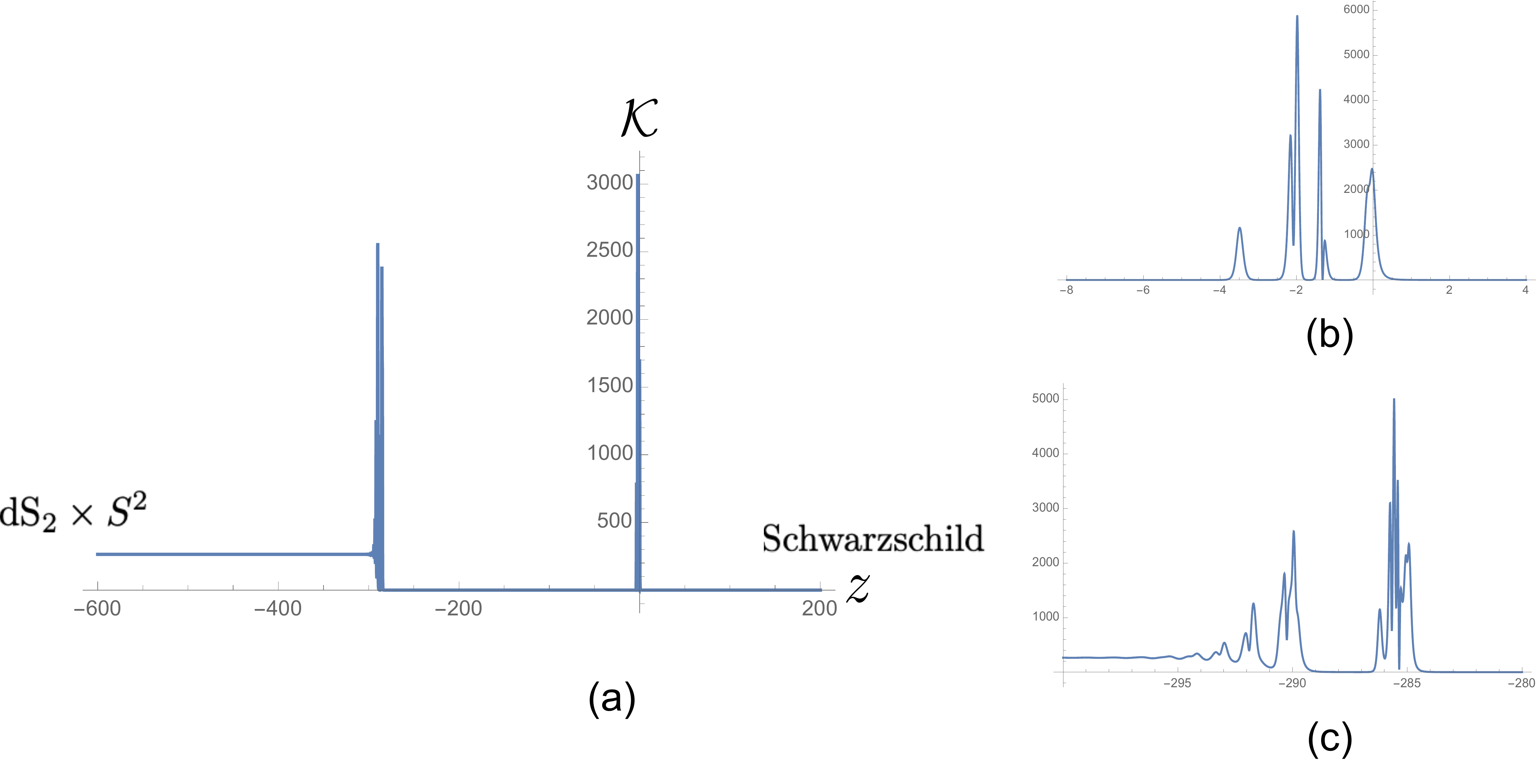} 
 \caption{(a) The Kretschmann invariant $\ck=R_{\mu\nu\rho\sig}R^{\mu\nu\rho\sig}$ evaluated on the spacetime as a solution of EOMs. The classical singularity at $z=x-t=0$ (the coordinate origin) is resolved with non-singular spacetime with finite $\ck$. The $z$-axis in the Lema\^{\i}tre-type coordinates is the spatial slice when fixing $t$, while parametrizing the time evolution if $x$ is fixed. Both $z\to\pm\infty$ are low curvature regimes, in which the solution reduces to Schwarzschild spacetime as $z\to\infty$, and reduces to ${\rm dS}_2\times S^2$ as $z\to-\infty$. In the high curvature regime demonstrated in the plot, the solution contains 2 groups of local maxima of $\ck$ on $z\in[-8,4]\equiv N_0$ and $z\in[-295,-280]\equiv N_<$. The Maxima of $\ck$ on this 2 intervals are $5913.2$ and $5075.1$ respectively. (b) Details of local maxima of $\ck$ on $z\in[-8,4]$. (c) Details of local maxima of $\ck$ on $z\in[-295,-280]$. Values of parameters for this solution are $z_0=3\times10^8$, $\Delta=0.1$, $\b=1$, $R_s=10^8$.}
\label{kres}
 \end{center}
\end{figure}

\begin{figure}[t]
\begin{center}
\includegraphics[width = 0.5\textwidth]{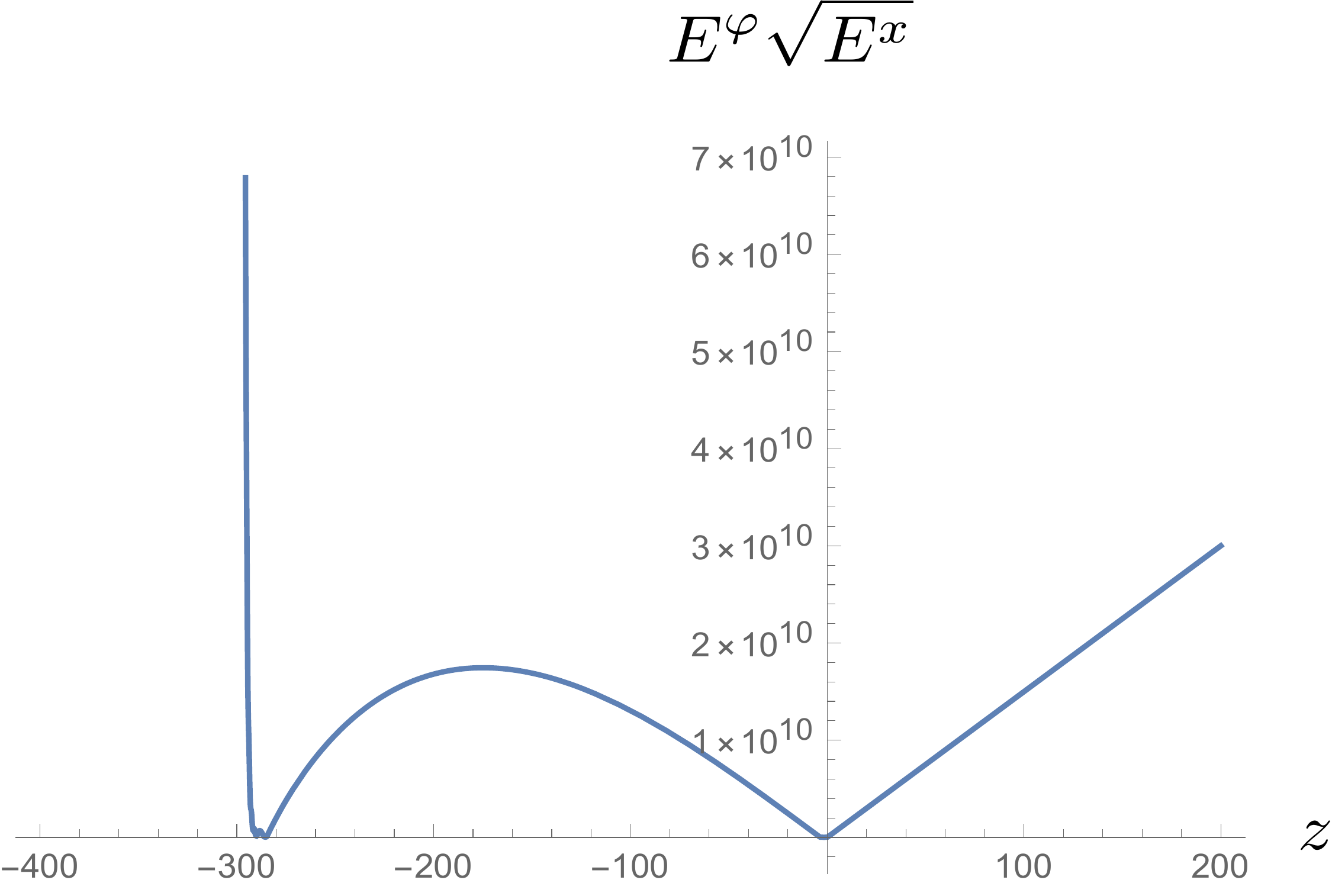} 
 \caption{Plot of spatial volume element $E^\varphi(z)\sqrt{E^x(z)}$ and its bounces. Values of parameters for this solution are $z_0=3\times10^8$, $\Delta=0.1$, $\b=1$, $R_s=10^8$.}
\label{volume111}
 \end{center}
\end{figure}

With the above boundary condition, we can obtain a family of numerically solutions satisfying the ansatz \Ref{ansztzz} to EOMs \Ref{effeom1} - \Ref{effeom4} using Mathematica or Julia (for higher precision). The solutions are labelled by different values of parameters $R_s,\Delta,z_0$. These solutions resolve the Schwarzschild black hole singularity with regular spacetime with finite but large curvature (see Figures \ref{kres} and \ref{volume111}).

\begin{figure}[t]
\begin{center}
\includegraphics[width = 0.9\textwidth]{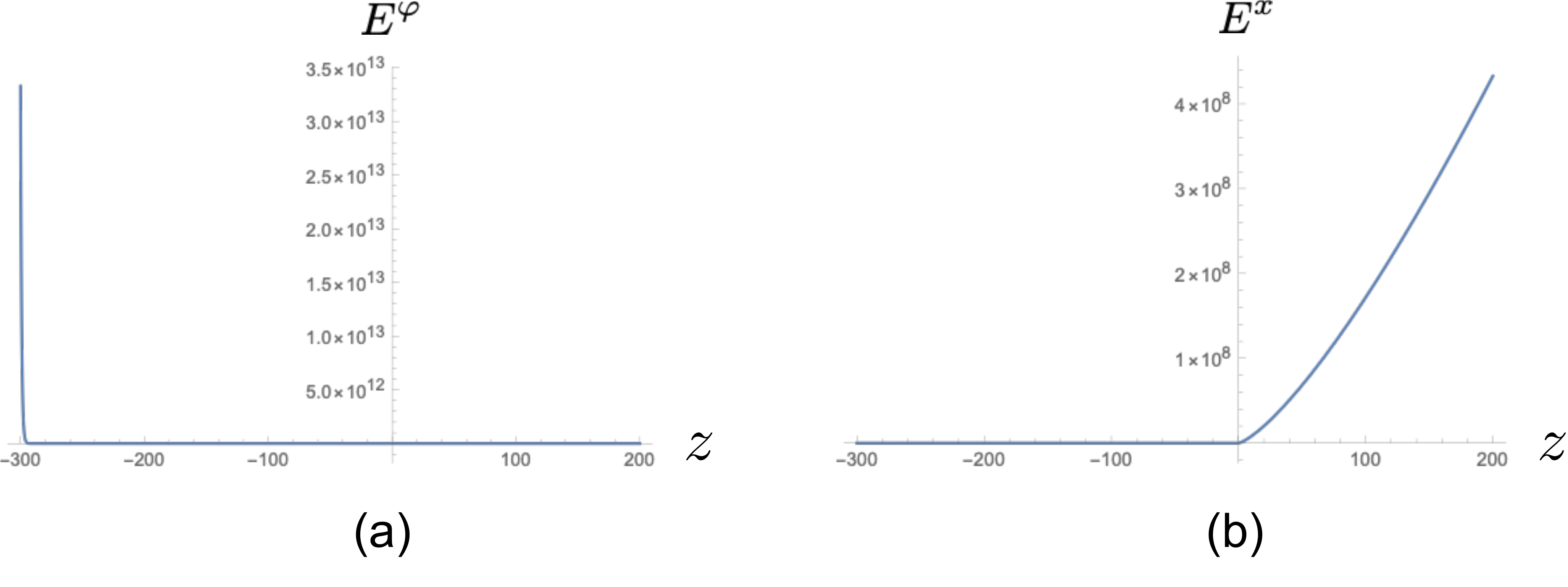} 
 \caption{(a) Plot of $E^\varphi(z)$. (b) Plot of $E^x(z)$. Values of parameters for this solution are $z_0=3\times10^8$, $\Delta=0.1$, $\b=1$, $R_s=10^8$.}
\label{Ephix}
 \end{center}
\end{figure}

The strategy of finding numerical solutions to Eq.\Ref{ODE1} is the following: Since both $E^x$ and $E^\varphi$ are large in the semiclassical regime $z\gg1$, and the initial condition is at $z_0\gg1$, it is more efficient to change variables:
\be
v_1(z)=\ln E^x(z),\quad v_2(z)=\ln E^\varphi(z),
\ee 
and express EOMs in terms of $K_1,K_2,v_1,v_2$:
\be
\frac{\rmd}{\rmd z}\left[\begin{array}{c}v_1 \\v_2  \\K_1  \\K_2 \end{array}\right]=\left[\begin{array}{c}F_1\lt(v_1,v_2,K_1,K_2\rt)\\ F_2\lt(v_1,v_2,K_1,K_2\rt)\\ G_1\lt(v_1,v_2,K_1,K_2\rt)\\ G_2\lt(v_1,v_2,K_1,K_2\rt)\end{array}\right].\label{ODE2}
\ee
Explicit expressions of (\ref{ODE2}) are given in \cite{github}. $v_1,v_2$ are not large in $z\gg1$, so less numerical errors are produced at the early stage of numerical evolution in $z$. In the $z$-evolution across $z=0$ and toward $-\infty$, $E^x$ is suppressed and stabilized at a nonzero constant value, while $E^\varphi$ grows exponentially as $-z$ goes large (see Figure \ref{Ephix}).

In the region where $E^\varphi$ is exponentially large, we can expand EOMs \Ref{ODE1} in $1/E^\varphi$, and  neglect $O(1/E^\varphi)$ to simplify the EOMs: 
\be
&&\frac{1}{4 E^x{} }+K_1{}' -\frac{\sin \left(2 \beta  \sqrt{\Delta } K_1{} \right) \sin \left(\beta  \sqrt{\Delta } K_2{} \right)}{2 \beta ^2 \Delta }-\frac{K_1{}  \sin \left(\beta  \sqrt{\Delta } K_2{} \right) \cos \left(\beta  \sqrt{\Delta } K_2{} \right)}{\beta  \sqrt{\Delta }}\nonumber\\
&&-\frac{K_1{}  \sin \left(2 \beta  \sqrt{\Delta } K_1{} \right) \cos \left(\beta  \sqrt{\Delta } K_2{} \right)}{\beta  \sqrt{\Delta }}+\frac{K_2{}  \sin \left(2 \beta  \sqrt{\Delta } K_1{} \right) \cos \left(\beta  \sqrt{\Delta } K_2{} \right)}{2 \beta  \sqrt{\Delta }}-\frac{\sin ^2\left(\beta  \sqrt{\Delta } K_2{} \right)}{4 \beta ^2 \Delta }\nonumber\\
&&+\frac{K_2{}  \sin \left(\beta  \sqrt{\Delta } K_2{} \right) \cos \left(\beta  \sqrt{\Delta } K_2{} \right)}{2 \beta  \sqrt{\Delta }}=0,\label{approx1}\\
&&-\frac{1}{2 E^x{} }-\frac{\sin \left(2 \beta  \sqrt{\Delta } K_1{} \right) \sin \left(\beta  \sqrt{\Delta } K_2{} \right)}{\beta ^2 \Delta }+\frac{2 K_1{}  \cos \left(2 \beta  \sqrt{\Delta } K_1{} \right) \sin \left(\beta  \sqrt{\Delta } K_2{} \right)}{\beta  \sqrt{\Delta }}\nonumber\\
&&-\frac{K_2{}  \cos \left(2 \beta  \sqrt{\Delta } K_1{} \right) \sin \left(\beta  \sqrt{\Delta } K_2{} \right)}{\beta  \sqrt{\Delta }}+K_2{}' -\frac{\sin ^2\left(\beta  \sqrt{\Delta } K_2{} \right)}{2 \beta ^2 \Delta }=0,\label{approx2}\\
&&E^x{}' +\frac{2 E^x{}  \cos \left(2 \beta  \sqrt{\Delta } K_1{} \right) \sin \left(\beta  \sqrt{\Delta } K_2{} \right)}{\beta  \sqrt{\Delta }}=0,\label{approx3}\\
&&\frac{\sin \left(2 \beta  \sqrt{\Delta } K_1{} \right) \cos \left(\beta  \sqrt{\Delta } K_2{} \right)}{\beta  \sqrt{\Delta }}+\frac{\sin \left(\beta  \sqrt{\Delta } K_2{} \right) \cos \left(\beta  \sqrt{\Delta } K_2{} \right)}{\beta  \sqrt{\Delta }}+v_2' =0,\label{approx4}
\ee
where $v_2=\ln E^\varphi$ only appears in the Eq.\Ref{approx4}. %Eqs.\Ref{approx1} - \Ref{approx3} can be numerically solved for $E^x,K_1,K_2$ with high precision. The solutions $K_1,K_2$ determines $E^\varphi$ by Eq.\Ref{approx4}.

%Eqs.\Ref{approx1} - \Ref{approx4} admit an interesting solution with constant $K_1,K_2,E^x\equiv r_0^2$. $E^x{}'=0$ and Eq.\Ref{aprox3} leads to 
%\be
%(a)\quad \cos \left(2 \beta  \sqrt{\Delta } K_1{} \right)=0\quad \text{or}\quad(b)\quad \sin \left(\beta  \sqrt{\Delta } K_2{} \right)=0
%\ee
%The option $(a)$ and Eq.\Ref{approx2} gives
%\be
%\frac{1}{2 E^x}=-\frac{\sin ^2\left(\beta  \sqrt{\Delta } K_2\right)}{2 \beta ^2 \Delta }-\frac{\sin \left(\beta  \sqrt{\Delta } K_2\right)}{\beta ^2 \Delta }\quad\text{or}\quad\frac{1}{2 E^x}=-\frac{\sin ^2\left(\beta  \sqrt{\Delta } K_2\right)}{2 \beta ^2 \Delta }+\frac{\sin \left(\beta  \sqrt{\Delta } K_2\right)}{\beta ^2 \Delta }
%\ee
%which is solved respectively by
%\be
%\sin \left(\beta  \sqrt{\Delta } K_2\right)= -\frac{\sqrt{E^x-\beta ^2 \Delta }}{\sqrt{E^x}}-1,\quad\text{or}\quad\sin \left(\beta  \sqrt{\Delta } K_2\right)= \frac{\sqrt{E^x-\beta ^2 \Delta }}{\sqrt{E^x}}-1
%\ee
%while the first solution less than $-1$ should be dropped. The option (b) reduces Eq.\Ref{approx2} to $1/E^x=0$ thus is unphysical. Solving Eqs.\Ref{approx2} and \Ref{approx3} gives
%\be
% K_1=\pm \frac{\pi}{4 \beta  \sqrt{\Delta }},\quad K_2=\frac{1}{\beta  \sqrt{\Delta } }\lt[\arcsin\lt(\frac{\sqrt{E^x-\beta ^2 \Delta }}{\sqrt{E^x}}-1\rt)+2k\pi\rt],\quad k={-1,0,1}
%\ee
%where $2 \beta  \sqrt{\Delta } K_1{},\beta  \sqrt{\Delta } K_2{}\in(-\pi,\pi]$.

In practice, the numerical $z$-evolution from $z_0$ to $z<0$ follows the EOMs \Ref{ODE2} until certain $z_{mid}<0$ where $E^\varphi$ is too large. Then the solution $E^x,E^\varphi,K_1,K_2$ of \Ref{ODE2} evaluated at $z_{mid}$ serves as the initial condition for Eqs.\Ref{approx1} - \Ref{approx4}, which can be further evolved to arbitrarily large $-z$. The approximation in Eqs.\Ref{approx1} - \Ref{approx4} by neglecting $O(1/E^\varphi)$ is consistent because the solution $E^\varphi$ keeps growing exponentially as $z\to-\infty$, while all other quantities are bounded by $O(1)$. A full solution of the EOMs is given by connecting 2 solutions of \Ref{ODE2} and \Ref{approx1} - \Ref{approx4} at $z_{mid}$.   

Note that the similar approximation with $E^{\tilde \varphi}\gg1$ can be applied to \Ref{effeomop1} - \Ref{effeomop4} and leads to 
\be
&&\frac{1}{4 E^{\tilde x}{}}+\partial_{\tilde z}\tilde{K}_1-\frac{\sin \left(2 \beta  \sqrt{\Delta } \tilde{K}_1\right) \sin \left(\beta  \sqrt{\Delta } \tilde{K}_2\right)}{2 \beta ^2 \Delta }-\frac{\tilde{K}_1 \sin \left(\beta  \sqrt{\Delta } \tilde{K}_2\right) \cos \left(\beta  \sqrt{\Delta } \tilde{K}_2\right)}{\beta  \sqrt{\Delta }}\nonumber\\
&&+\frac{\tilde{K}_1 \sin \left(2 \beta  \sqrt{\Delta } \tilde{K}_1\right) \cos \left(\beta  \sqrt{\Delta } \tilde{K}_2\right)}{\beta  \sqrt{\Delta }}+\frac{\tilde{K}_2 \sin \left(2 \beta  \sqrt{\Delta } \tilde{K}_1\right) \cos \left(\beta  \sqrt{\Delta } \tilde{K}_2\right)}{2 \beta  \sqrt{\Delta }}+\frac{\sin ^2\left(\beta  \sqrt{\Delta } \tilde{K}_2\right)}{4 \beta ^2 \Delta }\nonumber\\
&&-\frac{\tilde{K}_2 \sin \left(\beta  \sqrt{\Delta } \tilde{K}_2\right) \cos \left(\beta  \sqrt{\Delta } \tilde{K}_2\right)}{2 \beta  \sqrt{\Delta }}=0,\label{approxop1}\\
&&\frac{1}{2 E^{\tilde x}{}}+\frac{\sin \left(2 \beta  \sqrt{\Delta } \tilde{K}_1\right) \sin \left(\beta  \sqrt{\Delta } \tilde{K}_2\right)}{\beta ^2 \Delta }-\frac{2 \tilde{K}_1 \cos \left(2 \beta  \sqrt{\Delta } \tilde{K}_1\right) \sin \left(\beta  \sqrt{\Delta } \tilde{K}_2\right)}{\beta  \sqrt{\Delta }}\nonumber\\
&&-\frac{\tilde{K}_2 \cos \left(2 \beta  \sqrt{\Delta } \tilde{K}_1\right) \sin \left(\beta  \sqrt{\Delta } \tilde{K}_2\right)}{\beta  \sqrt{\Delta }}+\partial_{\tilde z}\tilde{K}_2-\frac{\sin ^2\left(\beta  \sqrt{\Delta } \tilde{K}_2\right)}{2 \beta ^2 \Delta }=0,\label{approxop2}\\
&&\partial_{\tilde z}E^{\tilde x}+\frac{2 E^{\tilde x}{} \cos \left(2 \beta  \sqrt{\Delta } \tilde{K}_1\right) \sin \left(\beta  \sqrt{\Delta } \tilde{K}_2\right)}{\beta  \sqrt{\Delta }}=0,\label{approxop3}\\
&&-\frac{\sin \left(2 \beta  \sqrt{\Delta } \tilde{K}_1\right) \cos \left(\beta  \sqrt{\Delta } \tilde{K}_2\right)}{\beta  \sqrt{\Delta }}+\frac{\sin \left(\beta  \sqrt{\Delta } \tilde{K}_2\right) \cos \left(\beta  \sqrt{\Delta } \tilde{K}_2\right)}{\beta  \sqrt{\Delta }}+\partial_{\tilde z}\tilde{v}_2=0.\label{approxop4}
\ee 
where $\tilde{v}_2=\ln E^{\tilde \varphi}$. Eqs.\Ref{approxop1}-\Ref{approxop4} relate to Eqs.\Ref{approx1}-\Ref{approx4} by $\tilde{z}=-z$ and transformations \Ref{transf01} - \Ref{transf04}.

\subsection{Properties of solutions}\label{Properties of solutions}

\begin{figure}[h]
\begin{center}
\includegraphics[width = 0.9\textwidth]{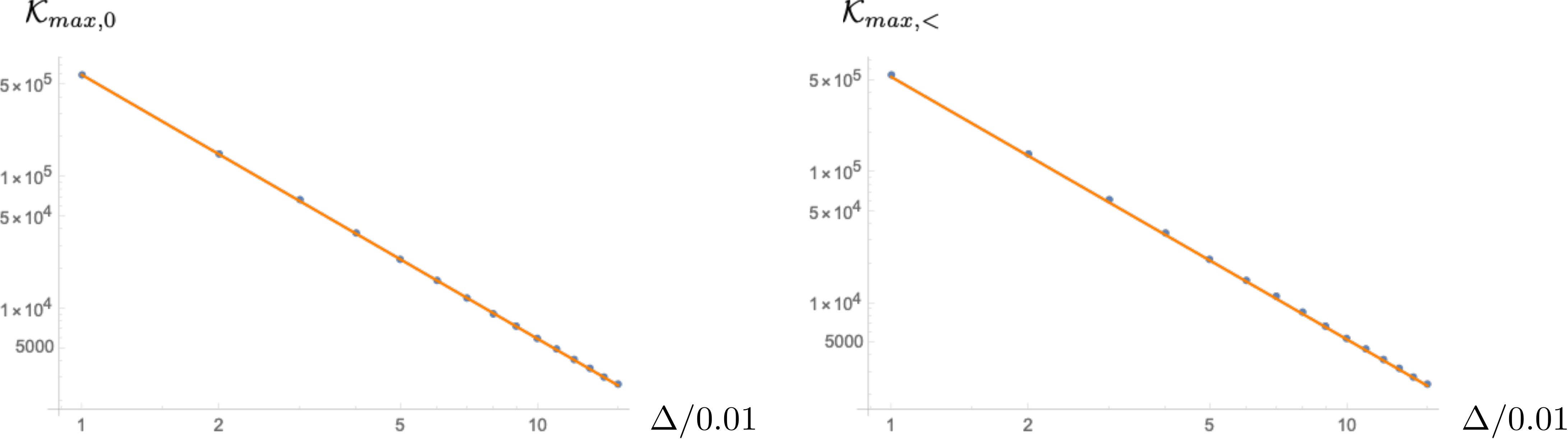} 
 \caption{The left panel is the Log-Log plot of $\ck_{max,0}$ versus $\Delta$. The right panel is the Log-Log plot of $\ck_{max,<}$ versus $\Delta$. The orange straight lines are $constant\times\Delta^{-2}$. Values of parameters are $z_0=3\times10^8$, $\Delta=0.01m$ $(m=1,\cdots,15)$, $\b=1$, $R_s=10^9$.}
\label{kresDelta}
 \end{center}
\end{figure}

\begin{figure}[h]
\begin{center}
\includegraphics[width = 0.9\textwidth]{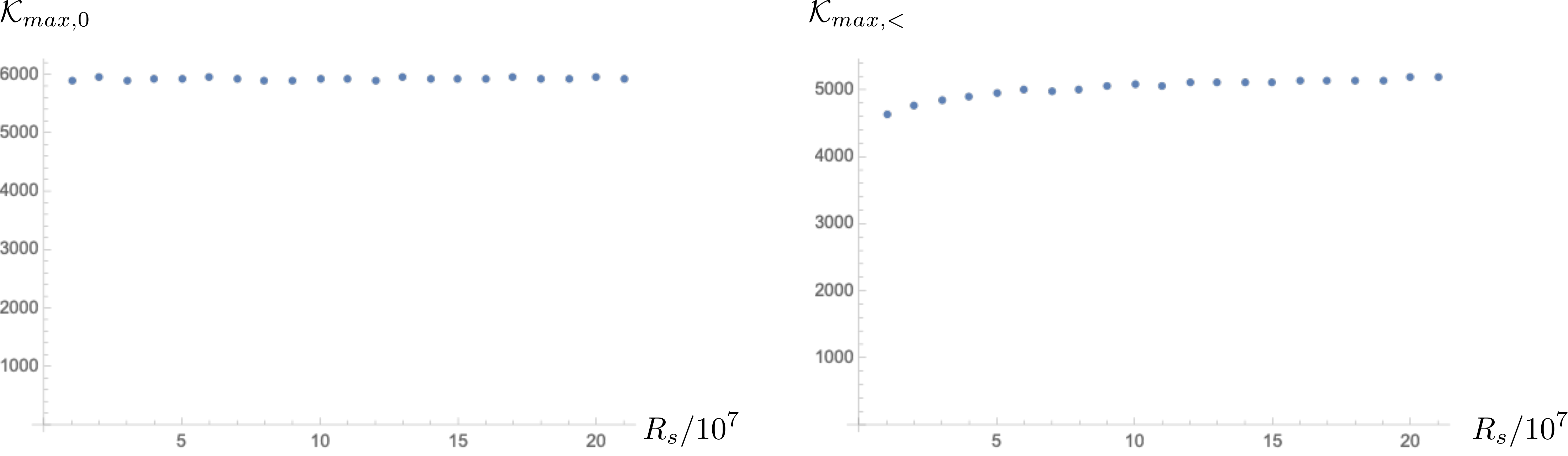} 
 \caption{The left panel is the plot of $\ck_{max,0}$ versus $R_s$. The right panel is the plot of $\ck_{max,<}$ versus $R_s$. Values of parameters are $z_0=3\times10^8$, $\Delta=0.1$, $\b=1$, $R_s=10^7m$ $(m=1,\cdots,21)$.}
\label{kresRs}
 \end{center}
\end{figure}

The numerical solutions exhibit following properties: The spacetime curvature is finite on the entire range of $z$ (from $z_0\gg1$ to negative $z$ with arbitrarily large $|z|$). The classical singularity at $z=0$ is resolved. The entire spacetime is nonsingular and has large but finite curvature at $z=0$. Figure \ref{kres} plots the Kretschmann invariant $\ck=R_{\mu\nu\rho\sig}R^{\mu\nu\rho\sig}$ of the solution. It demonstrates 2 groups of local maxima of $\ck$ located respectively in the neighborhood $N_0$ of $z=0$ (Figure \ref{kres}(b)) and in a neighborhood $N_<$ of $z<0$ (Figure \ref{kres}(c)). The oscillatory $\ck$ in these 2 neighborhoods indicates strong quantum fluctuations in these regimes. We denote by $\ck_{max,0}$ and $\ck_{max,<}$ the maximal $\ck$ in $N_0$ and $N_<$ respectively, and test their dependences with respect to $\Delta$ and horizon radius $R_s$ (we fix $\b=1$). The numerics demonstrate that both $\ck_{max,0}$ and $\ck_{max,<}$ are proportional to $\Delta^{-2}$  (see Figure \ref{kresDelta})
\be
\ck_{max,0}\big|_{\b=1}\simeq\frac{1}{\Delta^2}k_0(R_s,\Delta),\quad \ck_{max,<}\big|_{\b=1}\simeq \frac{1}{\Delta^2}k_<(R_s,\Delta),
\ee
The behavior has qualitative similarity with results in \cite{Gambini:2020nsf,Ashtekar:2018cay}. The behavior of Kretschmann scalar $\ck\sim \Delta^{-2}$ motivates us to understand $\Delta\sim \ell_P^2$ such that the singularity resolution and bounce of spatial volume (Fig.\ref{volume111}) happen at the Planckian curvature. Both $\ck_{max,0}$ and $\ck_{max,<}$ are Planckian curvatures. In models of LQC and LQG black holes, $\Delta$ is chosen to be the minimal nonzero eigenvalue of the LQG area operator. Here $\ck_{max,0}>\ck_{max,<}$, and $k_0,k_<$ have mild dependence on $R_s$  and $\Delta$ (see Figs.\ref{kresDelta} and \ref{kresRs}). The $R_s$ and $\Delta$ dependences in $k_0,k_<$ are subleading corrections. Asymptotically for large negative $z$, $\ck$ approaches to be $z$-independent constant whose dependence on $\Delta$ is still $\sim 1/\Delta^2$. We come back to this asymptotic behavior shortly. 

\begin{figure}[h]
\begin{center}
\includegraphics[width = 0.5\textwidth]{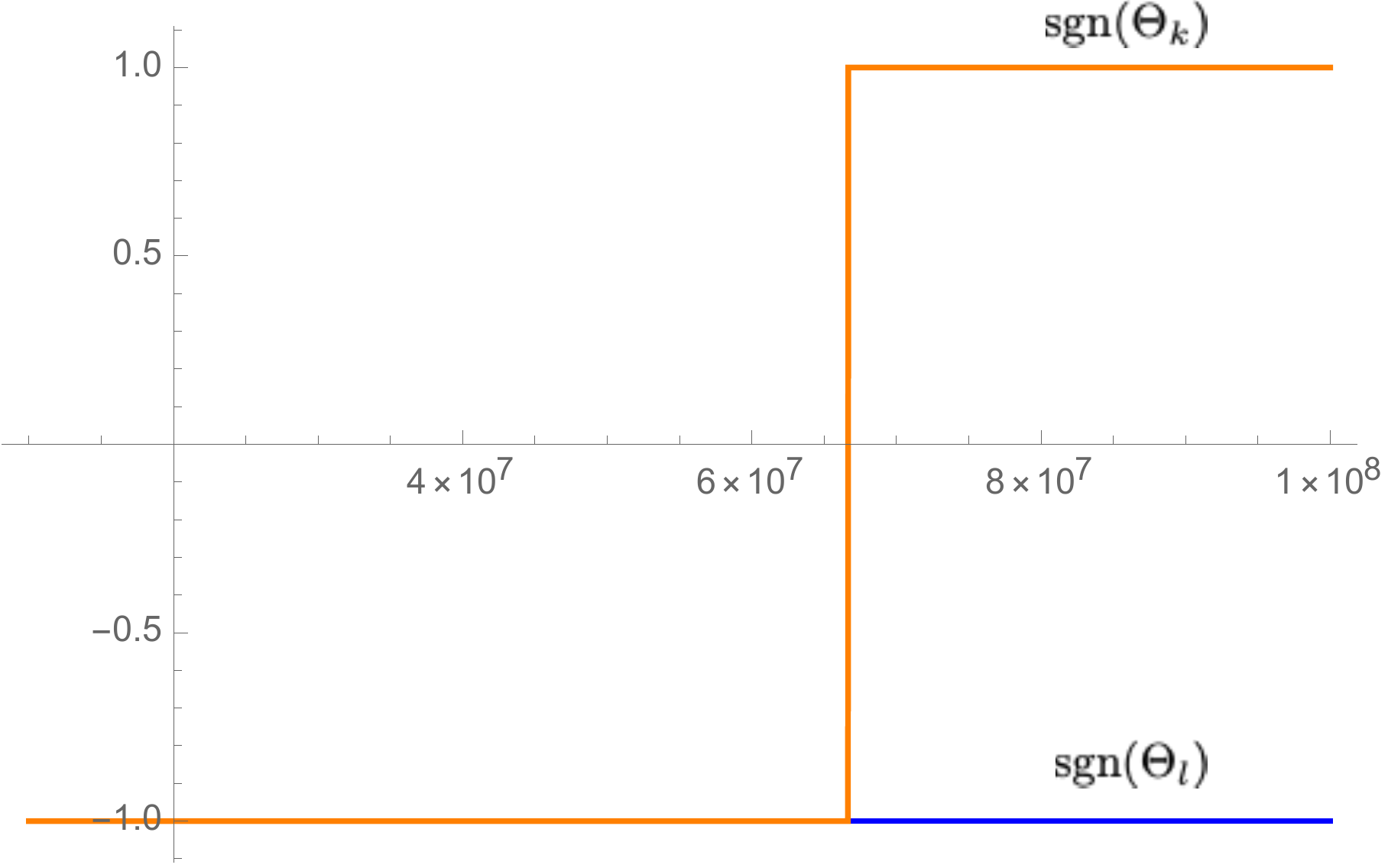} 
 \caption{Plots of $\sgn(\Theta_k)$ (orange) and $\sgn(\Theta_l)$ (blue). $\Theta_k=0$ is at $z_s\simeq 6.66666\times10^7\simeq \frac{2}{3}R_s$. The relative correction $\frac{|z_s-\frac{2}{3}R_s|}{\frac{2}{3}R_s}\sim 10^{-8}$ (from numerics using Mathematica) is small. Values of parameters for this solution are $z_0=3\times10^8$, $\Delta=0.1$, $\b=1$, $R_s=10^8$.}
\label{expansions}
 \end{center}
\end{figure}

In the regime $z>0$ and far away from $N_0$, the solution is semiclassical and reduces to the Schwarzschild spacetime in Lema\^{\i}tre coordinates. The quantum effect is negligible in this regime. It is clear from EOMs \Ref{effeom1} - \Ref{effeom4} that as far as $K_x,K_\varphi$ do not blow up, the classical Schwarzschild geometry is approximately a solution to \Ref{effeom1} - \Ref{effeom4} up to corrections of $O(\sqrt{\Delta})$. In particular, the numerical solution indicates that the quantum correction at the event horizon $z=\frac{2}{3}R_s$ is negligible. $z=\frac{2}{3}R_s$ is a marginal trapped surface with $\Theta_k=0$ and $\Theta_l<0$ where $\Theta_k$ and $\Theta_l$ are outward and inward null expansions (see Figure \ref{expansions}).

\begin{figure}[h]
\begin{center}
\includegraphics[width = 0.9\textwidth]{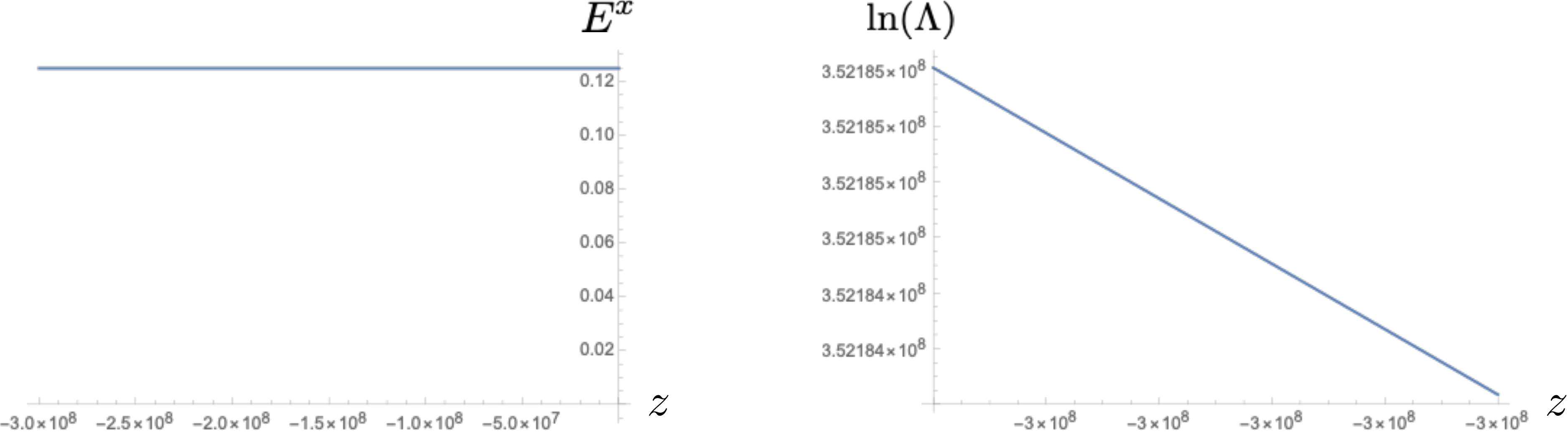} 
 \caption{The left panel plots the asymptotic behavior of $E^x$ as $-z$ goes large. The right panel plots the asymptotic behavior of $\ln(\L)$ which can be fit by the linear function $-371.252 - 1.17395 z$. Values of parameters are $z_0=3\times10^8$, $\Delta=0.1$, $\b=1$, $R_s=10^8$.}
\label{asymp}
 \end{center}
\end{figure}

The other asymptotic regime is in the opposite side where $z<0$ and $-z$ is large. In this regime $E^x$ approaches a constant, denoted by $r_0^2$, and $E^\varphi$ grows exponentially:
\be
E^x(z)\sim r_0^2, \quad \L(z)=\frac{E^\varphi(z)}{\sqrt{E^x(z)}}\sim e^{-\a_1-\alpha^{-1}_0 z},\quad z\to-\infty, \label{asympELambda}
\ee
$r_0,\a_0,\a_1$ can be obtained numerically (see Figure \ref{asymp}). $E^x(z)$ approaching to constant as $z\to-\infty$ ($x\to-\infty$ when fixing $t$) fulfills the boundary condition \Ref{otherboundary} discussed earlier.

\begin{figure}[h]
\begin{center}
\includegraphics[width = 0.9\textwidth]{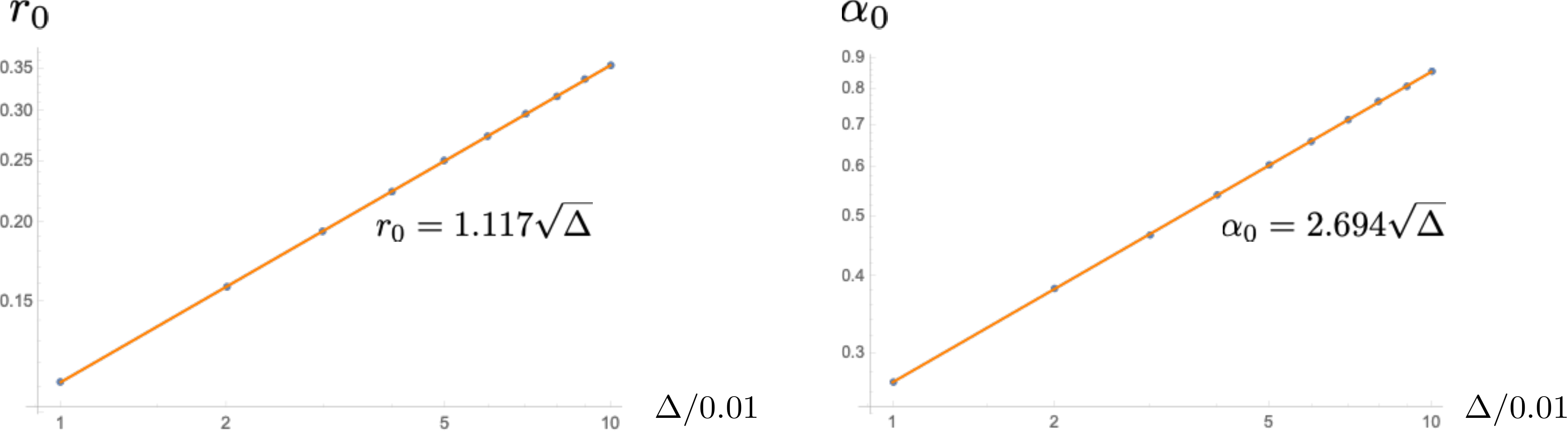} 
 \caption{The left panel is the Log-Log plot of $r_0$ versus $\Delta$. Blue dots are $r_0$ at different values of $\Delta$. The orange line is the fit function $r_0=1.117\sqrt{\Delta}$. The right panel is the Log-Log plot of $\a_0$ versus $\Delta$. Blue dots are $\a_0$ at different values of $\Delta$. The orange line is the fit function $\a_0=2.694\sqrt{\Delta}$. Values of parameters are $z_0=3\times10^8$, $\Delta=0.01m$ $(m=1,\cdots,10)$, $\b=1$, $R_s=10^8$.}
\label{radiusDelta}
 \end{center}
\end{figure}

\begin{figure}[h]
\begin{center}
\includegraphics[width = 0.9\textwidth]{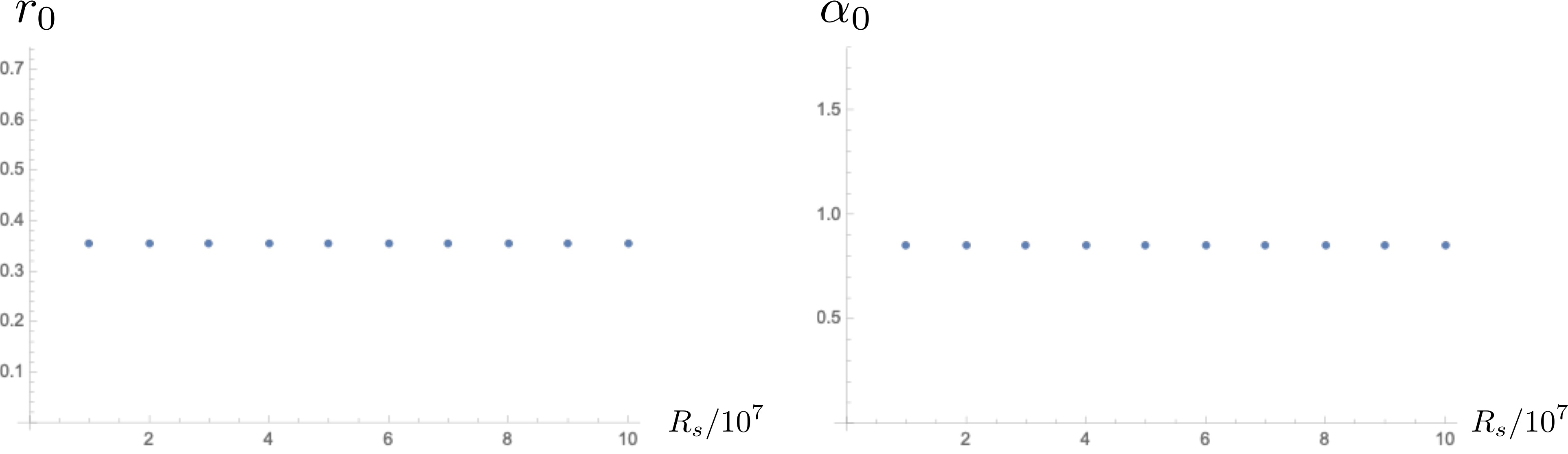} 
 \caption{The left panel plots of $r_0$ at different values of $R_s$. The right panel plots of $\a_0$ at different values of $R_s$. Both $\a_0$ and $r_0$ are constant in $R_s$. Values of parameters are $z_0=3\times10^8$, $\Delta=0.1$, $\b=1$, $R_s=10^7 m$ $(m=1,\cdots,10)$.}
\label{radiusRs}
 \end{center}
\end{figure}

\begin{figure}[h]
\begin{center}
\includegraphics[width = 0.9\textwidth]{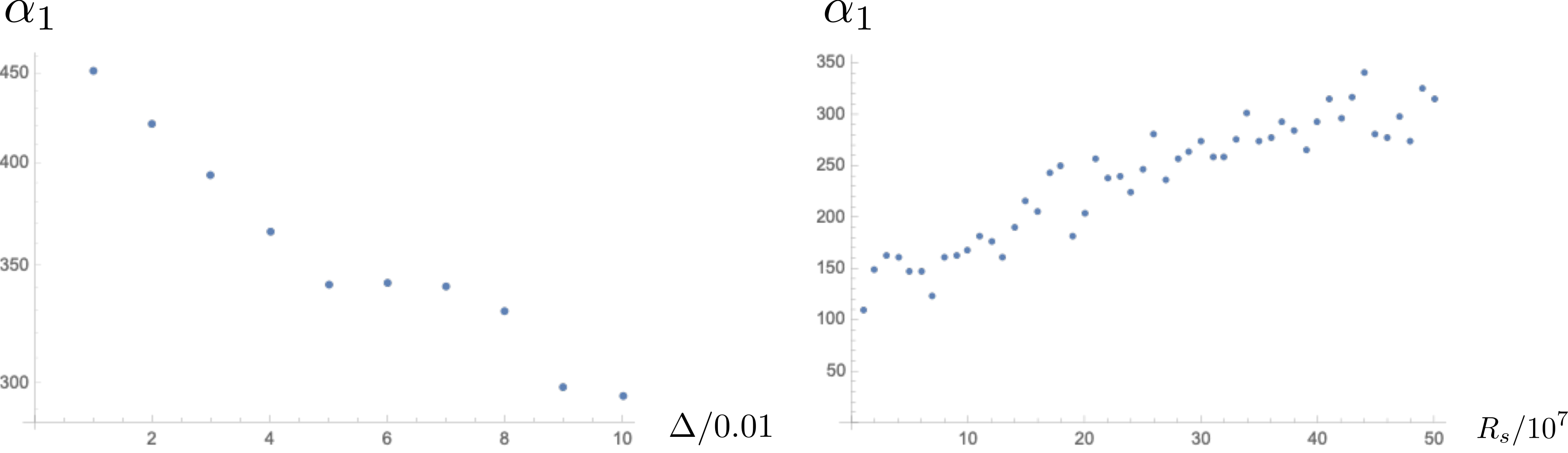} 
 \caption{Plots of $\a_1$ versus different values of $\Delta$ and $R_s$. Values of other parameters are $z_0=3\times10^8$ and $\b=1$.}
\label{alpha1}
 \end{center}
\end{figure}

Recall that the spacetime metric is given by \Ref{dustmetric}, the simple asymptotic behavior \Ref{asympELambda} indicates that at $z\to-\infty$ is the metric looks like ${\rm dS}_2\times S^2$: a product of 2-dimensional de Sitter (dS) spacetime and 2-sphere, 
\be
\rmd s^2\sim -\rmd t^2+e^{-2\a_1+2\alpha^{-1}_0 (t-x)}\rmd x^2+r_0^2\lt(\rmd\theta^2+\sin^2\theta\rmd\varphi^2\rt)\label{ds2s2}
\ee
A coordinate transformation $x\to\eta$ with $\rmd\eta=e^{-\a_1-\alpha^{-1}_0 x}\rmd x$ makes \Ref{ds2s2} as $\rmd s^2\sim -\rmd t^2+e^{2\alpha^{-1}_0 t}\rmd \eta^2+r_0^2\lt(\rmd\theta^2+\sin^2\theta\rmd\varphi^2\rt)$ where $(t,\eta)$ is the inflationary coordinate in the ${\rm dS}_2$ \footnote{It may also be the global coordinate since \Ref{ds2s2} is only the asymptotic behavior at $\t\to\infty$. The ${\rm dS}_2$ metric at $\t\to\infty$ does not distinguish between the global and inflationary coordinates. }. The dS radius is given by $\a_0$ and the $S^2$ sphere radius is constantly $r_0$. The numerical result indicates that $\a_0$ and $r_0$ are purely quantum effects: 
\be
r_0\simeq1.11724\Delta^{1/2},\quad \a_0\simeq 2.69371\Delta^{1/2},\quad (\text{at}\ \b=1),\label{alpha0r0}
\ee
and independent of $R_s$ (see Figures \ref{radiusDelta} and \ref{radiusRs}). The Kretschmann invariant $\ck$ approaches constant
\be
\ck\sim 4 \left(\frac{1}{\alpha_0 ^4}+\frac{1}{r_0^4}\right)\simeq\frac{2.64325}{\Delta ^2},
\ee
as indicated in Figure \ref{kres}. The asymptotical regime as $z\to-\infty$ still has Planckian curvature. $\a_1$ depends on both $\Delta$ and $R_s$ (see Figure \ref{alpha1}). Although the asymptotic ${\rm dS}_2\times S^2$ geometry is independent of $R_s$ or the mass of black hole, the dust coordinate $x$ in the geometry depends on $R_s$ since $\a_1$ depends on $R_s$.

\begin{figure}[h]
\begin{center}
\includegraphics[width = 0.5\textwidth]{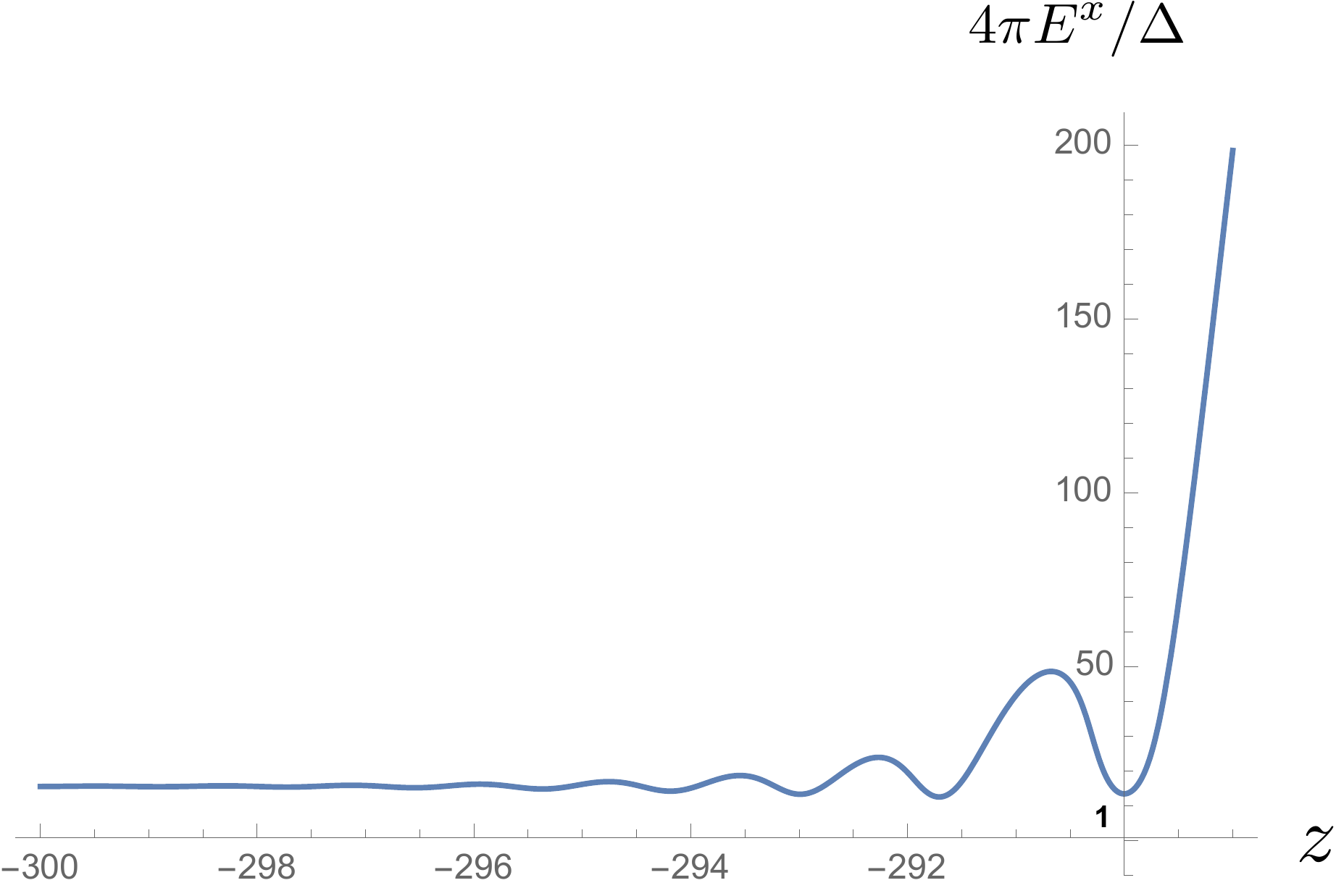} 
 \caption{Minima of the $S^2$ area $4\pi E^x$ divided by $\Delta$. The axis origin is $(-290,1)$. Values of parameters for this solution are $z_0=3\times10^8$, $\Delta=0.1$, $\b=1$, $R_s=10^8$.}
\label{s2area}
 \end{center}
\end{figure}

${\rm dS}_2\times S^2$ with $r_0\neq\a_0$ is also known as the charged Nariai geometry. It can be obtained as the near horizon limit of the (near) extremal Reissner-Nordstrom-de Sitter (RN-dS) spacetime where the cosmological horizon and event horizon coincide \cite{Bousso:1996pn,Hawking:1995ap}. %The cosmological constant $3/l_c^2$, charge $q$, and mass $\mu$ of the extremal RN-dS spacetime relate to $r_0,\a_0$ by
%\be
%l_c&=&\frac{\sqrt{6} r_0}{\sqrt{1+{r_0}^2/\a_0^2}}\simeq 2.52786 \sqrt{\Delta },\\
%q&=&\frac{\sqrt{r_0^2- r_0^4/\a_0^2}}{\sqrt{2G}}\simeq 0.718853 \sqrt{\Delta /G},\\
%\mu&=&\frac{1}{3G} \left(2 r_0- r_0^3/\a_0^2\right)\simeq 0.680762 \sqrt{\Delta }/G.
%\ee
%The proportionality to $\sqrt{\Delta}$ indicates that they are excited by quantum gravity effects. 
The relation between LQG black hole and the Nariai geometry has been proposed in earlier studies of effective dynamics using the Kantowski-Sachs foliation of the black hole interior \cite{Bohmer:2007wi,Boehmer:2008fz}. However as is pointed out in \cite{Ashtekar:2018cay} that the analysis in \cite{Bohmer:2007wi,Boehmer:2008fz} suffers from 2 problems: (1) their $\bar{\mu}$-scheme model of black hole interior produce large quantum effect near the event horizon which is of low curvature, and (2) the area of $S^2$ is even smaller than the minimal area gap $\Delta$ at certain stage of the time evolution, inconsistent with the $\bar{\mu}$-scheme treatment of holonomies ($S^2$ has no room for the $\bar{\mu}$-scheme holonomy since it has to be around an area of $\Delta$). Our analysis does not have these problems: Firstly the quantum effect is negligible in the low curvature regime including the event horizon as discussed above. Secondly, numerical results indicate that the area of $S^2$ is always larger than $\Delta$ during the evolution (see Figure \ref{s2area}).

It may be more proper to view the charged Nariai geometry ${\rm dS}_2\times S^2$ here as a quantum geometry, since both $E^x=r_0^2$ and the dS radius $\a_0$ is of $O(\sqrt{\Delta})$. ${\rm dS}_2\times S^2$ corresponds quantum states $|r_0,\a_0;\a_1\rangle$ such that when $ z\to -\infty$:
\be
\langle r_0,\a_0;\a_1|\hat{E}^x(z)|r_0,\a_0;\a_1\rangle=r_0^2,\quad
\langle r_0,\a_0;\a_1|\hat{E}^\varphi(z)|r_0,\a_0;\a_1\rangle=r_0 e^{-\a_1-\a_0^{-1}z}.\label{VEV1}
\ee
$r_0\sim \sqrt{\Delta}$ is of Planck size, so that quantum fluctuations of ${E}^x$ given by $|r_0,\a_0;\a_1\rangle$ must be small in order to trust this effective geometry. Therefore the charged Nariai geometry should have large quantum fluctuation of $K_x$ by the uncertainty principle.

%Why do we still trust the accuracy of the effective ${\rm dS}_2\times S^2$ geometry, given that ${E}^x(z)\sim \Delta$ is quantum? Let us exam the implication of quantum fluctuations from the uncertainty principle. Assuming $|r_0,\a_0;\a_1\rangle$ should give small fluctuations of ${E}^x(z),{E}^\varphi(z)$ so that Eqs.\Ref{VEV1} are accurate,
%\be
%\sig_{{E}^x(z)}\ll \Delta,\quad \sig_{{E}^\varphi(z)}\ll e^{-\a_1-\a_0^{-1}z}\Delta^{1/2}
%\ee
%where $\sig_\co =\sqrt{\langle\hat{\co}^2\rangle-\langle\hat{\co}\rangle^2}$ is the standard deviation. The uncertainty principle $\sig_{{E}^x(z)}\sig_{{K}_x(z)}\gtrsim \Delta$ and $\sig_{{E}^\varphi(z)}\sig_{{K}_\varphi(z)}\gtrsim \Delta$ implies
%\be
%\sig_{{K}_x(z)}\gg 1,\quad \sig_{{K}_\varphi(z)}\gg e^{\a_1+\a_0^{-1}z}\Delta^{1/2},\label{sigmaK}
%\ee
%Although $\sig_{{K}_x(z)}$ is large, but the variables that we actually use for solving effective dynamics is $K_1,K_2$ in Eq.\Ref{varK12}. Their fluctuations can be very small since \Ref{sigmaK} implies
%\be
%\sig_{{K}_1(z)}\gg e^{\a_1+\a_0^{-1}z},\quad \sig_{{K}_1(z)}\gg e^{\a_1+\a_0^{-1}z},
%\ee
%while $e^{\a_1+\a_0^{-1}z}\to0$ as $z\to-\infty$. Therefore the uncertainty principle allows to have small fluctuation of all $E^x,E^\varphi,K_1,K_2$ in the effective EOM \Ref{ODE1}.

It is interesting that these states $|r_0,\a_0;\a_1\rangle$ depend on the additional parameter $\a_1$ which indicates that ${\rm dS}_2\times S^2$ is not a single state, but has infinite degeneracy from the quantum point of view. Although different values of $\a_1$ correspond to the same spacetime geometry, and are related by diffeomorphisms, here they indeed correspond to different physical states since we start with the reduced phase space formulation where gravity is deparametrized by dust fields. All states $|r_0,\a_0;\a_1\rangle$ corresponding to ${\rm dS}_2\times S^2$ should span a Hilbert space $\ch_{\rm dS_2\times S^2}$. We come back to discussing more details of $\ch_{\rm dS_2\times S^2}$ in Section \ref{Infinitely many infrared states}.

%Null vectors lying in the $t$-$x$ plane are given up to rescalings by (recall \Ref{dustmetric})
%\be
%\lt(\frac{\partial}{\partial U}\rt)^a=\lt(\frac{\partial}{\partial t}\rt)^a+\frac{1}{\L}\lt(\frac{\partial}{\partial x}\rt)^a,\quad \lt(\frac{\partial}{\partial V}\rt)^a=\lt(\frac{\partial}{\partial t}\rt)^a-\frac{1}{\L}\lt(\frac{\partial}{\partial x}\rt)^a
%\ee
%where $\L={E^\varphi}/{\sqrt{|E^x|}}$. Due to the exponential growth of $\L$ as $z\to-\infty$, the normal vector $(\partial/\partial t)^a$ to the spatial slice $\cs$ is asymptotically null (see Figure \ref{nullray}). The cosmological horizon of the asymptotic ${\rm dS}_2$ geometry located at $z=x-t\to-\infty$ is a coordinate singularity of $(t,x,\theta,\varphi)$.  

\begin{figure}[h]
\begin{center}
\includegraphics[width = 0.5\textwidth]{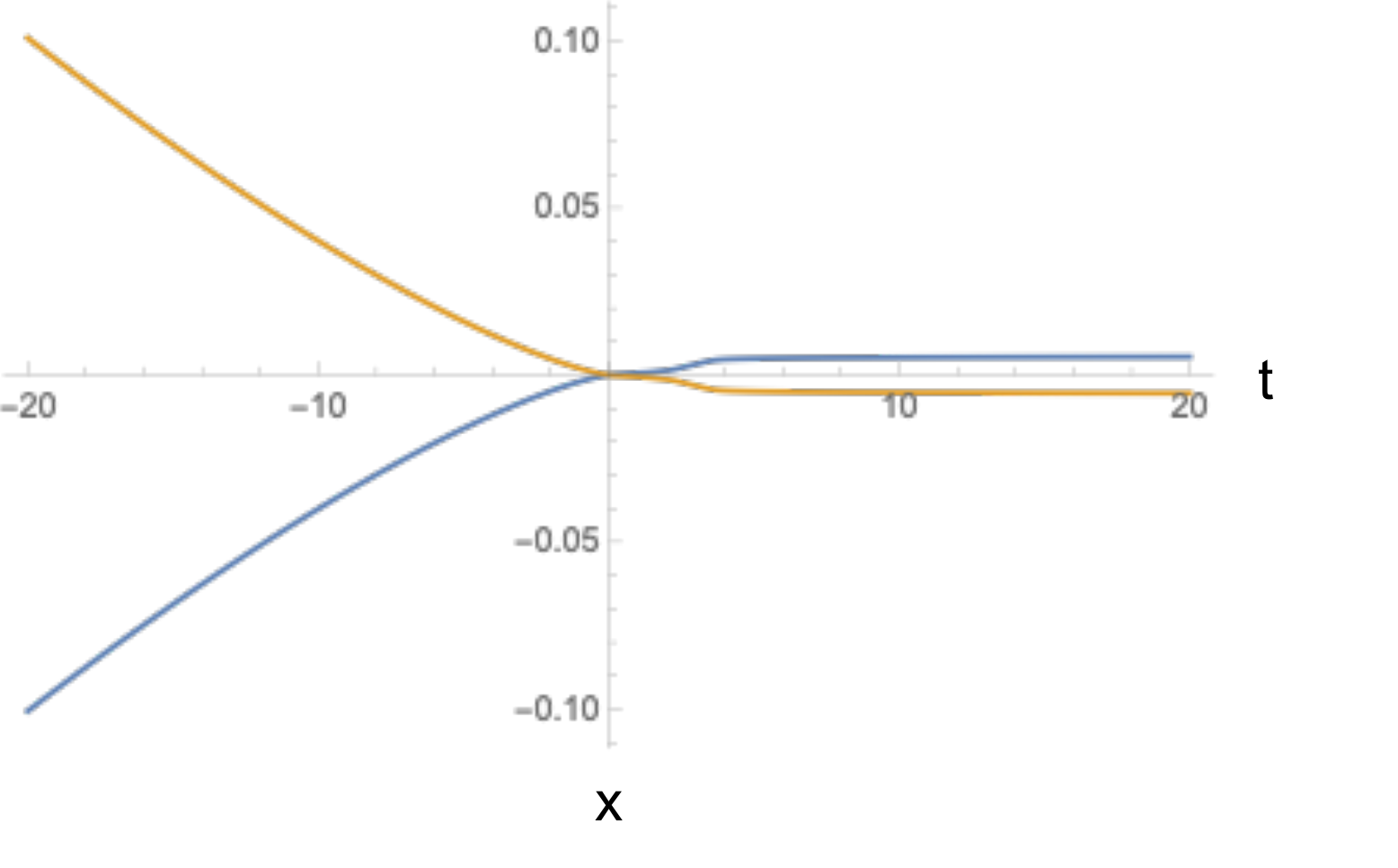} 
 \caption{Null rays viewed in the $t$-$x$ dust coordinates. The orange (blue) curve is generated by $U^\a$ ($V^\a$).}
\label{nullray}
 \end{center}
\end{figure}

\begin{figure}[h]
\begin{center}
\includegraphics[width = 0.6\textwidth]{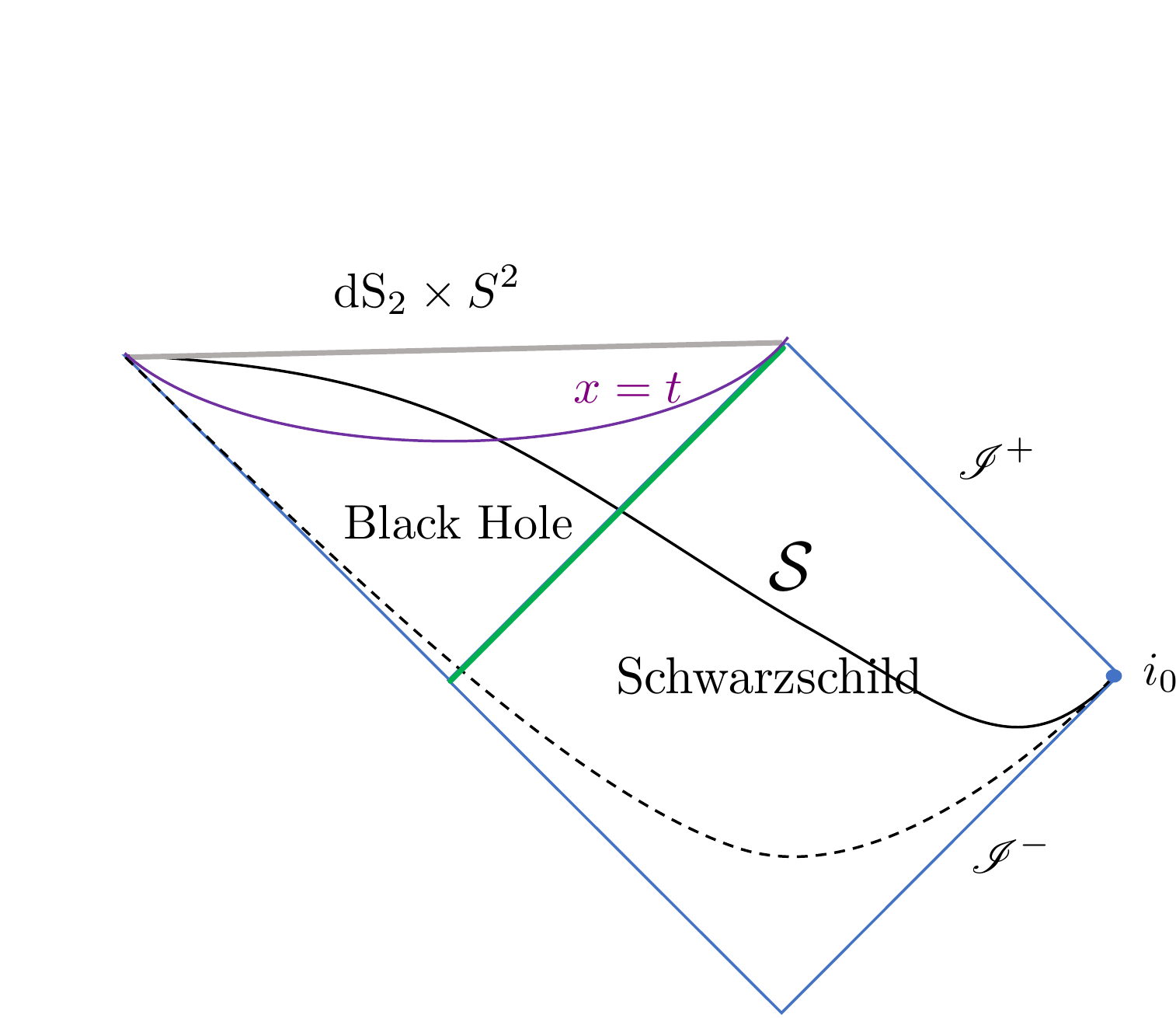} 
 \caption{The quantum effective black hole spacetime covered by $(t,x,\theta,\varphi)$ coordinate. $\cs$ (black curve) is a typical spatial slice with constant $t$. Dashed curves are another spatial slice in the far past. The green line illustrate the event horizon, which bounds the black hole region. The grey triangular region has strong quantum fluctuation and has Plankian curvature. It enclose the classical singularity at $z=x-t=0$ near its past boundary. At the future boundary of the patch, the asymptotic geometry is ${\rm dS}_2\times S^2$ with Planckian radii.}
\label{spacetime}
 \end{center}
\end{figure}

Viewing ${\rm dS}_2\times S^2$ to be the asymptotic geometry, we draw the Penrose diagram of the effective spacetime in Figure \ref{spacetime}. The resulting spacetime has a complete future infinity since ${\rm dS}_2\times S^2$ is complete. The dust time $t$ can extend to $t\to\infty$ in the spacetime as the inflationary coordinate in ${\rm dS}_2$. 

We compute the Einstein tensor of the solution and define the quantum effective stress-energy tensor $T^{\rm (eff)}_{\mu\nu}$ by $R_{\mu\nu}-\frac{1}{2}g_{\mu\nu}R =8\pi G T^{\rm (eff)}_{\mu\nu}$. We find $T^{\rm (eff)}_{\mu\nu}$ violate the average null energy condition, i.e. $\int T^{\rm (eff)}_{UU}\rmd U$ and $\int T^{\rm (eff)}_{VV}\rmd V $ are negative. In concrete, when parameters are e.g. $z_0=3\times10^8$, $\Delta=0.1$, $\b=1$, $R_s=10^8$, $\int T^{\rm (eff)}_{UU}\rmd U\simeq\int T^{\rm (eff)}_{VV}\rmd V\simeq-2.29749 $. the main contributions to $\int T^{\rm (eff)}_{UU}\rmd U,\int T^{\rm (eff)}_{VV}\rmd V$ are from the regions with local maxima of $\ck$. Here $T^{\rm (eff)}_{\mu\nu}$ does not correspond to any physical matter (the dust density is approximately zero in this solution), but rather is an effective account of the LQG effect in the black hole.

%\item The null expansion of $U^\a,V^\a$ are given by
%\be
%\Theta_U=\frac{\left(\sqrt{E^x(z)}-E^\varphi(z)\right) E^x{}'(z)}{2  E^x(z) E^\varphi(z)},\quad\Theta_V=\frac{\left(-\sqrt{E^x(z)}-E^\varphi(z)\right) E^x{}'(z)}{2  E^x(z) E^\varphi(z)}
%\ee
%$\Theta_U=0$ and $\Theta_V<0$ at the classical event horizon $x\simeq 6.67\times10^7$. 

The numerical errors can be tested by inserting numerical solutions back into the EOMs. We find the EOMs are satisfied by solutions up to numerical errors which are bounded by $\sim 10^{-21}$ with Julia and by $\sim 10^{-8}$ with Mathematica \cite{github}. %In addition, recall that $\cc_x(x)$ are conserved quantities, the solution reduces to Schwarzschild spacetime with $\cc_x(x)=0$ in the semiclassical regime, so $\cc_x(x)=0$ is preserved in the $z$-evolution by an exact solution of EOMs. The conservation of $\cc_x$ provides a different method to show numerical errors, we compute $\cc_x/\sqrt{\det(q)}$ where $\sqrt{\det q}=E^\varphi\sqrt{E^x}$ and find it bounded by $\sim 10^{-21}$ by Julia and by $\sim 10^{-10}$ with Mathematica. Note that $\cc_x$ is a density of wight-1, while the density factor $\sqrt{\det(q)}$ in $\cc_x$ grows exponentially at large $-z$ (Figure \ref{volume111}). Therefore a sensible quantity to truly show the numerical error is $\cc_x/\sqrt{\det(q)}$.

%\begin{figure}[h]
%\begin{center}
%\includegraphics[width = 0.9\textwidth]{nerrornew} 
% \caption{(a) Viewing Eqs.\Ref{effeom1} - \Ref{effeom4} as a vector equation $\vec{\mathrm{EOMs}}=0$ where $\vec{\mathrm{EOMs}}$ has 4 components, the norm $\|\mathrm{EOMs}\|$ shows the numerical error of the solution. (b) The numerical error shown by $\cc_x/\sqrt{\det(q)}$. }
%\label{nerrors}
% \end{center}
%\end{figure}

\subsection{Perturbation and stability}\label{Stability of dSS2}

\begin{figure}[t]
\begin{center}
\includegraphics[width = 0.9\textwidth]{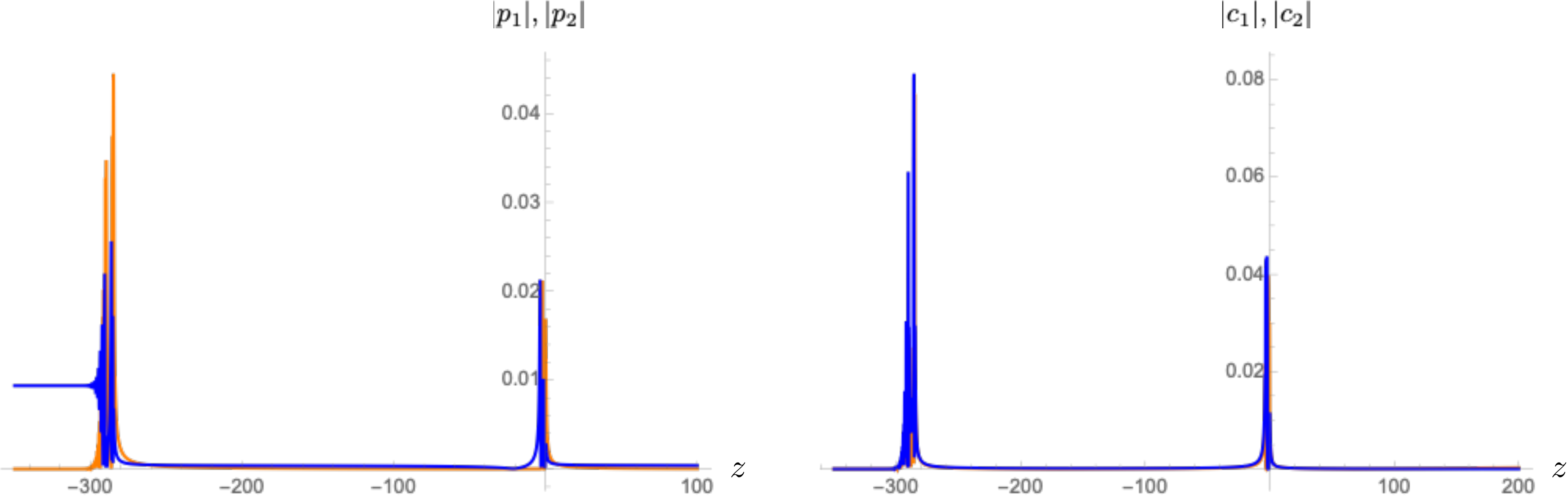} 
 \caption{Plots of absolute values $|p_1(\xi,k)|,|p_2(\xi,k)|,|c_1(\xi,k)|,|c_2(\xi,k)|$ of a typical solution of perturbations, at $k=1.0076$ and random initial conditions $\sim 10^{-5}$ (at $z_{initial}=10^3$). The orange curves are for $p_1,c_1$ while the blue curves are for $p_2,c_2$. Asymptotically as $z\to-\infty$, $p_1,c_1,c_2$ damp off while $p_2$ stabilizes with a finite tail, consistent with the analytic result \Ref{pertu1}-\Ref{p2final}.}
\label{perturbations}
 \end{center}
\end{figure}

In this subsection, we exam the stability of the asymptotic geometry ${\rm dS}_2\times S^2$ by turning on some perturbations. We still assume perturbations to satisfy the spherical symmetry, so we are going to linearize EOMs \Ref{effeom1} - \Ref{effeom4} at the asymptotic background ${\rm dS}_2\times S^2$. In practice, we make the change of variable from $(E^x,E^\varphi,K_x,K_\varphi)$ to $(E^x,E^\varphi,K_1,K_2)$ where
\be
K_1=\frac{\sqrt{E^x{}} K_x{}}{E^\varphi{}},\quad K_2=\frac{K_\varphi{}}{\sqrt{E^x{}}}.
\ee
and insert in the EOMs the perturbation ansatz: 
\be
K_1(t,x)&=&\mathring{K}_1(z)\lt[1+\eps c_1(t,x)\rt],\quad K_2(t,x)=\mathring{K}_2(z)\lt[1+\eps c_2(t,x)\rt]\\
E^x(t,x)&=&\mathring{E}^x(z)\lt[1+\eps p_1(t,x)\rt],\quad E^\varphi(t,x)=\mathring{E}^\varphi(z)\lt[1+\eps p_2(t,x)\rt]
\ee
where $\eps\ll1$ and $z=x-t$. On ${\rm dS}_2\times S^2$, $\mathring{E}^x(z)=r_0^2$, $\mathring{E}^\varphi(z)=r_0e^{-\a_1-\a_0^{-1}z}$, and $\mathring{K}_1,\mathring{K}_2,r_0,\a_0,\a_1$ are constants. For example, the numerical solution with parameters $\Delta=0.1,\ R_s=10^8,\ \b=1$ gives
\be
\mathring{K}_1= 2.48365,\quad \mathring{K}_2= -1.85699, \quad r_0^2= 0.124823,\quad \a_1 = 321.05,\quad \a^{-1}_0=1.17395.
\ee 
When inserting the ansatz and expand the EOMs in $\eps$, $O(\eps^0)$ is identically satisfied for the asymptotic background, and $O(\eps^1)$ gives
\be
&&0.0528952 p_1^{(1,0)}(t,x)+0.0450575 p_1^{(2,0)}(t,x)+0.2 p_1(t,x)+0.34073 p_2^{(1,0)}(t,x)=0,\\
&&-0.170365 p_1^{(1,0)}(t,x)-0.2 p_1(t,x)+0.0623492 p_2^{(1,0)}(t,x)+0.0531106 p_2^{(2,0)}(t,x)=0
\ee
after eliminating $c_1,c_2$ and neglecting terms that are exponentially suppressed as $t\to\infty$, since we are interested in the asymptotic behavior of perturbations as $t\to\infty$ ($z\to-\infty$) where ${\rm dS}_2\times S^2$ is located. The solution is given by
\be
{p_1}(t,x)&=& e^{-0.586975 t} \big(\sin (5.32462 t)\lt[
0.203424   f_1(x)
+0.187807   f_2(x)
-0.0290504   f_4(x)\rt]\nonumber\\
&&+\ \cos (5.32462 t)\lt[0.154682  f_1(x)
+0.263524  f_4(x)\rt]\big)\nonumber\\
&&+\ e^{-1.17395 t}\lt[0.845318 f_1(x)-0.263524 f_4(x)\rt],\label{pertu1}\\
{p_2}(t,x)&= &-0.291431 f_1(x)+0.111783 f_2(x)+ f_3(x)+0.131762 f_4(x)\nonumber\\
&&+\ e^{-0.586975 t}\big(\sin (5.32462 t)\lt[0.0787198   f_1(x)
-0.0123227   f_2(x)
+0.158757  f_4(x)\rt]\nonumber\\
&&-\ \cos (5.32462 t)\lt[0.131228   f_1(x)
+0.111783   f_2(x)\rt]\big)\nonumber\\
&&+\ e^{-1.17395 t}\lt[0.422659 f_1(x)
-0.131762 f_4(x)\rt]\label{pertu2}\\
{c_1}(t,x)&=& 0.181674 {p_1}^{(1,0)}(t,x),\quad {c_2}(t,x)= -0.572819 {p_2}^{(1,0)}(t,x),\label{pertu3}
\ee
where initial perturbations $f_1,\cdots,f_4$ are arbitrary functions of $x$. As $t\to\infty$, asymptotically $p_1,c_2,c_2$ exponentially damp off while 
\be
p_2\to-0.291431 f_1(x)+0.111783 f_2(x)+ f_3(x)+0.131762 f_4(x) \label{p2final}
\ee 
is controlled by the initial perturbation. We conclude that the asymptotic geometry ${\rm dS}_2\times S^2$ is stable with linear perturbations. The perturbation of $p_2$ in \Ref{p2final} modifies $\a_1$ in \Ref{asympELambda} and effectively defines a transformation in $\ch_{\rm dS_2\times S^2}$ from $|r_0,\a_0;\a_1\rangle$ to $|r_0,\a_0;\a_1'\rangle$. We are going to come back to this point in Section \ref{Infinitely many infrared states}. Although ${\rm dS}_2\times S^2$ is stable with linear perturbations, as we are going to see in Section \ref{Evidence of quantum tunneling}, this ${\rm dS}_2\times S^2$ geometry may be unstable by non-perturbative quantum effect, and have nontrivial quantum transit by tunneling effect.

We can extend the study of perturbations on the entire effective black hole spacetime. Since the background spacetime metric components only depend on $z$, we can rewrite the perturbations in terms of $\xi=-z=t-x$ and $x$
\be
K_1(t,x)&=&\mathring{K}_1(z)\lt[1+\eps c_1(\xi,x)\rt],\quad K_2(t,x)=\mathring{K}_2(z)\lt[1+\eps c_2(\xi,x)\rt]\\
E^x(t,x)&=&\mathring{E}^x(z)\lt[1+\eps p_1(\xi,x)\rt],\quad E^\varphi(t,x)=\mathring{E}^\varphi(z)\lt[1+\eps p_2(\xi,x)\rt]
\ee
and make Fourier transformations of perturbations along $x$, e.g. 
\be
p_1(\xi,x)=\int\frac{\rmd k}{\sqrt{2\pi}}\, p_1(\xi,k)e^{-i k x}.
\ee
We numerically solve $p_1(\xi,k),p_2(\xi,k),c_1(\xi,k),c_2(\xi,k)$ from linearizing EOMs \Ref{effeom1} - \Ref{effeom4} on the entire effective black hole spacetime studied in the lasted subsection. By choosing randomly $k$ and initial conditions at $\xi\to-\infty$, numerical experiments of generating solutions of perturbations indicate that $p_1(\xi,k),p_2(\xi,k),c_1(\xi,k),c_2(\xi,k)$ are always bounded from above in the time evolution, with upper bounds controlled by initial values. A typical example is shown in Figure \ref{perturbations}. The asymptotic behaviors of perturbations as $\xi\to\infty $ give $p_1(\xi,k),c_1(\xi,k),c_2(\xi,k)\to0$ and $p_2(\xi,k)$ approaching to constant, consistent with the above analytic result in ${\rm dS}_2\times S^2$.

\section{Picture of black hole evaporation}\label{Picture of black hole evaporation}

\begin{figure}[t]
\begin{center}
\includegraphics[width = 0.7\textwidth]{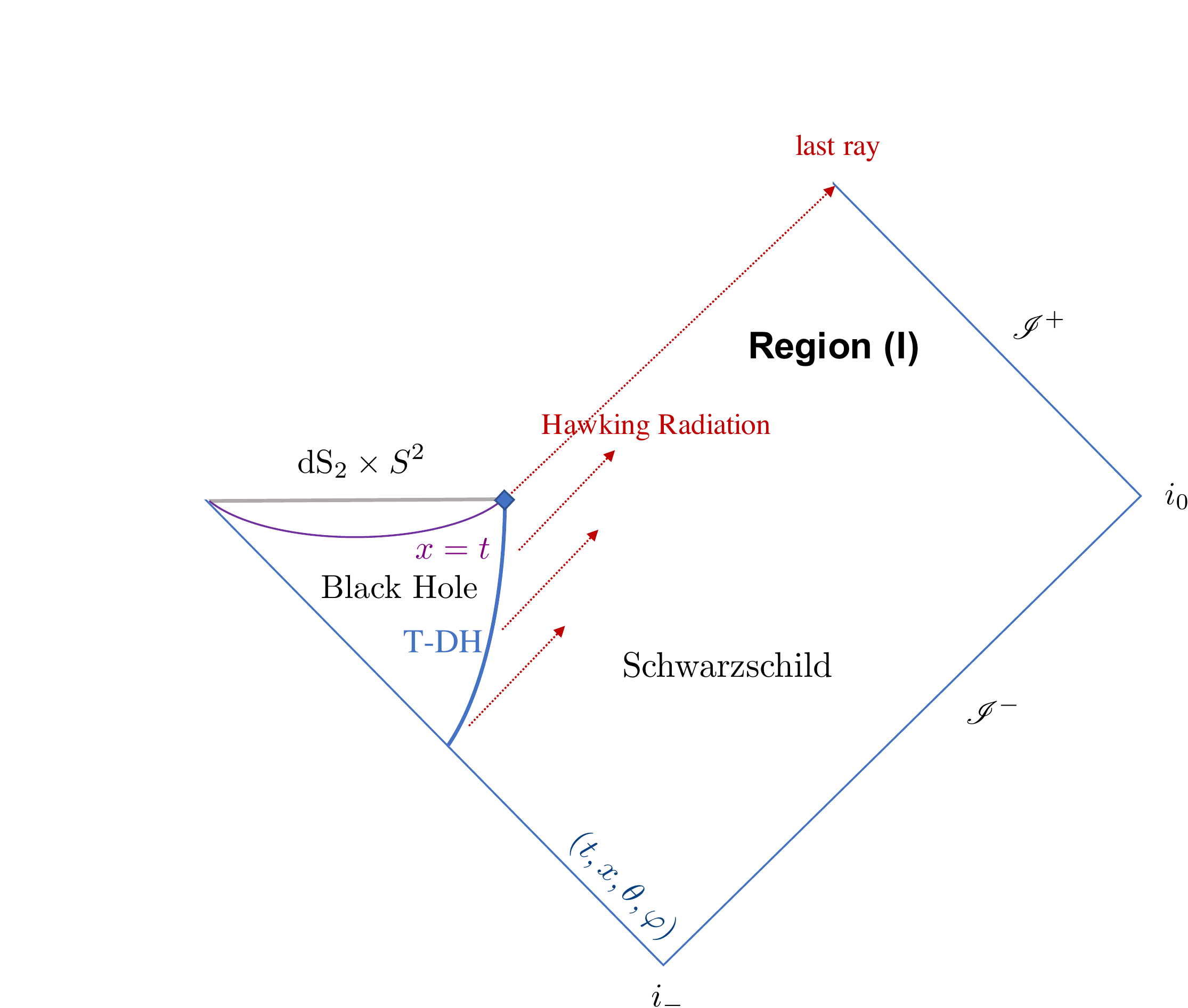} 
 \caption{The picture of quantum effective black hole by taking into account the back-reaction from Hawking radiation. The blue curve is the timelike T-DH of the evaporating black hole. The blue diamond illustrates the regime where the horizon area is comparable to $\Delta$.}
\label{hawking}
\end{center}
\end{figure}

\begin{figure}[t]
\begin{center}
\includegraphics[width = 1\textwidth]{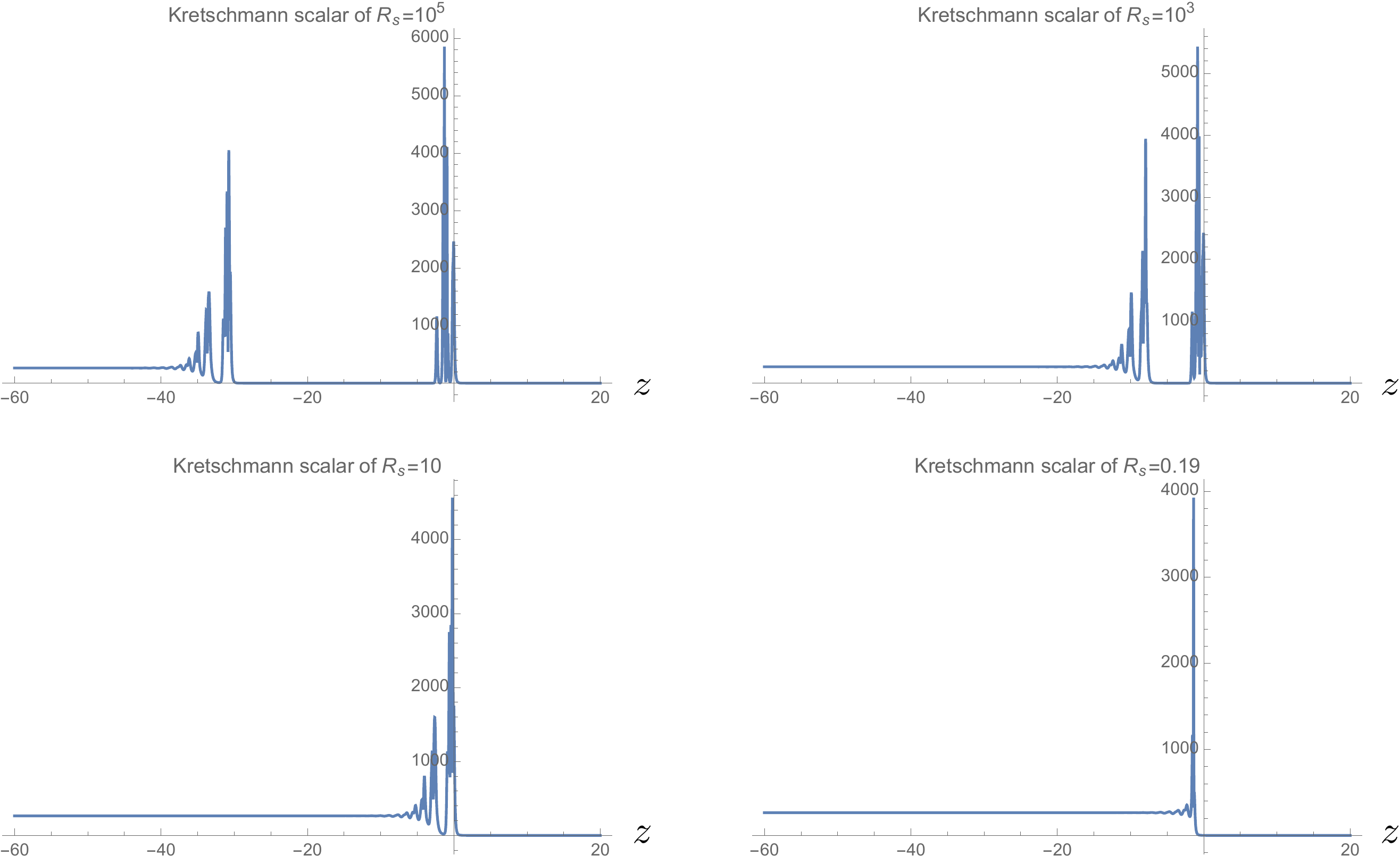} 
 \caption{$N_<$ and $N_0$ become closer as $R_s$ decreasing and finally merge. Other parameters for these solutions are $z_0=3\times10^5$, $\Delta=0.1$, $\b=1$. The maximum of $\ck$ slowly decreases as $R_s$ becoming small. }
\label{smallrs}
\end{center}
\end{figure}

The quantum correction at the event horizon $z=\frac{2}{3}R_s$ is negligible in the black hole solution of effective EOMs. The geometry near and outside the horizon almost has no difference from the classical Schwarzschild spacetime. By turning on quantum field perturbations, the Hawking's original derivation of black hole evaporation happening near the horizon carries over to the black hole spacetime obtained here. The quantum correction from nonzero $\Delta$ to Hawking's derivation is negligible. The back-reaction from Hawking radiation reduces the black hole mass and causes the horizon to shrink. Then the event horizon should be replaced by the trapping dynamical horizon (T-DH). A T-DH is a 3-dimensional time-like submanifold foliated by 2-dimensional surfaces $\fh$ with 2-sphere topology, so that at each leaf $\fh$, $\Theta_k=0$ and $\Theta_l<0$ where $\Theta_k$ and $\Theta_l$ are expansions of outward and inward null normals of $\fh$ \cite{Ashtekar:2003hk}. Moreover, for the semiclassical spacetime outside and far from the black hole, the future null infinity $\mathscr{I}^+$ is extended until the ``last ray'': the last Hawking particle radiated from the black hole before the evaporation stops. The picture of quantum effective black hole spacetime is illustrated in Figure \ref{hawking}. The black hole evaporation results in the existence of the classical asymptotic flat regime which we call the Region (I). All future null rays from points in the Region (I) does not intersect with the black hole horizon, in contrast to the spacetime in Figure \ref{spacetime} where the inward future null ray always cross the horizon.

\begin{figure}[t]
\begin{center}
\includegraphics[width = 0.5\textwidth]{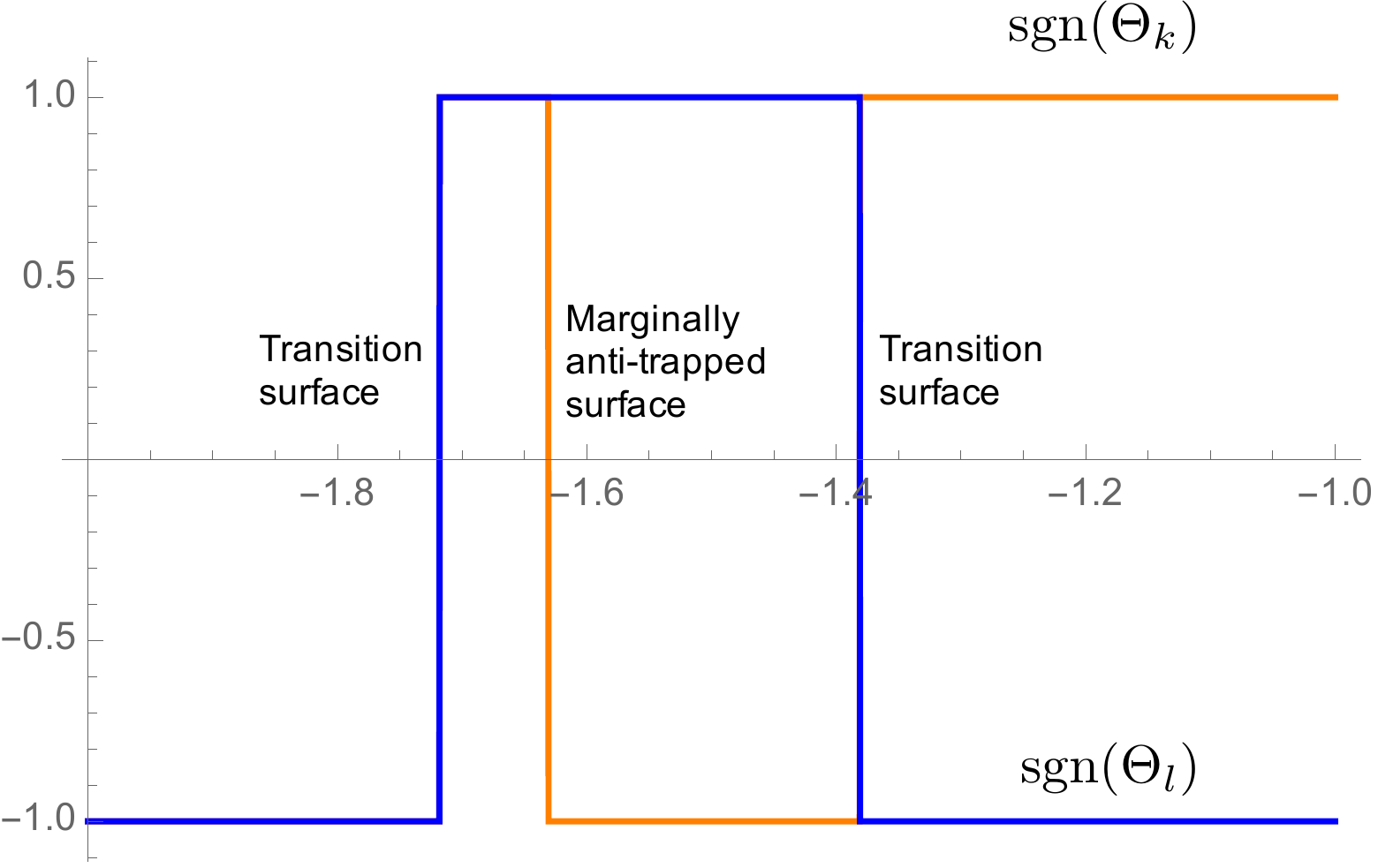} 
 \caption{At the instance with $R_s=0.19$, the spatial slice $\cs_t$ has no marginal trapped surface but has transition surfaces with $\Theta_k=\Theta_l=0$ and a marginal anti-trapped surface (null surface) with $\Theta_k=0,\Theta_l>0$. }
\label{traps}
\end{center}
\end{figure}

\begin{figure}[h]
\begin{center}
\includegraphics[width = 0.9\textwidth]{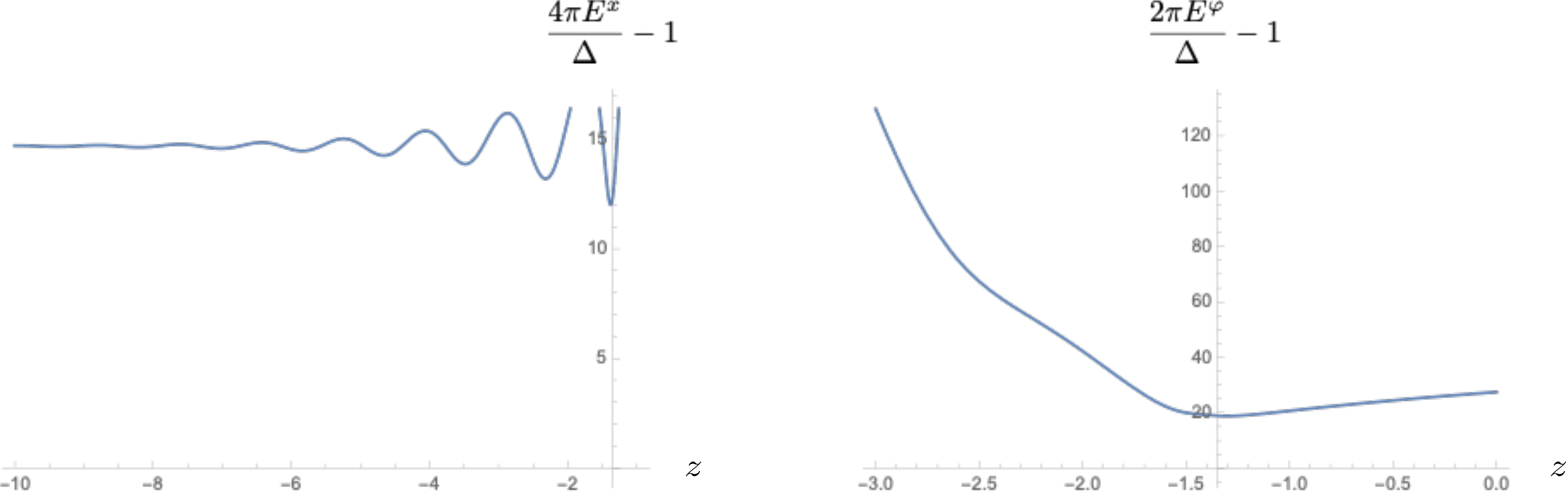} 
 \caption{$4\pi E^x >\Delta,\ 2\pi E^\varphi >\Delta$ is satisfied at $R_s=0.19$.}
\label{areabound}
\end{center}
\end{figure}

%The picture Figure \ref{hawking} can be supported by numerical solutions with different $R_s$. 

Here we assume the evaporation process is sufficiently slow such that at every instance the spacetime can be approximately described by the solution of effective EOMs with a fixed $R_s$. Foliating the spacetime with solutions with different $R_s$ approximates the dynamical black hole spacetime.  

As an advantage of solutions following the anzatz \Ref{ansztzz}, when we can set a constant $t=t_0$, the solution $E^x(x-t_0),E^\varphi(x-t_0),K_x(x-t_0),K_\varphi(x-t_0)$ describe the geometry on a constant $t=t_0$ spatial slices $\cs$. The 2-sphere $\fh\subset\cs$ given by $x-t_0=\frac{2}{3}R_s$ is a marginal trapped surface with $\Theta_k=0$ and $\Theta_l<0$ where $\Theta_k$ and $\Theta_l$ are outward and inward null expansions (see Figure \ref{expansions}). We may implement two different numerical solutions of different parameters $R_s(t_0), R_s(t_1)$ at different spatial slices $\cs_{t_0},\cs_{t_1}$ at two instances $t=t_0,t_1$. The marginal trapped surfaces $\fh(t_0)$ and $\fh(t_1)$ are at $x-t_0=\frac{2}{3}R_s(t_0)$ and $x-t_1=\frac{2}{3}R_s(t_1)$ On $\cs_{t_0}$ and $\cs_{t_1}$ respectively. If the evaporation is slow enough, the dynamical black hole can be approximated by a large number of spatial slices $\cs_{t}$ carrying different solutions with different horizon radii $R_s(t)$ which is monotonically decrease as $t$ growing. The T-DH is foliated by the set of marginal trapped surfaces $\{\fh(t)\}_t$, $\fh(t)\subset\cs_t$. 

We run numerical experiments of solving effective EOMs (Eqs.\ref{ODE2} by the ansatz \Ref{ansztzz}) with smaller and smaller $R_s$ ($R_s$ is implemented by the initial condition \Ref{bc0} and \Ref{bc1}). From the results, we find that the asymptotic ${\rm dS}_2\times S^2$ geometry is invariant under changing $R_s$, although details of curvature fluctuations and bounces between semiclassical Schwarzschild and ${\rm dS}_2\times S^2$ can change. In particular two neighborhoods $N_<$ and $N_0$ (of 2 groups of local maxima of $\ck$) become closer as $R_s$ decreasing, and finally merge when the horizon area $4\pi R^2_s$ is comparable to $\Delta$ (see Figure \ref{smallrs}). When $N_<$ and $N_0$ merge, the bounce near $z=0$ looks like a domain wall separating the semiclassical (low curvature) Schwarzschild spacetime and the quantum (high curvature) ${\rm dS}_2\times S^2$ spacetime.

Remarkably, when $R_s$ is small such that $4\pi R^2_s$ is comparable to $\Delta$, e.g. in the case that $R_s=0.19$ in Figure \ref{smallrs}, the T-DH disappears and is replaced by a spacelike transition surface with both $\Theta_k=0$ and $\Theta_l=0$ (see Figure \ref{traps}). It is followed by a marginal anti-trapped null surface with $\Theta_k=0,\Theta_l>0$ on the left (at smaller $x$), then followed by anti-trapped and trapped regions (due to small fluctuations of geometry after the bounce, as seen in Figure \ref{smallrs}) before approaching to ${\rm dS}_2\times S^2$. Since the T-DH disappears, Hawking's derivation of black hole evaporation fails to valid in this regime. We expect the quantum gravity effect to be strong in the region of transition surface since the spacetime curvature at the surface is Planckian. The last ray of Hawking radiation should happen before this instance.

Note that the small $R_s$ doesn't break $4\pi E^x >\Delta,\ 2\pi E^\varphi >\Delta$ (see Figure \ref{areabound}) as long as $4\pi R_s^2 >\Delta$ \footnote{The situation $4\pi R_s^2 \leq\Delta$ is not tested since the evolution in this situation reaches the singularity of the ODEs. But $4\pi R_s^2 \leq\Delta$ is inconsistent with the $\bar{\mu}$-scheme regularization.}, so our results are self-consistent with the starting point of $\bar{\mu}$-scheme regularization.

The region where the Hawking radiation stops is the blue diamond in Figure \ref{hawking}. This region is a neighborhood of the transition surface in Figure \ref{traps}, and has the strong quantum gravity effect \cite{Haggard:2014rza,Ashtekar:2020ifw} since the curvature is Planckian. This region should also be strongly dynamical. Our analysis of the effective dynamics and approximation using foliations of solutions with $R_s(t)$ should be not adequate although it provides a preliminary picture of this phase. The rigorous analysis of this region should apply the full theory of LQG. The study on this aspect is beyond the scope of the present paper.

\section{Black hole to white hole transition}\label{Black hole to white hole transition}

\begin{figure}[t]
\begin{center}
\includegraphics[width = 1\textwidth]{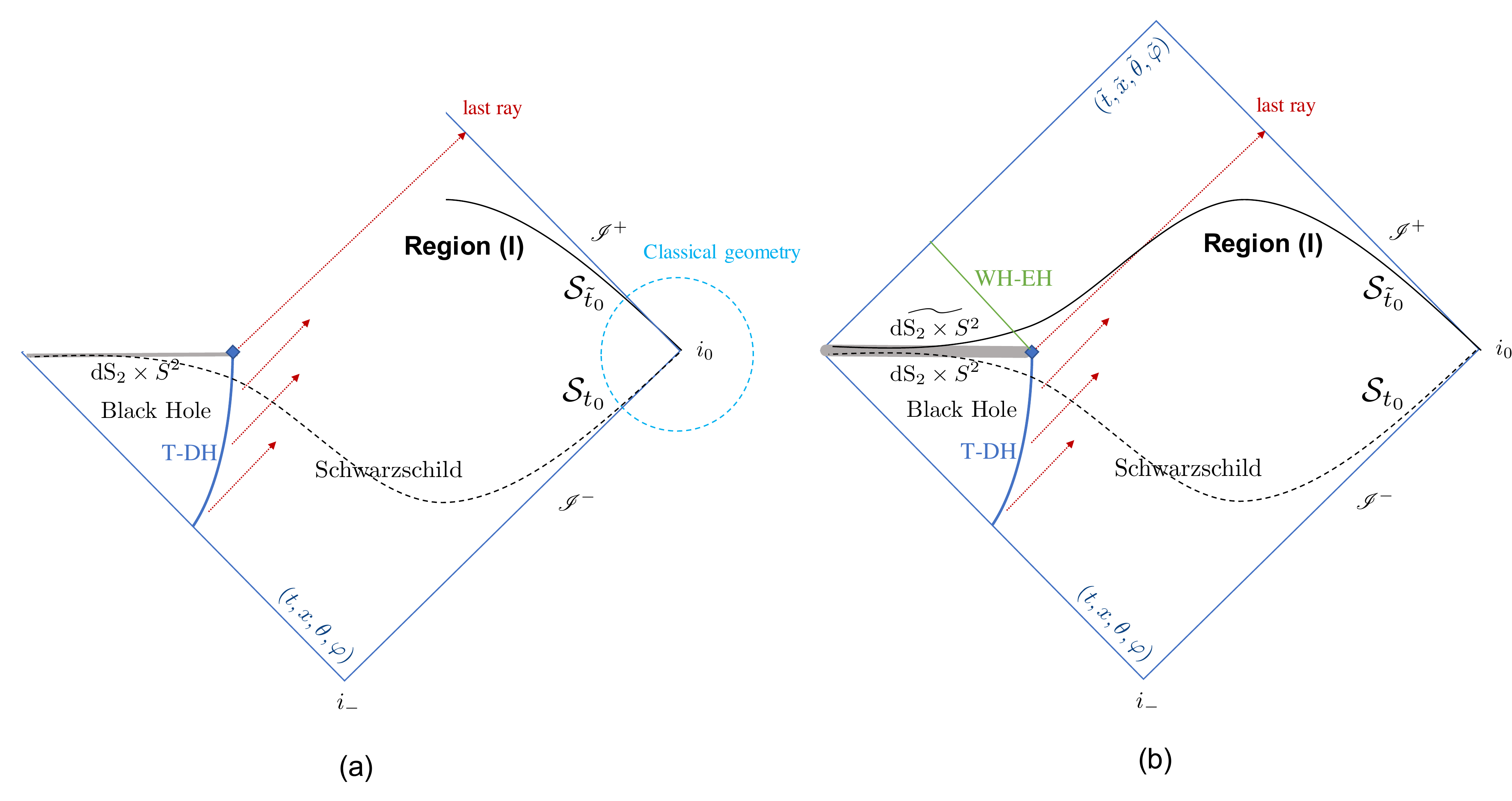} 
 \caption{(a) The picture of Region (I) extended from the Schwarzschild spacetime in the past: Assuming the dynamics near $i_0$ (enclosed by the light blue dashed circle) can be approximated by classical asymptotically flat geometry with negligible dynamical effect or back-reaction from Hawking radiation. The internal and external geometries of $\cs_{\tilde{t}_0}$ near $i_0$ are consistent with the slice $\cs_{{t}_0}$ in the close past of the blue diamond; (b) An extended effective spacetime geometry beyond the last ray suggests the black-hole-to-white-hole transition. There is a ${\rm dS}_2\times S^2$ geometry (denoted by $\widetilde{{\rm dS}_2\times S^2}$) behind the white hole event horizon (green line). The grey region between ${\rm dS}_2\times S^2$ and $\widetilde{{\rm dS}_2\times S^2}$ contains a quantum tunneling. The entire spacetime includs two coordinate patches $({t},{x},{\theta},{\varphi})$ and $(\tilde{t},\tilde{x},\tilde{\theta},\tilde{\varphi})$}.
\label{BHWH}
 \end{center}
\end{figure}

The black hole spacetime in Figure \ref{hawking} is incomplete due to the existence of the Region (I). The future null infinity $\mathscr{I}^+$ should be extended beyond the last ray of Hawking radiation (see \cite{Ashtekar:2010qz} for an earlier study of extension in 2d dilaton black hole). We are going to use the effective EOMs of ${\bf H}_\Delta$ to derive the extension. We make the following assumptions of Region (I) in order to obtain boundary conditions for the effective EOMs:

\begin{enumerate}

\item The quantum dynamics in Region (I) near the spatial infinity $i_0$ can be well approximated by the quantum field theory on classical background spacetime. The dynamical effect is weak near $i_0$. The spacetime near $i_0$ is Schwarzschild with certain ADM mass.

\item For all spatial slices Region (I), their asymptotic geometries (metrics and extrinsic curvatures) near the spatial infinity $i_0$ are classical and asymptotically flat. Their geometries are continuous extensions from geometries in the past. Namely there exists a slice $\cs_{\tilde{t}_0}$ in Region (I) such that its asymptotic geometry near $i_0$ should be consistent with the asymptotic geometry of $\cs_{t_0}$ in the close past of the blue diamond (see Figure \ref{BHWH}(a)). In particular, their ADM masses are approximately the same. 

\end{enumerate}

We find above assumptions are physically reasonable and should be an excellent approximation of the full quantum dynamics. We note that in the most rigorous treatment where the back-reaction from Hawking radiation to Region (I) are taken into account, the ADM mass of $\cs_{\tilde{t}_0}$ should in principle include the energy of Hawking radiation, since all Hawking particles register in  $\cs_{\tilde{t}_0}$. However in the present work, we assume the energy density of the Hawking radiation is small and ignore its back-reaction to the geometry.

Now we focus on the asymptotic geometry of Region (I) near the spatial infinity $i_0$: The neighborhood of $\cs_{\tilde{t}_0}$ near infinity has the semiclassical Schwarzschild geometry with $R_s=2 M_r$ obtained from extending the geometry from the past, i.e. from $\cs_{t_0}$. Here $M_r$ is the remnant black hole mass before Hawking radiation stops. We may draw infinitely many spatial slices $\cs_{\tilde{t}}$ in Region (I) such that all $\cs_{\tilde{t}}$ with $\tilde{t}>\tilde{t}_0$ end at $i_0$ and live in the future of $\cs_{\tilde{t}_0}$. For simplicity of the present model, we do not consider to vary the ADM mass in the future of $\cs_{\tilde{t}_0}$. Neighborhoods of all these slices carry the same semiclassical Schwarzschild spacetime geometry near $i_0$. The feature of asymptotic flat geometry is preserved when $\tilde{t}\to\infty$. %This is in contrast to the early time Schwarzschild geometry (in the region between T-DH and $\mathscr{I}^-$) which is asymptotically flat when $t\to -\infty$ (or $z\to\infty$, see Eqs.\Ref{bc0} and \Ref{bc1}). 

The foliation $({t},{x},{\theta},{\varphi})$ cannot be extend to Region (I) due to the strong quantum dynamical effect in the blue diamond region, thus a new foliation by $(\tilde{t},\tilde{x},\tilde{\theta},\tilde{\varphi})$ is necessary for studying the dynamics of the spacetime including $\cs_{\tilde{t}_0}$ and its future. We impose the following boundary conditions near $i_0$ corresponding to $\tilde{z}=\tilde{x}-\tilde{t}\to-\infty$ for slices in Region (I):
\be
&&E^{\tilde{x}}(\tilde{z})\sim  -\left(\frac{3}{2}\sqrt{R_s}
  (- \tilde{z})\right)^{4/3},\quad E^{\tilde{\varphi}}(\tilde{z})\sim  \sqrt{R_s}
   \lt(\frac{3}{2}\sqrt{R_s} (-\tilde{z})\rt)^{1/3},\label{newbc0}\\
 &&  K_{\tilde x}(z)\sim \frac{R_s}{3\times 2^{2/3} {3}^{1/3}
   \left(\sqrt{R_s} (-\tilde{z})\right)^{4/3}},\quad K_{\tilde \varphi}(\tilde{z})\sim\frac{\lt(\frac{2}{3}\rt)^{1/3}
   \sqrt{R_s}}{\lt({\sqrt{R_s} (-\tilde{z})}\rt)^{1/3}},\label{newbc1}
% &&  E^x{}'(z_0)=\frac{3 \lt(\frac{3}{2}\rt)^{1/3} R_s^{2/3} z_0^{4/3} \sin \left(\frac{2 \beta  \sqrt{\Delta }}{3 z_0}\right) \cos \left(\frac{\beta  \sqrt{\Delta }}{3 z_0}\right)}{\beta \sqrt{\Delta }}\label{bc2}
\ee  
%These boundary conditions may be obtained from Eqs.\Ref{bc0} and \Ref{bc1}) by the reflection $x\to -\tilde{x}$ and $t\to -\tilde{t}$. 
where $R_s$ equals $2M_r$. We are going to apply the boundary conditions \Ref{newbc0} and \Ref{newbc1} to the effective EOMs of ${\bf H}_\Delta$. The resulting solution extends the effective spacetime beyond the last ray.

\begin{figure}[t]
\begin{center}
\includegraphics[width = 0.9\textwidth]{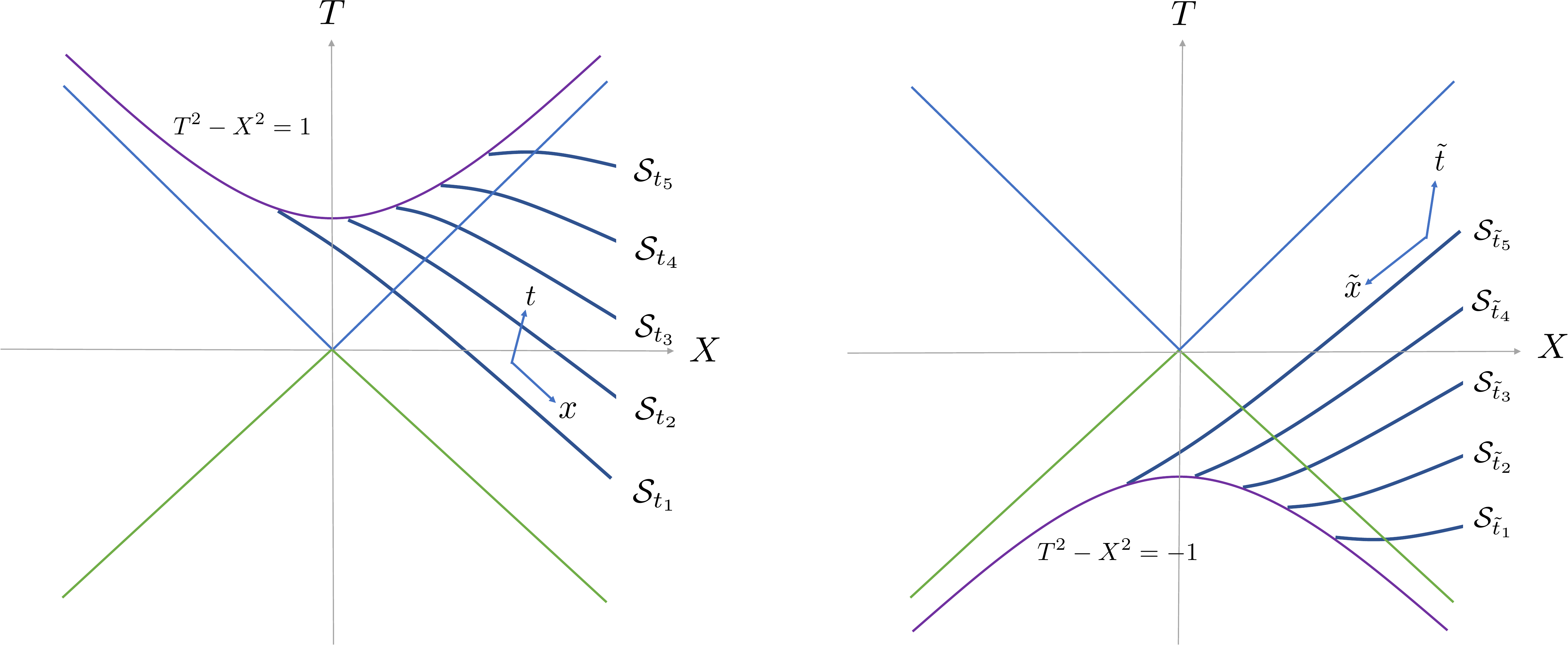} 
 \caption{Lema\^{\i}tre foliations $({t},{x},{\theta},{\varphi})$ (left) and $(\tilde{t},\tilde{x},\tilde{\theta},\tilde{\varphi})$ (right) in Kruskal coordinates. Both $t_1,\cdots,t_5$ and $\tilde{t}_1,\cdots,\tilde{t}_5$ are monotonically increasing sequences of time.}
\label{GP}
 \end{center}
\end{figure}

The boundary conditions \Ref{newbc0} and \Ref{newbc1} are obtained as follows: In Region (I) near spatial infinity, the foliation $\cs_{\tilde{t}}$ is obtained from $\cs_{{t}}$ by the time reflection symmetry $T\to -T$ of the Schwarzschild-Kruskal geometry (see Figure \ref{GP} for illustration of spatial slices $\cs_t$ and $\cs_{\tilde{t}}$ when they are in the classical Schwarzschild spacetime). We define diffeomorphism $\mathscr{R}$ maps $\cs_t$ to $\cs_{\tilde{t}}$, and relates the coordinate basis by\footnote{We denote the standard Schwarzschild coordinate by $(\t,r,\theta,\phi)$. Its transformation to Lemaitre coordinate is $
d t =d \t+\sqrt{\frac{r_{s}}{r}}\left(1-\frac{r_{s}}{r}\right)^{-1} d r,\ 
d x =d \t+\sqrt{\frac{r}{r_{s}}}\left(1-\frac{r_{s}}{r}\right)^{-1} d r
$. We firstly use the time refection $\t\to \t'=-\t$ maps the foliation $\cs_{{t}}$ to $\cs_{\tilde{t}}$, then define the new coordinate by $
-d \tilde{t} =-d \t'+\sqrt{\frac{r_{s}}{r}}\left(1-\frac{r_{s}}{r}\right)^{-1} d r,\ 
-d \tilde{x} =-d \t'+\sqrt{\frac{r}{r_{s}}}\left(1-\frac{r_{s}}{r}\right)^{-1} d r
$ in terms of the Schwarzschild coordinate $({\t}',r,\theta,\phi)$ in the time-reflected spacetime}
\be
\mathscr{R}_*(\partial/\partial {t})^\a=-(\partial/\partial\tilde{t})^\a,&& \mathscr{R}_*(\partial/\partial {x})^\a=-(\partial/\partial\tilde{x})^\a,\label{xtpushfor}\\
\mathscr{R}_*(\partial/\partial{\theta})^\a=(\partial/\partial \tilde{\theta})^\a,&& \mathscr{R}_*(\partial/\partial {\varphi})^\a=(\partial/\partial\tilde{\varphi})^\a,\\
\text{equivallently},\quad\tilde{t}({p})=-t(\mathscr{R}^{-1}(p)),&&  \tilde{x}(p)=-x(\mathscr{R}^{-1}(p)),\\
\tilde{\theta}(p)=\theta(\mathscr{R}^{-1}(p)),&& \tilde{\varphi}(p)=\varphi(\mathscr{R}^{-1}(p)),
\ee 
for any point $p$ in any $\cs_{\tilde{t}}$. Restricting $\mathscr{R}$: $\cs_t\to\cs_{\tilde{t}}$, $E^{\tilde{x}},E^{\tilde{\varphi}}$ in \Ref{newbc0} and \Ref{newbc1} are obtained by push-forward $\mathscr{R}^*E^\a|_{\cs_{\tilde{t}}}=E^\a|_{\cs_{{t}}}$:
\be
E^{\tilde{x}}(p)&=&(\mathscr{R}_* E)^\a(p)(\rmd \tilde{x})_\a=E^{\a}(\mathscr{R}^{-1}(p))\mathscr{R}^*(\rmd \tilde{x})_\a=-E^{x}(\mathscr{R}^{-1}(p))\\
E^{\tilde{\varphi}}(p)&=&(\mathscr{R}_* E)^\a(p)(\rmd \tilde{\varphi})_\a=E^{\a}(\mathscr{R}^{-1}(p))\mathscr{R}^*(\rmd \tilde{\varphi})_\a=E^{\varphi}(\mathscr{R}^{-1}(p))
\ee
for any point $p\in\cs_{\tilde{t}}$. In terms of coordinates, we obtain that as functions, $E^{\tilde{x}}(\tilde{t},\tilde{x})=-E^{{x}}(-\tilde{t},-\tilde{x})$ and $E^{\tilde{\varphi}}(\tilde{t},\tilde{x})=E^{{\varphi}}(-\tilde{t},-\tilde{x})$, which give \Ref{newbc0}. Since $\mathscr{R}$ leaves the 4d metric invariant, the 4-metric in $(\tilde{t},\tilde{x},\tilde{\theta},\tilde{\varphi})$ is given by
\be
&&\rmd\tilde{s}^2=-\mathrm{d} \tilde{t}^{2}+\Lambda(\tilde{t}, \tilde{x})^{2} \mathrm{d} \tilde{x}^{2}+R(\tilde{t}, \tilde{x})^{2}\left[\mathrm{d} \tilde{\theta}^{2}+\sin ^2\tilde{\theta} \mathrm{d} \tilde{\varphi}^{2}\right], \quad \Lambda=\frac{E^{\tilde \varphi}}{\sqrt{\left|E^{\tilde x}\right|}}, \quad R=\sqrt{\left|E^{\tilde x}\right|}. %\\
%&&\quad E^{\tilde{x}}(\tilde{t}, \tilde{x})=E^{\tilde{x}}(-t,-x)=-E^{x}(t, x), \quad E^{\tilde{\varphi}}(\tilde{t}, \tilde{x})=E^{\tilde{\varphi}}(-t,-x)=E^{\varphi}(t, x)
\ee
The boundary conditions \Ref{newbc1} are given by solving the classical EOMs of ${\bf H}_0$ (in the coordinates $(\tilde{t},\tilde{x},\tilde{\theta},\tilde{\varphi})$) with $E^{\tilde{x}}(\tilde{t},\tilde{x}),E^{\tilde{\varphi}}(\tilde{t},\tilde{x})$ given by \Ref{newbc0}, or equivalently, they can be obtained by computing the extrinsic curvature of $\cs_{\tilde{t}}$ \footnote{Recall that $K_x$ is $1/2$ of the extrinsic curvature $K_j^a$ along the $x$ direction.}.

The extension of black hole exterior geometry in Region (I) can be viewed as the geometry of white hole exterior, as illustrated by Figure \ref{GP}. This aspect is similar to \cite{Haggard:2014rza}.

In order to extend the geometry beyond the last ray from Region (I) to the causal future of black hole interior, we apply the effective EOMs of the improved Hamiltonian ${\bf H}_\Delta$ and implement the boundary conditions \Ref{newbc0} and \Ref{newbc1}. The effective EOMs are given by Eqs.\Ref{effeomop1} - \Ref{effeomop4} since we have $E^x<0$ by the boundary condition (due to Eq.\Ref{xtpushfor}).

Recall that the EOMs \Ref{effeomop1} - \Ref{effeomop4} in $(\tilde{t},\tilde{x},\tilde{\theta},\tilde{\varphi})$ relate to the EOMs \Ref{effeom1} - \Ref{effeom4} in $(t,x,\theta,\varphi)$ by identifying $\tilde{x}=-x$, $\tilde{t}=-t$ (thus $z\to \tilde{z}=-z$) and \Ref{tran01} - \Ref{tran04}) which consistently match the relation between the boundary conditions \Ref{newbc0} and \Ref{newbc1} of $E^{\tilde{x}},E^{\tilde{\varphi}},{K}_{\tilde{x}},{K}_{\tilde{\varphi}}$ at $\tilde{z}\to-\infty$ and \Ref{bc0} and \Ref{bc1} of $E^{{x}},E^{{\varphi}},{K}_x,{K}_\varphi$ at $z\to\infty$. We again apply the ansatz $E^{\tilde{x}}(\tilde{t},\tilde{x})=E^{\tilde{x}}(\tilde{z}),\ E^{\tilde{\varphi}}(\tilde{t},\tilde{x})=E^{\tilde{\varphi}}(\tilde{z}),\ K_{\tilde{x}}(\tilde{t},\tilde{x})=K_{\tilde{x}}(\tilde{z}),\ K_{\tilde{\varphi}}(\tilde{t},\tilde{x})=K_{\tilde{\varphi}}(\tilde{z})$ to reduce the PDEs to ODEs (Eqs.\Ref{ODEop1}), so that the solution is uniquely determined by the boundary conditions \Ref{newbc0} and \Ref{newbc1}. As a result, the solution $E^{\tilde{x}},E^{\tilde{\varphi}},K_{\tilde{x}},K_{\tilde{\varphi}}$ in $(\tilde{t},\tilde{x},\tilde{\theta},\tilde{\varphi})$ is the spacetime inversion of the solution $E^{{x}},E^{{\varphi}},K_{{x}},K_{{\varphi}}$ in $({t},{x},{\theta},{\varphi})$ by the mapping \Ref{tran01} - \Ref{tran04}. All properties of solutions discussed in Section \Ref{Properties of solutions} is carried over to solutions in $(\tilde{t},\tilde{x},\tilde{\theta},\tilde{\varphi})$ by \Ref{tran01} - \Ref{tran04}. In particular, the solution gives asymptotically ${\rm dS}_2\times S^2$ geometry as $\tilde{z}\to\infty$ ($\tilde{t}\to-\infty$ with fixed $\tilde{x}$ or $\tilde{x}\to\infty$ with fixed $\tilde{t}$)
\be
&E^{\tilde{x}}(\tilde{z})\sim -r_0^2,\quad E^{\tilde{\varphi}}(\tilde{z})\sim r_0e^{-\a_1+\a_0^{-1}\tilde{z}},&%\quad \partial_{\tilde z}E^{\tilde{x}}\sim0,%\quad \partial_{\tilde{z}} E^{\tilde{\varphi}}\sim r_0\a_0^{-1}e^{-\a_1+\a_0^{-1}\tilde{z}},\quad\text{as}\quad \tilde{z}\to\infty,
\label{initialEEprime}\\
&\rmd \tilde{s}^2\sim -\rmd \tilde{t}^2+e^{-2\a_1+2\alpha^{-1}_0(\tilde{x}- \tilde{t})}\rmd \tilde{x}^2+r_0^2\lt(\rmd\tilde{\theta}^2+\sin^2\tilde{\theta}\rmd\tilde{\varphi}^2\rt),&
\ee
and constant $\tilde{K}_1,\tilde{K}_2$ ($\tilde{K}_1=\frac{\sqrt{-E^{\tilde{x}}} K_{\tilde{x}}}{E^{\tilde{\varphi}}},\  \tilde{K}_2=\frac{K_{\tilde{\varphi}}}{\sqrt{-E^{\tilde{x}}}}$), i.e. 
\be
\tilde{K}_1(\tilde{z}\to\infty)={K}_1({z}\to-\infty),\label{KK1}\\
\tilde{K}_2(\tilde{z}\to\infty)=-{K}_2({z}\to-\infty),\label{KK2}
\ee
are constant. The extended spacetime is illustrated by Figure \ref{BHWH}(b). $\a_0,r_0$ are independent of $R_s$ and the same as in \Ref{alpha0r0}, while $\a_1$ depends on $R_s$. When we compare to the ${\rm dS}_2\times S^2$ geometry obtained in Section \Ref{Properties of solutions} and attempt to glue this 2 version of ${\rm dS}_2\times S^2$, we find that, if we identify their spatial slices, they have the same $E^j_a$ (minus sign in $E^{\tilde{x}}$ is due to $\tilde{x}=-x$) when they are obtained from the same $R_s$, but they have opposite $K_j^a$ (no minus sign in $K_1$ is again due to $\tilde{x}=-x$), so classically they cannot be glued. By this reason, we denote this new ${\rm dS}_2\times S^2$ geometry by $\widetilde{{\rm dS}_2\times S^2}$. The time orientation of $\widetilde{{\rm dS}_2\times S^2}$ is opposite to ${\rm dS}_2\times S^2$. The discussion in Section \ref{Evidence of quantum tunneling} suggests that the transition from ${{\rm dS}_2\times S^2}$ to $\widetilde{{\rm dS}_2\times S^2}$ may be due to the quantum tunneling. 

\begin{figure}[h]
\begin{center}
\includegraphics[width = 0.5\textwidth]{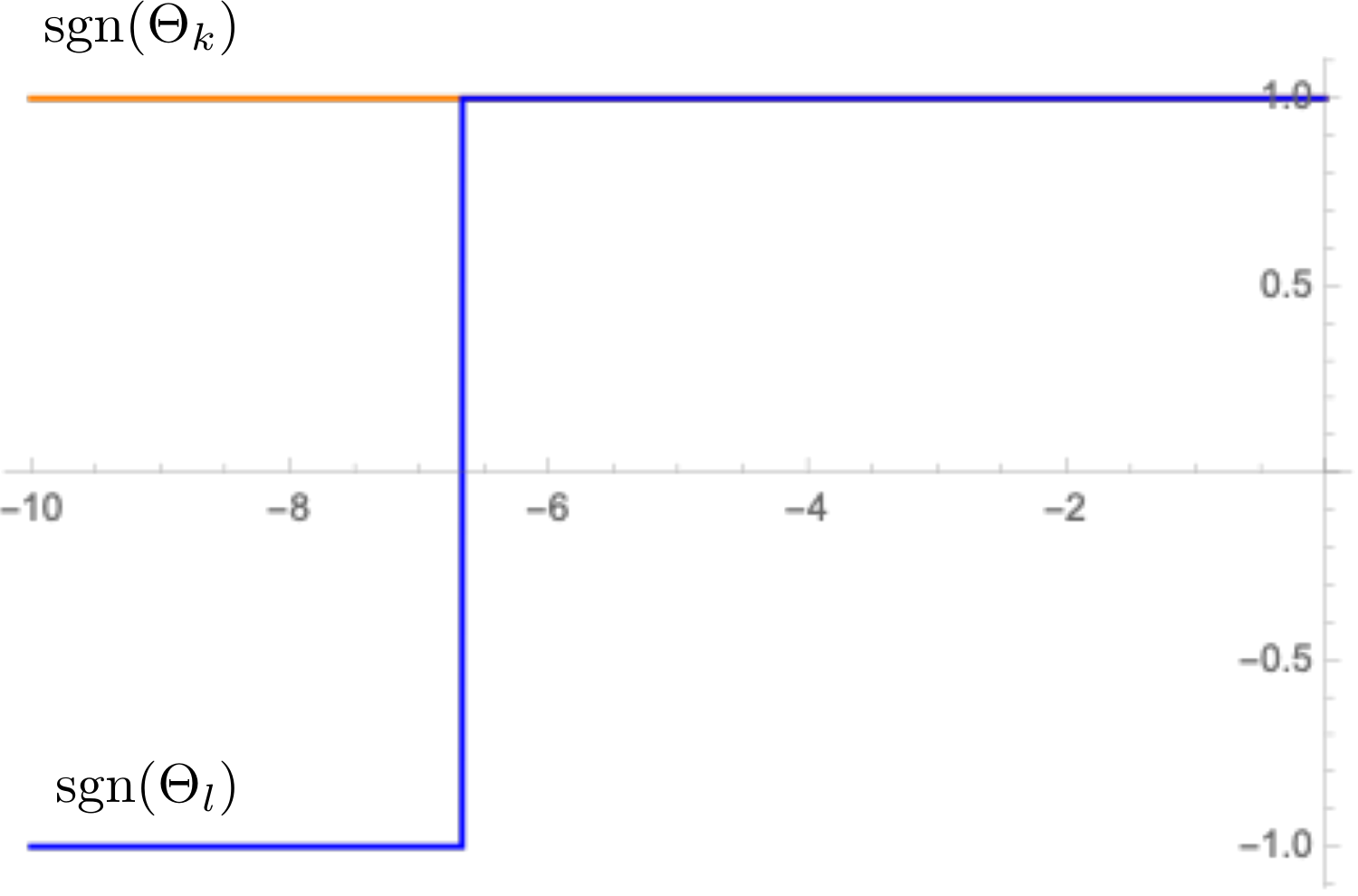} 
 \caption{The white hole horizon ($\Theta_l=0$ and $\Theta_k>0$) at $\tilde{z}=\tilde{z}_s=-6.66614$ based on the numerical solution with $R_s=10,\ \Delta=0.1$. The relative correction $|\frac{\tilde{z}_s-(-\frac{2}{3}R_s)}{-\frac{2}{3}R_s}|\simeq 7\times 10^{-5}$.}
\label{antitrap}
 \end{center}
\end{figure}

The solution $E^{\tilde{x}},E^{\tilde{\varphi}},K_{\tilde{x}},K_{\tilde{\varphi}}$ extends the spacetime geometry beyond the last ray. When fixing $\tilde{t}=\tilde{t}_0$, the solution $E^{\tilde{x}}(\tilde{x}-\tilde{t}_0),E^{\tilde{\varphi}}(\tilde{x}-\tilde{t}_0),K_{\tilde{x}}(\tilde{x}-\tilde{t}_0),K_{\tilde{\varphi}}(\tilde{x}-\tilde{t}_0)$ extends the internal and external geometries of $\cs_{\tilde{t}_0}$ from Region (I) to the future of the blue diamond. Given the numerical solution, we find at $\tilde{z}=\tilde{x}-\tilde{t}_0=-\frac{2}{3}R_s$ the marginal anti-trapped surface where $\Theta_l=0$ and $\Theta_k>0$, while $\tilde{z}>-\frac{2}{3}R_s$ is the anti-trapped region with both $\Theta_k,\Theta_l>0$ (see Figure \ref{antitrap}). $\Theta_l$ and $\Theta_k$ are inward and outward null expansion. If $R_s=2M_r$ is fixed and we do not consider the loss of white hole mass, $\tilde{z}=-\frac{2}{3}R_s$ is a null surface being the white hole event horizon. The white hole event horizon may become an spacelike anti-trapping dynamical horizon if the dynamics further reduces $M_r$.

%The extension depends on a choice of slice $\cs_{t_0}$ in the close past of the blue diamond region, and the corresponding remnant mass $M_r$ of the black hole. There are several choices due to the existence of several phases

%\textcolor{red}{no need to assume the last ray is semiclassical. the semiclassical infinity of $\cs_{\tilde{t}_0}$ is enough.}

\section{Evidence of quantum tunneling}\label{Evidence of quantum tunneling}

The discussion in the last section leaves a question about how ${\rm dS}_2\times S^2$ and $\widetilde{{\rm dS}_2\times S^2}$ can be glued to make the black hole interior transit to the white hole interior. To understand this transition, we compute the improved Hamiltonian density at ${\rm dS}_2\times S^2$ 
\be
\rho=-\frac{\cc_\Delta}{E^\varphi\sqrt{E^x}}= \frac{1}{2 E^x}+\frac{\sin \left(2 \beta  \sqrt{\Delta } K_1\right) \sin \left(\beta  \sqrt{\Delta } K_2\right)}{\beta ^2 \Delta }+\frac{\sin ^2\left(\beta  \sqrt{\Delta } K_2\right)}{2 \beta ^2 \Delta },
\ee
where we have implemented that $\partial_xE^x=\partial^2_xE^x=0$. We notice that ${\bf H}_\Delta$ has a symmetry of reflection $K_x\to -K_x$ and $K_\varphi\to -K_\varphi$ which precisely is the relation between ${\rm dS}_2\times S^2$ and $\widetilde{{\rm dS}_2\times S^2}$. $\rho$ is also the density of Gaussian dust by Eqs.\Ref{rho111} and \Ref{cc111}.

\begin{figure}[h]
\begin{center}
\includegraphics[width = 0.9\textwidth]{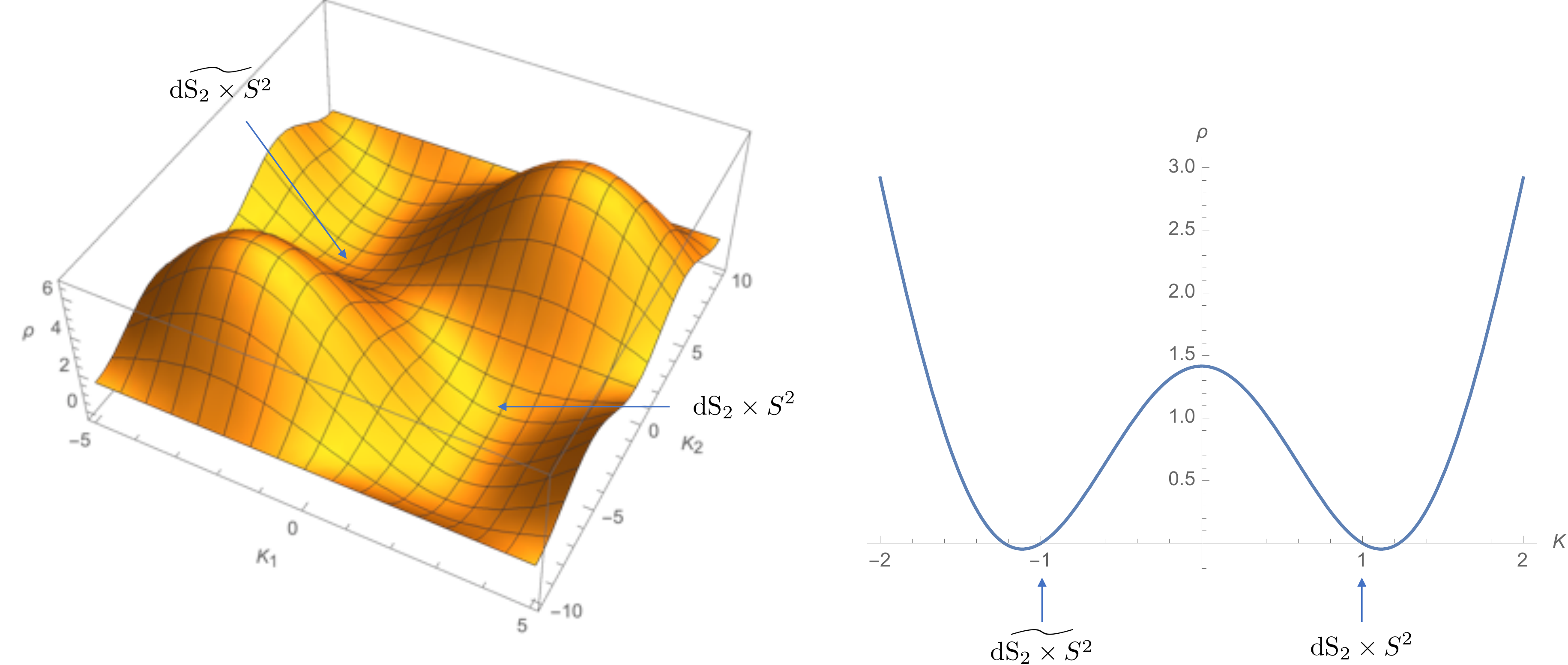} 
 \caption{The left: Plots of the Hamiltonian density $\rho$ as a function of $K_1,K_2$ in the region $2\b\sqrt{\Delta} K_1\in[-\pi,\pi]$, $\b\sqrt{\Delta} K_2\in[-\pi,\pi]$. The right: the cross section of the left along $K_1=2.4836 K, K_2= -1.8570 K$. ${\rm dS}_2\times S^2$ is located at $K=1$ while $\widetilde{{\rm dS}_2\times S^2}$ is at $K=-1$.}
\label{mexhat}
 \end{center}
\end{figure}

When fixing $E^x=r_0^2$, $\rho$ as a function of $K_1,K_2$ is plotted in Figure \ref{mexhat}, where we find that $\rho$ is a double-well effective potential. The values of constant $K_1,K_2$ for ${\rm dS}_2\times S^2$ and $\widetilde{{\rm dS}_2\times S^2}$ are located at two zeros of $\rho$ respectively. They are two points in the space of $K_1,K_2$ located respectively in two potential wells. Both ${\rm dS}_2\times S^2$ and $\widetilde{{\rm dS}_2\times S^2}$ are not the ground state of $\rho$ since they are not exactly at the minima of $\rho$, although they are close to the minima. The minima of $\rho$ corresponds to $\a_0\to\infty$ (by computing $\partial_{K_1}\rho=\partial_{K_2}\rho=0$ and applying Eq.\Ref{effeom4}), which is not possible to approach by the effective dynamics.

Figure \ref{mexhat} suggests an analog with the double-well potential in quantum mechanics, which is the standard example of demonstrating quantum tunneling from one potential well to the other. The low energy states of the double-well model are linear combinations of wave packets located respectively in 2 wells.

Both ${\rm dS}_2\times S^2$ and $\widetilde{{\rm dS}_2\times S^2}$ should be understood as quantum states (or wave packets) since $r_0,\a_0\sim\sqrt{\Delta}$, they should have nonzero quantum fluctuation in $K_x,K_\varphi$. In particular their $K_x$ fluctuation should be large since the fluctuation of $E^x$ should be small. The analogy with the double-well potential suggests that there should be  a quantum tunneling effect transiting from ${\rm dS}_2\times S^2$ to $\widetilde{{\rm dS}_2\times S^2}$. We conjecture that the quantum state at the gray region in Figure \ref{BHWH}(b) should be a sum over ${\rm dS}_2\times S^2$ and $\widetilde{{\rm dS}_2\times S^2}$, namely the state $|r_0,\a_0;\a_1\rangle$ in Eq.\Ref{VEV1} should be
\be
|r_0,\a_0;\a_1\rangle=\frac{1}{\sqrt{2}}\lt(|r_0,\a_0;\a_1\rangle_{{\rm dS}_2\times S^2}+|r_0,\a_0;\a_1\rangle_{\widetilde{{\rm dS}_2\times, S^2}}\rt).
\ee
which is well-posed since both ${\rm dS}_2\times S^2$ and $\widetilde{{\rm dS}_2\times S^2}$ share the same values of $\a_0,\a_1,r_0$. It is $|r_0,\a_0;\a_1\rangle$ that exists as the final state of the black hole and the initial state of the white hole.

%\footnote{$\langle r_0,\a_0;\a_1|\hat{E}^\varphi(t,x)|r_0,\a_0;\a_1\rangle=r_0 e^{-\a_1-\a_0^{-1}(x-t)}+r_0 e^{-\a_1-\a_0^{-1}(x+t)}$ where $x=-\tilde{x}$ and $t=\tilde{t}$ in the second term.}.

\begin{figure}[t]
\begin{center}
\includegraphics[width = 0.9\textwidth]{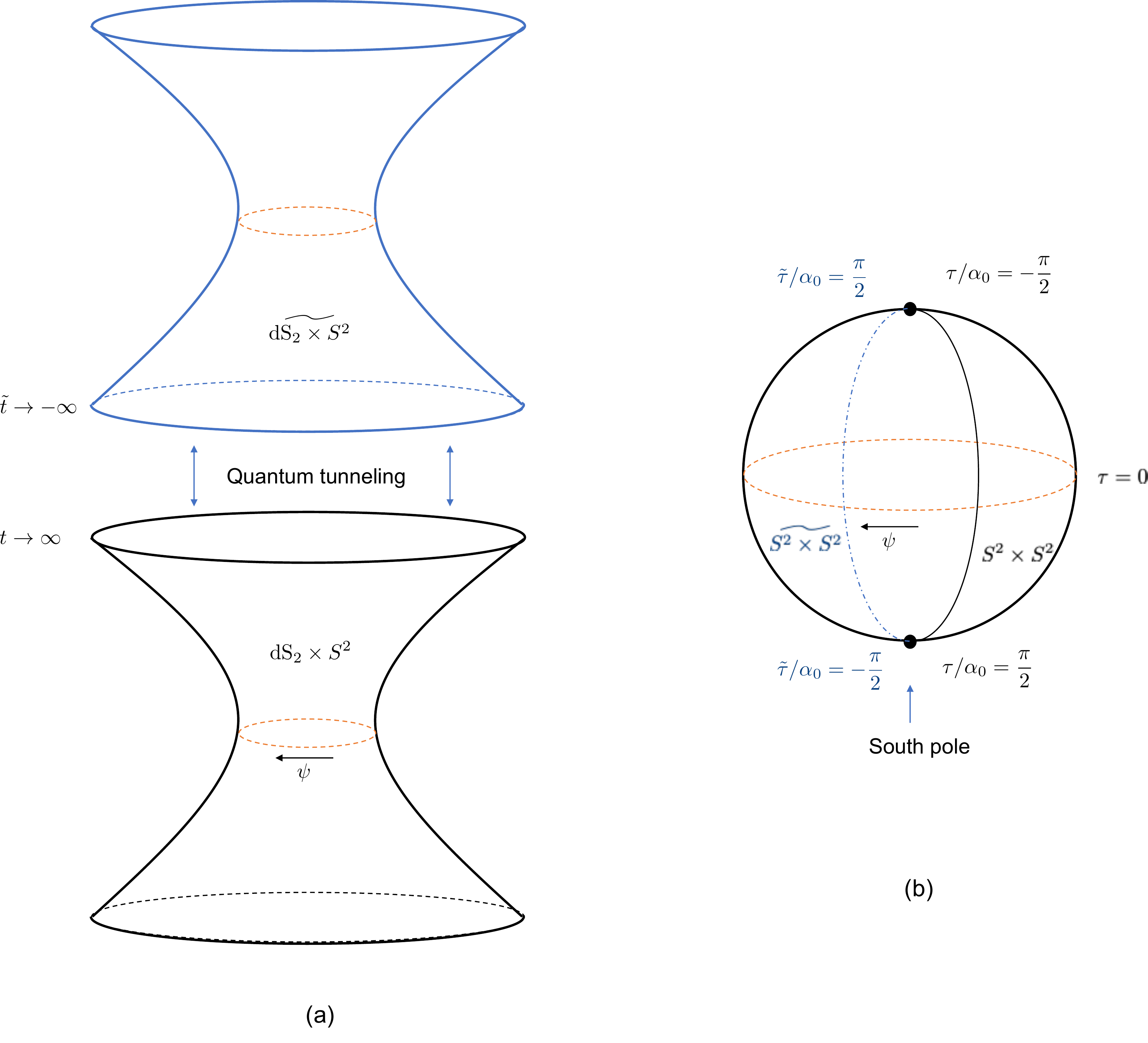} 
 \caption{(a) Gluing $t\to\infty$ of ${\rm dS}_2\times S^2$ to $\tilde{t}\to-\infty$ of $\widetilde{{\rm dS}_2\times S^2}$ is understood as the quantum tunneling (the $S^2$ factor is suppressed in this figure); (b) Both ${\rm dS}_2\times S^2$ and $\widetilde{{\rm dS}_2\times S^2}$ are analytic continued to the Euclidean $S^2\times S^2$. The $S^2\times S^2$ from ${\rm dS}_2\times S^2$ has $\tau/\a_0\in[-\pi/2,\pi/2]$ and $\psi\in[0,2\pi)$ (black half-circle is the $\t$-obit), while the $\widetilde{S^2\times S^2}$ from $\widetilde{{\rm dS}_2\times S^2}$ has $\tilde{\tau}/\a_0\in[-\pi/2,\pi/2]$ (blue dashed half-circle is the $\tilde{\t}$-obit). The transition from $S^2\times S^2$ to $\widetilde{S^2\times S^2}$ is at the south pole where the geometry is smooth.}
\label{Euclid}
 \end{center}
\end{figure}

This quantum tunneling may also be described by analytic continuing ${\rm dS}_2\times S^2$ to Euclidean signature (see e.g. \cite{Bousso:1999ms,Bousso:1996wz} for early works on quantum tunneling in the Nariai limit). We write the ${\rm dS}_2\times S^2$ metric in the global coordinate $\rmd s^2=-\rmd t^2+\a_0^2\cosh^2(t/\a_0)\rmd\psi^2+r_0^2[\rmd\theta^2+\sin^2(\theta)\rmd\varphi^2]$ where $\psi$ relates to $x$ by $\rmd\psi=\a_0^{-1}e^{-\a_1-\a_0^{-1}x}\rmd x$ (as $t\to\infty$). The analytic continuation $t\to i\tau$ gives 
\be
\rmd s^2\to\rmd s^2_E=\rmd \t^2+\a_0^2\cos^2(\t/\a_0)\rmd\psi^2+r_0^2\lt[\rmd\theta^2+\sin^2(\theta)\rmd\varphi^2\rt],
\ee
The geometry of the Euclidean metric is $S^2\times S^2$ whose radii are $\a_0$ and $r_0$. %\footnote{The charged Nariai spacetime ${\rm dS}\times S^2$ is the extremal limit of RN-dS whose outer black hole horizon radius approaches the radius of the cosmological horizon. The cosmological constant $\L=3/l_c^2$, charge $q$, and mass $\mu$ of the extremal RN-dS relate to $r_0,\a_0$ by $l_c=\frac{\sqrt{6} r_0}{\sqrt{1+{r_0}^2/\a_0^2}}$, $q=\frac{\sqrt{r_0^2- r_0^4/\a_0^2}}{\sqrt{2G}}$, $\mu=\frac{1}{3G} \left(2 r_0- r_0^3/\a_0^2\right)$. The dynamics in the black hole interior as $t\to\infty$ has an effective action $S_{eff}=\frac{1}{\kappa}\int R-2\L-F_{\mu\nu}F^{\mu\nu}$ while ${\rm dS}\times S^2$ is a saddle point. }. 
The coordinate transformation $\cos(\t/\a_0)=\sin(\chi)$ with $\t/\a_0\in[-\frac{\pi}{2},\frac{\pi}{2}]$ makes $\rmd s^2_E=\a_0^2\lt[\rmd \chi^2+\sin^2(\chi)\rmd\psi^2\rt]+r_0^2\lt[\rmd\theta^2+\sin^2(\theta)\rmd\varphi^2\rt]$. In the black hole interior, ${\rm dS}_2\times S^2$ and $\widetilde{{\rm dS}_2\times S^2}$ cannot be glued classically because the future boundary $t\to\infty$ slice of the former and the past boundary $\tilde{t}\to-\infty$ slice of the latter cannot be glued smoothly. However when we analytic continue to the Euclidean signature and denote their analytic continuation to be $S^2\times S^2$ and $\widetilde{S^2\times S^2}$, transiting from $S^2\times S^2$ to $\widetilde{S^2\times S^2}$ is a coordinate transformation $\tilde{\tau}/\a_0=-{\tau}/\a_0$, $\tilde{\psi}=-\psi$ in the first $S^2$. The ``future boundary'' of $S^2\times S^2$ at $\t/\a_0=\pi/2$ and ``past boundary'' of $\widetilde{S^2\times S^2}$ at $\tilde{\t}/\a_0=-\pi/2$ are glued at the the south pole where the geometry is smooth (see Figure \ref{Euclid}).

The above argument is based on the effective theory. At present we still do not have a derivation of this quantum tunneling from the quantum theory. An analysis of the quantization of ${\bf H}_\Delta$ is currently undergoing which is expected to provide more details of the quantum tunneling. %The result of this analysis will be reported elsewhere. 

\section{Chaos}\label{Chaos}

We study linear perturbations in $\widetilde{{\rm dS}_2\times S^2}$: 
\be
\tilde{K}_1(\tilde{t},\tilde{x})&=&\tilde{K}_1(\tilde{z}\to\infty)\lt[1+\eps \tilde{c}_1(\tilde{t},\tilde{x})\rt],\quad \tilde{K}_2(\tilde{t},\tilde{x})=\tilde{K}_2(\tilde{z}\to\infty)\lt[1+\eps \tilde{c}_2(\tilde{t},\tilde{x})\rt]\\
E^{\tilde{x}}(\tilde{t},\tilde{x})&=&-r_0^2\lt[1+\eps \tilde{p}_1(\tilde{t},\tilde{x})\rt],\quad E^{\tilde{\varphi}}(\tilde{t},\tilde{x})=r_0e^{-\a_1+\a_0^{-1}(\tilde{x}-\tilde{t})}\lt[1+\eps \tilde{p}_2(\tilde{t},\tilde{x})\rt]
\ee
where $\eps\ll1$. These perturbations satisfy the EOMs \Ref{effeomop1} - \Ref{effeomop4}, and can be obtained from perturbations in the earlier patch $({t},{x},{\theta},{\varphi})$ by the transformation \Ref{tran01} - \Ref{tran04}:
\be
\tilde{c}_1(\tilde{t},\tilde{x})&=&\tilde{c}_1(-{t},-{x})={c}_1({t},{x}),\label{cptrans1}\\
\tilde{c}_2(\tilde{t},\tilde{x})&=&\tilde{c}_2(-{t},-{x})={c}_2({t},{x}),\\
\tilde{p}_1(\tilde{t},\tilde{x})&=&\tilde{p}_1(-{t},-{x})={p}_1({t},{x}),\\
\tilde{p}_2(\tilde{t},\tilde{x})&=&\tilde{p}_2(-{t},-{x})={p}_2({t},{x}).\label{cptrans4}
\ee

If we apply to the numerical solution with parameters $\Delta=0.1,\ R_s=10^8,\ \b=1$, from Eqs.\Ref{pertu1} - \Ref{pertu3}, we obtain 
\be
\tilde{p}_1(\tilde{t},\tilde{x})&=& e^{0.586975 \tilde{t}} \big(\sin (-5.32462 \tilde{t})\lt[
0.203424   \tilde{f}_1(\tilde{x})
+0.187807   \tilde{f}_2(\tilde{x})
-0.0290504   \tilde{f}_4(\tilde{x})\rt]\nonumber\\
&&+\ \cos (5.32462 \tilde{t})\lt[0.154682  \tilde{f}_1(\tilde{x})
+0.263524  \tilde{f}_4(\tilde{x})\rt]\big)\nonumber\\
&&+\ e^{1.17395 \tilde{t}}\lt[0.845318 \tilde{f}_1(\tilde{x})-0.263524 \tilde{f}_4(\tilde{x})\rt],\\
\tilde{p}_2(\tilde{t},\tilde{x})&= &-0.291431 \tilde{f}_1(\tilde{x})+0.111783 \tilde{f}_2(\tilde{x})+ \tilde{f}_3(\tilde{x})+0.131762 \tilde{f}_4(\tilde{x})\nonumber\\
&&+\ e^{0.586975 \tilde{t}}\big(\sin (-5.32462 \tilde{t})\lt[0.0787198   \tilde{f}_1(\tilde{x})
-0.0123227   \tilde{f}_2(\tilde{x})
+0.158757  \tilde{f}_4(\tilde{x})\rt]\nonumber\\
&&-\ \cos (5.32462 \tilde{t})\lt[0.131228   \tilde{f}_1(\tilde{x})
+0.111783   \tilde{f}_2(\tilde{x})\rt]\big)\nonumber\\
&&+\ e^{1.17395 \tilde{t}}\lt[0.422659 \tilde{f}_1(\tilde{x})
-0.131762 \tilde{f}_4(\tilde{x})\rt],\\
\tilde{c}_1(\tilde{t},\tilde{x})&=& -0.181674 \partial_{\tilde{t}}\tilde{p}_1(\tilde{t},\tilde{x}),\quad \tilde{c}_2(\tilde{t},\tilde{x})= 0.572819 \partial_{\tilde{t}}\tilde{p}_2(\tilde{t},\tilde{x}),
\ee
where $\tilde{f}_1,\cdots,\tilde{f}_4$ are arbitrary functions of $\tilde{x}$. If initial perturbations are placed at any finite $\tilde{t}=\tilde{t}_0$, the time evolution of perturbations is chaotic, i.e. perturbations grow exponentially 
\be
\Big(\tilde{c}_1(\tilde{t}),\tilde{c}_2(\tilde{t}),\tilde{p}_1(\tilde{t}),\tilde{p}_2(\tilde{t})\Big)\sim e^{\l (\tilde{t}-\tilde{t}_0)}\Big(\tilde{c}_1(\tilde{t}_0),\tilde{c}_2(\tilde{t}_0),\tilde{p}_1(\tilde{t}_0),\tilde{p}_2(\tilde{t}_0)\Big),\label{chaospert1}
\ee
where $\l$ is the Lyapunov exponent and relates to the dS$_2$ radius $\a_0$ 
\be
\l\simeq \alpha_0^{-1}.
\ee
Numerically $\l\simeq 1.173954$ in this example with paramters $\Delta=0.1,\ R_s=10^8,\ \b=1$. The above relation between $\l$ and $\a_0$ is confirmed by various numerical tests with random choices of parameters. 

%The exponential growth of the perturbations does not contradict with the stability of ${\rm dS}_2\times S^2$ established in Section \ref{Stability of dSS2}. When initial perturbations is placed in the patch $(t,x,\theta,\varphi)$ at the early time of the black hole, propagations of perturbations remain bounded in $(\tilde{t},\tilde{x},\tilde{\theta},\tilde{\varphi})$ if initial perturbations are bounded, since the time evolution in $(\tilde{t},\tilde{x},\tilde{\theta},\tilde{\varphi})$ is just the spacetime inversion of the evolution in $(t,x,\theta,\varphi)$ by \Ref{cptrans1} - \Ref{cptrans4}. The time evolution of perturbations is the same as in Figure \ref{perturbations} but evolving in reverse direction. The exponential growth lasts only in a small period.

Because the computation that leads to the chaotic dynamics \Ref{chaospert1} is based on the effective dynamics which takes into account the quantum gravity effect, this result should indicate the quantum chaos on $\widetilde{{\rm dS}_2\times S^2}$. It should reflect that for certain expectation value of the squared commutator, e.g. 
\be
C(\tilde{t})&:=&-\lag\lt[\hat{E}^{\tilde{x}}(\tilde{t}),\hat{K}_{\tilde{x}}(\tilde{t}_0)\rt]^2\rag\\
&\sim &\hbar^2\lt\{{E}^{\tilde{x}}(\tilde{t}),{K}_{\tilde{x}}(\tilde{t}_0)\rt\}^2=G^2\hbar^2\lt(\frac{\delta {E}^{\tilde{x}}(\tilde{t})}{\delta{E}^{\tilde{x}}(\tilde{t}_0)}\rt)^2\sim G^2\hbar^2 e^{2\l (\tilde{t}-\tilde{t}_0)}.\label{Ct111}
\ee
The quantity $C(\tilde{t})$ is often called the out-of-time-order correlator, and considered as diagnosis of chaos in quantum systems \cite{Maldacena:2015waa,Roberts:2014ifa,Hosur:2015ylk}. Expectation values of commutators should relate to Poisson brackets in the effective dynamics \cite{Dapor:2017rwv,Alesci:2019pbs}, while the corresponding Poisson bracket gives the dependence of the final perturbation $\delta {E}^{\tilde{x}}(\tilde{t})$ on small changes in the initial perturbation $\delta{E}^{\tilde{x}}(\tilde{t}_0)$. The exponential grows in Eq.\Ref{Ct111} is the expected behavior of $C(\tilde{t})$ in a quantum chaotic system in early time (before the Ehrenfest time $\tilde{t}_s-\tilde{t}_0\sim \frac{1}{\l}\ln\frac{1}{G\hbar}$).

The dS temperature (the Hawking temperature at the cosmological horizon) relates to $\a_0$ by $T_{\rm dS}=\frac{1}{2\pi \a_0}$ \cite{Figari:1975km}. We relate the Lyapunov exponent to the dS temperature by
\be
\l\simeq 2\pi T_{\rm dS}.
\ee
This relation resembles the AdS/CFT black hole butterfly effect where the Lyapunov exponent $\l_{\rm CFT}$ of the boundary CFT relates to the black hole Hawking temperature by $\l_{\rm CFT}\simeq2\pi T_{\rm bh}$ \cite{Shenker:2013pqa,Maldacena:2015waa}. %Since here the dS temperature is Planckian $\a_0\sim\sqrt{\Delta}$, the Lyapunov exponent $\l\sim\Delta^{-1/2} $ is much larger than the one in the AdS/CFT.

%The exponential growth of the perturbations does not contradict with the stability of ${\rm dS}_2\times S^2$ established in Section \ref{Stability of dSS2}. When initial perturbations is placed in the patch $(t,x,\theta,\varphi)$ at the early time of the black hole, propagations of perturbations remain bounded in $(\tilde{t},\tilde{x},\tilde{\theta},\tilde{\varphi})$ if initial perturbations are bounded, since the time evolution in $(\tilde{t},\tilde{x},\tilde{\theta},\tilde{\varphi})$ is just the spacetime inversion of the evolution in $(t,x,\theta,\varphi)$ by \Ref{cptrans1} - \Ref{cptrans4}. The time evolution of perturbations is the same as in Figure \ref{perturbations} but evolving in reverse direction. 

The exponential growth lasts only in a small period near $\widetilde{{\rm dS}_2\times S^2}$. The time evolution in $(\tilde{t},\tilde{x},\tilde{\theta},\tilde{\varphi})$ is just the spacetime inversion of the evolution in $(t,x,\theta,\varphi)$ by \Ref{cptrans1} - \Ref{cptrans4}. The evolution of perturbations is the same as in Figure \ref{perturbations} but evolving in reverse direction.

\section{Infinitely many infrared states in ${\rm dS}_2\times S^2$}\label{Infinitely many infrared states}

Recall that the asymptotic ${\rm dS}_2\times S^2$ should be understood as a Hilbert space $\ch_{{\rm dS}_2\times S^2}$ spanned by states $|r_0,\a_0;\a_1\rangle$. By Eqs.\Ref{pertu1} - \Ref{p2final}, turning on perturbations changes the value of $\a_1$ although it leaves ${\rm dS}_2\times S^2$ geometry invariant. The perturbation defines an operator $\hat{O}_\eps$ on $\ch_{{\rm dS}_2\times S^2}$ by
\be
\hat{O}_\eps|r_0,\a_0;\a_1\rangle=|r_0,\a_0;\a_1+\eps \delta\a_1(x)\rangle,\quad\delta\a_1(x)= -\eps\lim_{t\to\infty}p_2(t,x)
\ee
The perturbation indicates that the state label $\a_1$ is generally a function $\a_1(x)$, although the background geometry has a constant $\a_1$. 

The numerics shows that the background geometry has approximately vanishing dust density $\rho$ throughout the evolution, given that the initial condition \Ref{bc0} and \Ref{bc1} corresponds to the vacuum Schwarzschild spacetime (there exists small numerical error, and $\rho$ is bounded by $\sim 10^{-6}$ throughout the evolution). But perturbations can make the dust density nonvanishing in principle. 

Even though perturbations can turn on the dust density, $\rho$ vanishes asymptotically in ${\rm dS}_2\times S^2$ no matter if perturbations are turned on or not. Indeed let's consider the PDEs \Ref{effeom1} - \Ref{effeom4} and the solution with constant $K_1,K_2,E^x\equiv r_0^2$ (these cannot be changed by perturbations). $\partial_tE^x{}=0$ and Eq.\Ref{effeom3} leads to 
\be
(a)\quad \cos \left(2 \beta  \sqrt{\Delta } K_1{} \right)=0\quad \text{or}\quad(b)\quad \sin \left(\beta  \sqrt{\Delta } K_2{} \right)=0
\ee
The option ($b$) is dropped since it reduces Eq.\Ref{effeom2} to $1/E^x=0$. The option $(a)$ and Eq.\Ref{effeom2} gives
\be
&&-\frac{1}{2 E^x{} }-\frac{\sin \left(2 \beta  \sqrt{\Delta } K_1{} \right) \sin \left(\beta  \sqrt{\Delta } K_2{} \right)}{\beta ^2 \Delta } -\frac{\sin ^2\left(\beta  \sqrt{\Delta } K_2{} \right)}{2 \beta ^2 \Delta }=0.
\ee
$\rho=\cc_\Delta/(|E^\varphi|\sqrt{|E^x|})$ (see Eqs.\Ref{rho111} and \Ref{cc111} in the dust coordinate) in ${\rm dS}_2\times S^2$ reduces to the above left-hand side by ignoring $\partial_xE^x{}$. On the other hand, in ${\rm dS}_2\times S^2$, we have $W_j=0$ since $W_j={P_j}/{\sqrt{\det(q)}}=\cc_j/{\sqrt{\det(q)}}=(\cc_x,0,0)/(|E^\varphi|\sqrt{|E^x|})$, and by constant $K_2,E^x$,
\be
\frac{\mathcal{C}_{x}}{|E^\varphi|\sqrt{|E^x|}}=\partial_xK_2+\frac{1}{2}\lt(K_2-2K_1\rt) \frac{\partial_x E^{x}}{ E^{x}}=0.
\ee
As a result, the dust stress-energy tensor always vanishes asymptotically 
\be
T_{\mu\nu}=0,\quad \text{in ${\rm dS}_2\times S^2$}.
\ee

%which is solved respectively by
%\be
%\sin \left(\beta  \sqrt{\Delta } K_2\right)= -\frac{\sqrt{E^x-\beta ^2 \Delta }}{\sqrt{E^x}}-1,\quad\text{or}\quad\sin \left(\beta  \sqrt{\Delta } K_2\right)= \frac{\sqrt{E^x-\beta ^2 \Delta }}{\sqrt{E^x}}-1
%\ee
%while the first solution less than $-1$ should be dropped. Solving Eqs.\Ref{approx2} and \Ref{approx3} gives
%\be
% K_1=\pm \frac{\pi}{4 \beta  \sqrt{\Delta }},\quad K_2=\frac{1}{\beta  \sqrt{\Delta } }\lt[\arcsin\lt(\frac{\sqrt{E^x-\beta ^2 \Delta }}{\sqrt{E^x}}-1\rt)+2k\pi\rt],\quad k={-1,0,1}
%\ee
%where $2 \beta  \sqrt{\Delta } K_1{},\beta  \sqrt{\Delta } K_2{}\in(-\pi,\pi]$.

\begin{figure}[h]
\begin{center}
\includegraphics[width = 0.6\textwidth]{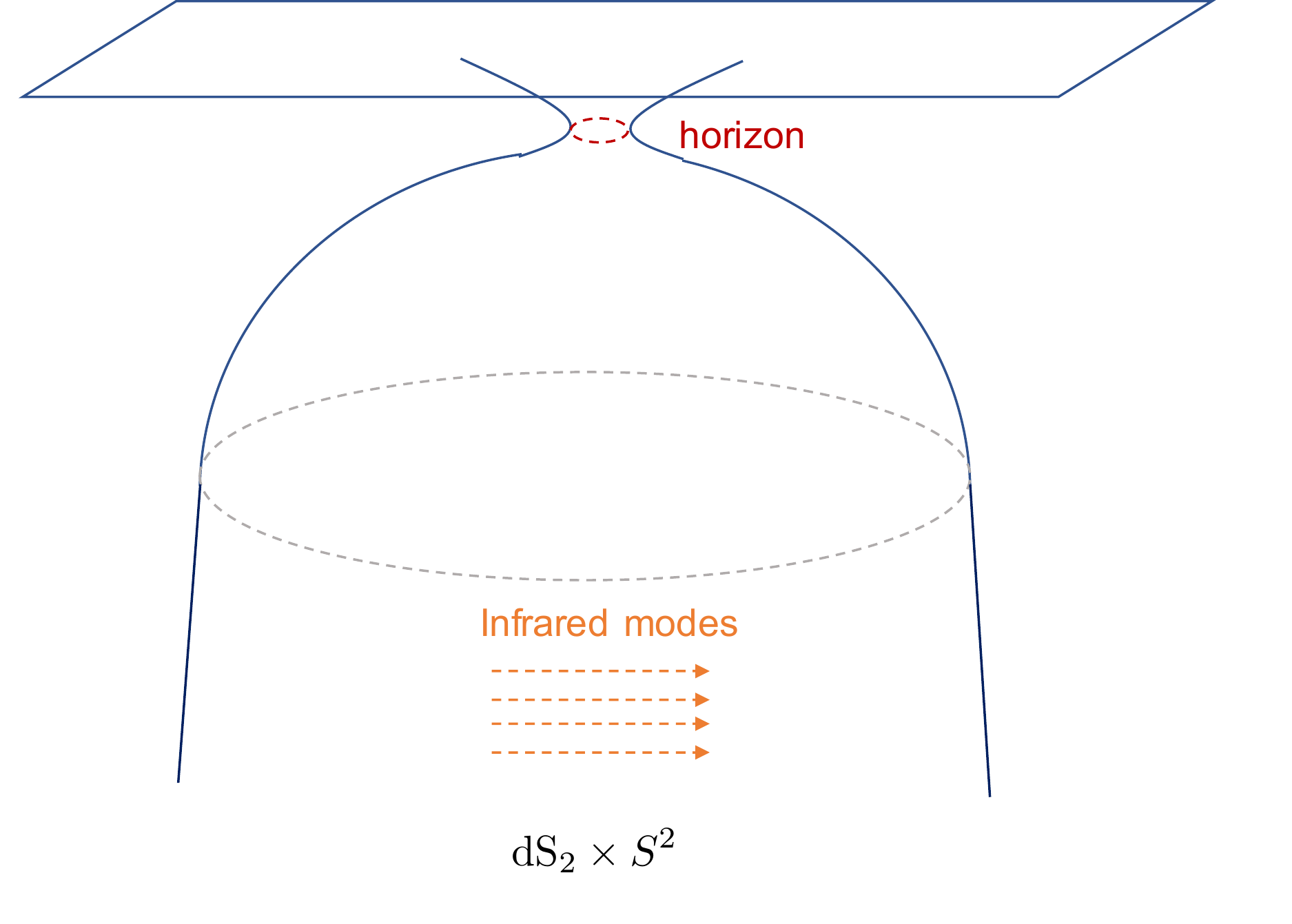} 
 \caption{The late time spatial slice (such as in Fig.\ref{smallrs} with small $R_s$) is Wheeler's bag of gold made by ${\rm dS}_2\times S^2$ containing infinitely many infrared modes.}
\label{baggold}
 \end{center}
\end{figure}

In the dynamics on the reduced phase space, $\rho=\cc_\Delta/(|E^\varphi|\sqrt{|E^x|})$ is the physical energy density since ${\bf H}_\Delta=\int\rmd x\, \cc_\Delta$ is the physical Hamiltonian. In terms of states $|r_0,\a_0;\a_1\rangle\in\ch_{\rm dS_2\times S^2}$, $\rho=0$ is understood as expectation values $\langle r_0,\a_0;\a_1|\hat{\rho}|r_0,\a_0;\a_1\rangle=0$, which means that all states $|r_0,\a_0;\a_1\rangle$ in $\ch_{\rm dS_2\times S^2}$ are infrared soft modes. In particular they have no back-reaction to ${\rm dS}_2\times S^2$. The black hole interior containing infinitely many infrared states are anticipated in existing studies of quantum black holes (see e.g. \cite{Ashtekar:2020ifw} for a summary). $\rho\to0$ relates to the exponentially large spatial volume in ${\rm dS}_2\times S^2$ as $t\to\infty$ (recall Eq.\Ref{ds2s2}). But the spatial slice has a very narrow throat, since $S^2$ area is small in the middle see Figures \ref{s2area} and \ref{areabound} and the horizon radius $R_s$ becomes small at late time. ${\rm dS}_2\times S^2$ gives an example of Wheeler's bag of gold (see Figure \ref{baggold}).

The diffeomorphisms in $x$-space leaves ${\rm dS}_2\times S^2$ invariant but changes $\a_1$. In quantum notation, the diffeomorphisms generated by $\cv(N)=\int\rmd x\, N(x)\cc(x)$ define operators acting on infrared states
\be
\widehat{e^{i\eps\cv(N)}}|r_0,\a_0;\a_1\rangle=|r_0,\a_0;\a'_1\rangle,\quad \a'_1(x)=\a_1(x)+\eps\a_0^{-1}N(x)+\eps\partial_x N(x)
\ee
where $\eps\in\R$ is an infinitesimal parameter. $\a'_1(x)$ comes from the coordinate transformation $x\to x'=x+\eps N(x)$ and $\rmd\eta =e^{-\a_1-\a_0^{-1} x}\rmd x=e^{-\a_1'-\a_0^{-1} x'}\rmd x'$. In the reduced phase space formulation, the diffeomorphisms are not gauge redundancy but symmetries of the theory. We find the Hilbert space $\ch_{\rm dS_2\times S^2}$ of infrared states is a representation space of the group of 1-dimensional diffeomorphisms. The $x$-space is $S^1$ in ${\rm dS_2}\times {S^2}$ as $t\to\infty$. %since the time coordinates in the inflationary and global coordinates of ${\rm dS}_2$ approach to coincide as $t\to\infty$ (so their time slices approach to coincide)\footnote{In the global coordinate $t_g,\theta$, the ${\rm dS}_2$ metric reads $d s^{2}=-d t_g^{2}+\ell^{2} \cosh ^{2}(t / \a_0)d \theta^{2}$, and in the inflationary coordinate $(t,\eta)$, $d s^{2}=-d t^{2}+e^{2 t / \a_0}d \eta^{2}$. The coordinate transformation is $e^{t / \a_0}=\cos \theta \cosh \left(t_{g} / \a_0\right)+\sinh \left(t_{g} / \a_0\right), \  \frac{\a_0}{\eta} e^{t / \a_0}=\sin \theta \cosh \left(t_{g} / \a_0\right)$.  }. 
Therefore $\ch_{\rm dS_2\times S^2}$ carries a representation of ${\rm Diff}(S^1)$ or equivalently $\ch_{\rm dS_2\times S^2}$ carries a representation of Witt algebra:
\be
\lt[L_m,L_n\rt]=(m-n)L_{m+n},\quad L_n=-ie^{i n \theta}\frac{\partial}{\partial\theta},
\ee 
or Virasoro algebra if we generally allow nontrivial central extension. %As a representation of Witt or Virasoro algebra, $\ch_{\rm dS_2\times S^2}$ is the Hilbert space of a 2-dimensional conformal field theory. 
$L_n$ as infinitesimal diffeomorphisms give infinitely many conserved charges (recall Eq.\Ref{charge1111}). 

%The Hilbert space of infrared states and infinite conserved charges might have relation with black hole's soft hair \cite{Hawking:2016sgy}, since the origin of soft hair is diffeomorphism at the spacetime boundary, while here the time slice in $\ch_{\rm dS_2\times S^2}$ as $t\to\infty$ is part of the future boundary of the effective spacetime. 

%If we view ${\rm dS}_2\times S^2$ is a ``final state'' of the black hole, it has infinite degeneracy due to the existence of $\cf(\ch_{{\rm dS}_2\times S^2})$ including perturbations without backreaction. States in $\cf(\ch_{{\rm dS}_2\times S^2})$ might be seen as analog of Goldstone modes on the degenerate ground state. 

%It indicates that the dust stress-energy tensor always vanishes in ${\rm dS}_2\times S^2$ although perturbations can turn on dust density elsewhere. In general, linear perturbations here relate to both the perturbation of dust field and gravity shock wave. Since perturbations doesn't affect the asymptotic ${\rm dS}_2\times S^2$ geometry, we conclude that all these perturbations give vanishing stress-energy tensor in ${\rm dS}_2\times S^2$.

\section{Outlook}\label{outlook}

The analysis of this work opens new windows of developments in 3 phases of quantum black hole dynamics. These 3 phases are (1) from black hole to the Nariai limit, (2) near the Nariai limit, and (3) from the Nariai limit to white hole. The future analysis of these 3 phases needs the upgrade from the present effective dynamics to a quantum operator formulation, which is a research undergoing.  

As an advantage of the reduced phase space formulation, a proper quantization of the physical Hamiltonian ${\bf H}_\Delta$ generates manifestly unitary time evolution. In the phase (1) from black hole to the Nariai limit, it is interesting to investigate the thermalization predicted from ${\bf H}_\Delta$, for instance, questions like whether we can find local observables that thermalize after the formation of the black hole, and if they relates to the Hawking radiation. A standard formalism of addressing these question is the Eigenstate Thermalization Hypothesis (ETH), whose purpose is to explain how (local) thermal equilibrium can be achieved by quantum evolutions from initially far-from-equilibrium states. A wide variety of many-body systems are shown to satisfy ETH, suggesting thermalization should be a generic feature for interacting quantum system (see e.g.\cite{Rigol_2008}). We expect that the ${\bf H}_\Delta$ should lead to thermalization of certain local observables. Moreover, thermalization often combines the quantum chaos and information scrambling \cite{Srednicki_1999} which are other perspectives to be investigated. In addition, ${\bf H}_\Delta$ may be related to the recent studies in \cite{Liu:2020jsv} on equilibrated pure states after long-time unitary evolution, in order to understand if the long-time unitary evolution of ${\bf H}_\Delta$ can be approximately typical and lead equilibrated pure states as outputs. The entanglement entropy of the final state might give an explain of the replica wormholes and page curves following the line of \cite{Liu:2020jsv}.

In the phase (2), it is important to carry out careful analysis for the expected quantum tunneling from the viewpoint of the quantum ${\bf H}_\Delta$, as mentioned earlier. On the other hand, the infrared modes in the Nariai limit and the Hilbert space $\ch_{\rm{dS}_2\times S^2}$ should be analyzed in more rigorous manner. It is interesting to describe $\ch_{\rm{dS}_2\times S^2}$ in terms of representations of $\mathrm{Diff}(S^1)$, and also understand the states as resulting from the unitary evolution of ${\bf H}_\Delta$. We may also looking for their relation with the equilibrated states. In addition, $\ch_{\rm{dS}_2\times S^2}$ might be embedded in the language of quantum error correcting code, as the code subspace similar as in \cite{Penington:2019kki}. It might also relate to states of baby universes \cite{Marolf:2020rpm}. There are debates about whether the modes in the Wheeler's bag of gold form infinitely dimensional Hilbert space, or these infrared modes inside the black hole are linearly dependent (The recent progresses from the AdS/CFT suggests that the Hilbert space may be actually 1-dimensional, see e.g. \cite{Hsin:2020mfa}). The unitary dynamics of ${\bf H}_\Delta$ should help to clarify the dimension of $\ch_{\rm{dS}_2\times S^2}$ of the infrared modes.

Rigorously speaking, the existence of the phase (3) relies on the precise description of the phase (2). We expect that the Nariai geometry ${\rm dS}_2\times S^2$ is not stable at the quantum level. Its decay and relation to the white hole requires to be further analyzed in detail. It is also interesting to investigate the dynamics of the infrared modes in ${\rm dS}_2\times S^2$ toward the white hole and the asymptotically flat regime. As discussed earlier, we expect this dynamics to be highly chaotic. It is interesting to understand the chaos in the full quantum theory instead of the effective theory. The detailed analysis of the black-hole-to-white-hole transition should shed light on the resolution of information paradox, given that our discussion is based on the unitary evolution of ${\bf H}_\Delta$.

Another important aspect is the strong quantum dynamical regime in the blue diamond in Figure \ref{BHWH}. The analysis may be carried by the quantization of ${\bf H}_\Delta$, or may require the full theory of LQG (see \cite{DAmbrosio:2020mut} for a discussion based on spinfoams). We plan to apply the full theory of LQG to black holes, preferably using the new path integral formulation similar to the recent works on cosmology \cite{Han:2019vpw,Han:2020chr,Han:2020iwk}.

%\begin{acknowledgments}

\section*{Acknowledgements}

%The author acknowledges useful discussions with Andrea Dapor and Hongguang Liu. 

This work receives support from the National Science Foundation through grant PHY-1912278. 

%Mathematica computations in this work are carried out on the HPC server at Fudan University and the KoKo HPC server at Florida Atlantic University. The authors acknowledge Ling-Yan Hung for sharing the computational resource at Fudan University. 
	
%\end{acknowledgments}

\bibliographystyle{jhep}

\bibliography{muxin}

\providecommand{\href}[2]{#2}\begingroup\raggedright\begin{thebibliography}{10}

\bibitem{Bohmer:2007wi}
C.~G. Boehmer and K.~Vandersloot, {\it {Loop Quantum Dynamics of the
  Schwarzschild Interior}},  {\em Phys. Rev. D} {\bf 76} (2007) 104030,
  [\href{http://arxiv.org/abs/0709.2129}{{\tt arXiv:0709.2129}}].

\bibitem{Ashtekar:2005qt}
A.~Ashtekar and M.~Bojowald, {\it {Quantum geometry and the Schwarzschild
  singularity}},  {\em Class. Quant. Grav.} {\bf 23} (2006) 391--411,
  [\href{http://arxiv.org/abs/gr-qc/0509075}{{\tt gr-qc/0509075}}].

\bibitem{Modesto:2005zm}
L.~Modesto, {\it {Loop quantum black hole}},  {\em Class. Quant. Grav.} {\bf
  23} (2006) 5587--5602, [\href{http://arxiv.org/abs/gr-qc/0509078}{{\tt
  gr-qc/0509078}}].

\bibitem{Ashtekar:2010qz}
A.~Ashtekar, F.~Pretorius, and F.~M. Ramazanoglu, {\it {Evaporation of
  2-Dimensional Black Holes}},  {\em Phys. Rev. D} {\bf 83} (2011) 044040,
  [\href{http://arxiv.org/abs/1012.0077}{{\tt arXiv:1012.0077}}].

\bibitem{Chiou:2012pg}
D.-W. Chiou, W.-T. Ni, and A.~Tang, {\it {Loop quantization of spherically
  symmetric midisuperspaces and loop quantum geometry of the maximally extended
  Schwarzschild spacetime}},  \href{http://arxiv.org/abs/1212.1265}{{\tt
  arXiv:1212.1265}}.

\bibitem{Gambini:2013hna}
R.~Gambini, J.~Olmedo, and J.~Pullin, {\it {Quantum black holes in Loop Quantum
  Gravity}},  {\em Class. Quant. Grav.} {\bf 31} (2014) 095009,
  [\href{http://arxiv.org/abs/1310.5996}{{\tt arXiv:1310.5996}}].

\bibitem{Bianchi:2018mml}
E.~Bianchi, M.~Christodoulou, F.~D'Ambrosio, H.~M. Haggard, and C.~Rovelli,
  {\it {White Holes as Remnants: A Surprising Scenario for the End of a Black
  Hole}},  {\em Class. Quant. Grav.} {\bf 35} (2018), no.~22 225003,
  [\href{http://arxiv.org/abs/1802.04264}{{\tt arXiv:1802.04264}}].

\bibitem{DAmbrosio:2020mut}
F.~D'Ambrosio, M.~Christodoulou, P.~Martin-Dussaud, C.~Rovelli, and F.~Soltani,
  {\it {The End of a Black Hole\textquoteright{}s Evaporation \textendash{}
  Part I}},  \href{http://arxiv.org/abs/2009.05016}{{\tt arXiv:2009.05016}}.

\bibitem{Olmedo:2017lvt}
J.~Olmedo, S.~Saini, and P.~Singh, {\it {From black holes to white holes: a
  quantum gravitational, symmetric bounce}},  {\em Class. Quant. Grav.} {\bf
  34} (2017), no.~22 225011, [\href{http://arxiv.org/abs/1707.07333}{{\tt
  arXiv:1707.07333}}].

\bibitem{Ashtekar:2018cay}
A.~Ashtekar, J.~Olmedo, and P.~Singh, {\it {Quantum extension of the Kruskal
  spacetime}},  {\em Phys. Rev.} {\bf D98} (2018), no.~12 126003,
  [\href{http://arxiv.org/abs/1806.02406}{{\tt arXiv:1806.02406}}].

\bibitem{Bojowald:2018xxu}
M.~Bojowald, S.~Brahma, and D.-h. Yeom, {\it {Effective line elements and
  black-hole models in canonical loop quantum gravity}},  {\em Phys. Rev.} {\bf
  D98} (2018), no.~4 046015, [\href{http://arxiv.org/abs/1803.01119}{{\tt
  arXiv:1803.01119}}].

\bibitem{Bodendorfer:2019cyv}
N.~Bodendorfer, F.~M. Mele, and J.~M{\"u}nch, {\it {Effective Quantum Extended
  Spacetime of Polymer Schwarzschild Black Hole}},  {\em Class. Quant. Grav.}
  {\bf 36} (2019), no.~19 195015, [\href{http://arxiv.org/abs/1902.04542}{{\tt
  arXiv:1902.04542}}].

\bibitem{Alesci:2019pbs}
E.~Alesci, S.~Bahrami, and D.~Pranzetti, {\it {Quantum gravity predictions for
  black hole interior geometry}},  {\em Phys. Lett. B} {\bf 797} (2019) 134908,
  [\href{http://arxiv.org/abs/1904.12412}{{\tt arXiv:1904.12412}}].

\bibitem{Assanioussi:2019twp}
M.~Assanioussi, A.~Dapor, and K.~Liegener, {\it {Perspectives on the dynamics
  in a loop quantum gravity effective description of black hole interiors}},
  {\em Phys. Rev. D} {\bf 101} (2020), no.~2 026002,
  [\href{http://arxiv.org/abs/1908.05756}{{\tt arXiv:1908.05756}}].

\bibitem{Kelly:2020lec}
J.~G. Kelly, R.~Santacruz, and E.~Wilson-Ewing, {\it {Black hole collapse and
  bounce in effective loop quantum gravity}},
  \href{http://arxiv.org/abs/2006.09325}{{\tt arXiv:2006.09325}}.

\bibitem{Gambini:2020nsf}
R.~Gambini, J.~Olmedo, and J.~Pullin, {\it {Spherically symmetric loop quantum
  gravity: analysis of improved dynamics}},
  \href{http://arxiv.org/abs/2006.01513}{{\tt arXiv:2006.01513}}.

\bibitem{Ashtekar:2020ifw}
A.~Ashtekar, {\it {Black Hole evaporation: A Perspective from Loop Quantum
  Gravity}},  {\em Universe} {\bf 6} (2020), no.~2 21,
  [\href{http://arxiv.org/abs/2001.08833}{{\tt arXiv:2001.08833}}].

\bibitem{Bojowald:2001xe}
M.~Bojowald, {\it {Absence of singularity in loop quantum cosmology}},  {\em
  Phys. Rev. Lett.} {\bf 86} (2001) 5227--5230,
  [\href{http://arxiv.org/abs/gr-qc/0102069}{{\tt gr-qc/0102069}}].

\bibitem{Ashtekar:2006wn}
A.~Ashtekar, T.~Pawlowski, and P.~Singh, {\it {Quantum Nature of the Big Bang:
  Improved dynamics}},  {\em Phys. Rev.} {\bf D74} (2006) 084003,
  [\href{http://arxiv.org/abs/gr-qc/0607039}{{\tt gr-qc/0607039}}].

\bibitem{Rovelli:2013zaa}
C.~Rovelli and E.~Wilson-Ewing, {\it {Why are the effective equations of loop
  quantum cosmology so accurate?}},  {\em Phys. Rev. D} {\bf 90} (2014), no.~2
  023538, [\href{http://arxiv.org/abs/1310.8654}{{\tt arXiv:1310.8654}}].

\bibitem{Bousso:1996pn}
R.~Bousso, {\it {Charged Nariai black holes with a dilaton}},  {\em Phys. Rev.
  D} {\bf 55} (1997) 3614--3621,
  [\href{http://arxiv.org/abs/gr-qc/9608053}{{\tt gr-qc/9608053}}].

\bibitem{Kuchar:1990vy}
K.~V. Kuchar and C.~G. Torre, {\it {Gaussian reference fluid and interpretation
  of quantum geometrodynamics}},  {\em Phys. Rev.} {\bf D43} (1991) 419--441.

\bibitem{Giesel:2012rb}
K.~Giesel and T.~Thiemann, {\it {Scalar Material Reference Systems and Loop
  Quantum Gravity}},  {\em Class. Quant. Grav.} {\bf 32} (2015) 135015,
  [\href{http://arxiv.org/abs/1206.3807}{{\tt arXiv:1206.3807}}].

\bibitem{Giesel:2009jp}
K.~Giesel, J.~Tambornino, and T.~Thiemann, {\it {LTB spacetimes in terms of
  Dirac observables}},  {\em Class. Quant. Grav.} {\bf 27} (2010) 105013,
  [\href{http://arxiv.org/abs/0906.0569}{{\tt arXiv:0906.0569}}].

\bibitem{Munch:2020czs}
J.~M\"unch, {\it {Effective Quantum Dust Collapse via Surface Matching}},
  \href{http://arxiv.org/abs/2010.13480}{{\tt arXiv:2010.13480}}.

\bibitem{Hawking:1995ap}
S.~Hawking and S.~F. Ross, {\it {Duality between electric and magnetic black
  holes}},  {\em Phys. Rev. D} {\bf 52} (1995) 5865--5876,
  [\href{http://arxiv.org/abs/hep-th/9504019}{{\tt hep-th/9504019}}].

\bibitem{Bousso:1999ms}
R.~Bousso, {\it {Quantum global structure of de Sitter space}},  {\em Phys.
  Rev. D} {\bf 60} (1999) 063503,
  [\href{http://arxiv.org/abs/hep-th/9902183}{{\tt hep-th/9902183}}].

\bibitem{Bousso:1996wz}
R.~Bousso and S.~W. Hawking, {\it {Pair creation and evolution of black holes
  in inflation}},  {\em Helv. Phys. Acta} {\bf 69} (1996) 261--264,
  [\href{http://arxiv.org/abs/gr-qc/9608008}{{\tt gr-qc/9608008}}].

\bibitem{Boehmer:2008fz}
C.~G. Boehmer and K.~Vandersloot, {\it {Stability of the Schwarzschild Interior
  in Loop Quantum Gravity}},  {\em Phys. Rev. D} {\bf 78} (2008) 067501,
  [\href{http://arxiv.org/abs/0807.3042}{{\tt arXiv:0807.3042}}].

\bibitem{Rovelli:2014cta}
C.~Rovelli and F.~Vidotto, {\it {Planck stars}},  {\em Int. J. Mod. Phys. D}
  {\bf 23} (2014), no.~12 1442026, [\href{http://arxiv.org/abs/1401.6562}{{\tt
  arXiv:1401.6562}}].

\bibitem{Shenker:2013pqa}
S.~H. Shenker and D.~Stanford, {\it {Black holes and the butterfly effect}},
  {\em JHEP} {\bf 03} (2014) 067, [\href{http://arxiv.org/abs/1306.0622}{{\tt
  arXiv:1306.0622}}].

\bibitem{Maldacena:2015waa}
J.~Maldacena, S.~H. Shenker, and D.~Stanford, {\it {A bound on chaos}},  {\em
  JHEP} {\bf 08} (2016) 106, [\href{http://arxiv.org/abs/1503.01409}{{\tt
  arXiv:1503.01409}}].

\bibitem{holst}
S.~Holst, {\it {Barbero's Hamiltonian derived from a generalized
  Hilbert-Palatini action}},  {\em Phys.Rev.} {\bf D53} (1996) 5966--5969,
  [\href{http://arxiv.org/abs/gr-qc/9511026}{{\tt gr-qc/9511026}}].

\bibitem{Dittrich:2004cb}
B.~Dittrich, {\it {Partial and complete observables for Hamiltonian constrained
  systems}},  {\em Gen. Rel. Grav.} {\bf 39} (2007) 1891--1927,
  [\href{http://arxiv.org/abs/gr-qc/0411013}{{\tt gr-qc/0411013}}].

\bibitem{Thiemann:2004wk}
T.~Thiemann, {\it {Reduced phase space quantization and Dirac observables}},
  {\em Class. Quant. Grav.} {\bf 23} (2006) 1163--1180,
  [\href{http://arxiv.org/abs/gr-qc/0411031}{{\tt gr-qc/0411031}}].

\bibitem{Giesel:2007wn}
K.~Giesel and T.~Thiemann, {\it {Algebraic quantum gravity (AQG). IV. Reduced
  phase space quantisation of loop quantum gravity}},  {\em Class. Quant.
  Grav.} {\bf 27} (2010) 175009, [\href{http://arxiv.org/abs/0711.0119}{{\tt
  arXiv:0711.0119}}].

\bibitem{Han:2020chr}
M.~Han and H.~Liu, {\it {Semiclassical limit of new path integral formulation
  from reduced phase space loop quantum gravity}},  {\em Phys. Rev. D} {\bf
  102} (2020), no.~2 024083, [\href{http://arxiv.org/abs/2005.00988}{{\tt
  arXiv:2005.00988}}].

\bibitem{Han:2019feb}
M.~Han and H.~Liu, {\it {Improved $\overline{\mu}$-scheme effective dynamics of
  full loop quantum gravity}},  {\em Phys. Rev. D} {\bf 102} (2020), no.~6
  064061, [\href{http://arxiv.org/abs/1912.08668}{{\tt arXiv:1912.08668}}].

\bibitem{Singh:2013ava}
P.~Singh and E.~Wilson-Ewing, {\it {Quantization ambiguities and bounds on
  geometric scalars in anisotropic loop quantum cosmology}},  {\em Class.
  Quant. Grav.} {\bf 31} (2014) 035010,
  [\href{http://arxiv.org/abs/1310.6728}{{\tt arXiv:1310.6728}}].

\bibitem{Bojowald:2005cb}
M.~Bojowald and R.~Swiderski, {\it {Spherically symmetric quantum geometry:
  Hamiltonian constraint}},  {\em Class. Quant. Grav.} {\bf 23} (2006)
  2129--2154, [\href{http://arxiv.org/abs/gr-qc/0511108}{{\tt gr-qc/0511108}}].

\bibitem{Ashtekar:2007em}
A.~Ashtekar, A.~Corichi, and P.~Singh, {\it {Robustness of key features of loop
  quantum cosmology}},  {\em Phys. Rev. D} {\bf 77} (2008) 024046,
  [\href{http://arxiv.org/abs/0710.3565}{{\tt arXiv:0710.3565}}].

\bibitem{github}
M.~Han. \url{https://github.com/LQG-Florida-Atlantic-University/black-holes},
  2020.

\bibitem{Ashtekar:2003hk}
A.~Ashtekar and B.~Krishnan, {\it {Dynamical horizons and their properties}},
  {\em Phys. Rev. D} {\bf 68} (2003) 104030,
  [\href{http://arxiv.org/abs/gr-qc/0308033}{{\tt gr-qc/0308033}}].

\bibitem{Haggard:2014rza}
H.~M. Haggard and C.~Rovelli, {\it {Quantum-gravity effects outside the horizon
  spark black to white hole tunneling}},  {\em Phys. Rev. D} {\bf 92} (2015),
  no.~10 104020, [\href{http://arxiv.org/abs/1407.0989}{{\tt
  arXiv:1407.0989}}].

\bibitem{Roberts:2014ifa}
D.~A. Roberts and D.~Stanford, {\it {Two-dimensional conformal field theory and
  the butterfly effect}},  {\em Phys. Rev. Lett.} {\bf 115} (2015), no.~13
  131603, [\href{http://arxiv.org/abs/1412.5123}{{\tt arXiv:1412.5123}}].

\bibitem{Hosur:2015ylk}
P.~Hosur, X.-L. Qi, D.~A. Roberts, and B.~Yoshida, {\it {Chaos in quantum
  channels}},  {\em JHEP} {\bf 02} (2016) 004,
  [\href{http://arxiv.org/abs/1511.04021}{{\tt arXiv:1511.04021}}].

\bibitem{Dapor:2017rwv}
A.~Dapor and K.~Liegener, {\it {Cosmological Effective Hamiltonian from full
  Loop Quantum Gravity Dynamics}},  {\em Phys. Lett.} {\bf B785} (2018)
  506--510, [\href{http://arxiv.org/abs/1706.09833}{{\tt arXiv:1706.09833}}].

\bibitem{Figari:1975km}
R.~Figari, R.~Hoegh-Krohn, and C.~Nappi, {\it {Interacting Relativistic Boson
  Fields in the de Sitter Universe with Two Space-Time Dimensions}},  {\em
  Commun. Math. Phys.} {\bf 44} (1975) 265--278.

\bibitem{Rigol_2008}
M.~Rigol, V.~Dunjko, and M.~Olshanii, {\it Thermalization and its mechanism for
  generic isolated quantum systems},  {\em Nature} {\bf 452} (Apr, 2008)
  854–858.

\bibitem{Srednicki_1999}
M.~Srednicki, {\it The approach to thermal equilibrium in quantized chaotic
  systems},  {\em Journal of Physics A: Mathematical and General} {\bf 32}
  (Jan, 1999) 1163–1175.

\bibitem{Liu:2020jsv}
H.~Liu and S.~Vardhan, {\it {Entanglement entropies of equilibrated pure states
  in quantum many-body systems and gravity}},
  \href{http://arxiv.org/abs/2008.01089}{{\tt arXiv:2008.01089}}.

\bibitem{Penington:2019kki}
G.~Penington, S.~H. Shenker, D.~Stanford, and Z.~Yang, {\it {Replica wormholes
  and the black hole interior}},  \href{http://arxiv.org/abs/1911.11977}{{\tt
  arXiv:1911.11977}}.

\bibitem{Marolf:2020rpm}
D.~Marolf and H.~Maxfield, {\it {Observations of Hawking radiation: the Page
  curve and baby universes}},  \href{http://arxiv.org/abs/2010.06602}{{\tt
  arXiv:2010.06602}}.

\bibitem{Hsin:2020mfa}
P.-S. Hsin, L.~V. Iliesiu, and Z.~Yang, {\it {A violation of global symmetries
  from replica wormholes and the fate of black hole remnants}},
  \href{http://arxiv.org/abs/2011.09444}{{\tt arXiv:2011.09444}}.

\bibitem{Han:2019vpw}
M.~Han and H.~Liu, {\it {Effective Dynamics from Coherent State Path Integral
  of Full Loop Quantum Gravity}},  {\em Phys. Rev.} {\bf D101} (2020), no.~4
  046003, [\href{http://arxiv.org/abs/1910.03763}{{\tt arXiv:1910.03763}}].

\bibitem{Han:2020iwk}
M.~Han, H.~Li, and H.~Liu, {\it {Manifestly Gauge-Invariant Cosmological
  Perturbation Theory from Full Loop Quantum Gravity}},  5, 2020.
\newblock \href{http://arxiv.org/abs/2005.00883}{{\tt arXiv:2005.00883}}.

\end{thebibliography}\endgroup

\end{document}